%% file: main.tex
\documentclass[11pt]{article}
\usepackage[numbers]{natbib}   %
\usepackage{fullpage}
\usepackage{enumitem}
\usepackage{url}
\usepackage[T1]{fontenc}

\def\mbafont{lora}

\makeatletter
\def\mbafont@lora{lora}\def\mbafont@lato{lato}\def\mbafont@times{times}
\ifx\mbafont\mbafont@lora
  \usepackage[lining,tabular]{font/Lora}
  \DeclareRobustCommand{\&}{{\upshape\char`\&}}  %
\else\ifx\mbafont\mbafont@lato
  \usepackage[default,lining,tabular]{lato}
\else\ifx\mbafont\mbafont@times
  \usepackage{newtxtext}
  \usepackage{newtxmath}
\else
  \PackageError{main}{Unknown \string\mbafont\space value `\mbafont'}{Use lora, lato or times.}
\fi\fi\fi
\ifx\mbafont\mbafont@lato\else
  \usepackage[lining,tabular]{lato}  %
\fi
\makeatother

\input{math_commands.tex}

\input{preamble}

\usepackage{fontawesome5}
\tcbuselibrary{skins}

\hypersetup{
  pdftitle={MBABenchV2: Evaluating AI Agents on Constructing Financial Spreadsheets From Scratch},
  linkcolor=mbaAccentStrong,
}

\definecolor{mbaBrand}{HTML}{1E375A}        %
\definecolor{mbaBrandDeep}{HTML}{14253F}    %
\definecolor{mbaAccent}{HTML}{3A78CD}       %
\definecolor{mbaAccentStrong}{HTML}{2F60A2} %
\definecolor{mbaHighlight}{HTML}{5FAFFF}    %
\definecolor{mbaBgSoft}{HTML}{F7F9FC}       %
\definecolor{mbaBorder}{HTML}{E1E8F1}       %
\definecolor{mbaBorderStrong}{HTML}{CDD8E6} %
\definecolor{mbaGrid}{HTML}{96AFCD}         %
\definecolor{mbaTextSoft}{HTML}{4A5A72}     %
\definecolor{mbaAnchor}{HTML}{E8830A}       %

\newcommand{\titleFont}{\fontfamily{lato-TLF}\fontseries{b}\selectfont}
\newcommand{\authorFont}{\fontfamily{lato-TLF}\selectfont}

\newcommand{\mbabenchmark}[1][2.4em]{%
  \resizebox{#1}{!}{%
  \begin{tikzpicture}[x=1mm, y=-1mm, line cap=butt]
    \begin{scope}
      \clip[rounded corners=8mm] (0,0) rectangle (32,32);
      \fill[mbaBgSoft]     (0,0)   rectangle (32,32);
      \fill[mbaAccent]     (0,0)   rectangle (32,8);
      \fill[mbaAccent]     (0,0)   rectangle (8,32);
      \fill[mbaHighlight]  (16,16) rectangle (24,24);
      \draw[mbaGrid, line width=1mm]
        (8,16) -- (32,16)  (8,24) -- (32,24)
        (16,8) -- (16,32)  (24,8) -- (24,32);
      \draw[mbaBrand, line width=1.5mm] (0,8) -- (32,8)  (8,0) -- (8,32);
    \end{scope}
  \end{tikzpicture}}}

\newlength{\sheetband}
\tcbset{
  spreadsheet frame/.style={
    enhanced,
    colback=mbaBgSoft,
    colframe=mbaBorderStrong,
    boxrule=0.6pt,
    arc=4pt,
    left=18pt+\sheetband, right=18pt, top=16pt+\sheetband, bottom=12pt,
    overlay unbroken and first={%
      \begin{tcbclipframe}
        \fill[mbaAccent] (frame.north west) rectangle ([yshift=-\sheetband]frame.north east);
        \fill[mbaAccent] (frame.north west) rectangle ([xshift=\sheetband]frame.south west);
        \draw[mbaBrand, line width=1.2pt]
          ([yshift=-\sheetband]frame.north west) -- ([yshift=-\sheetband]frame.north east)
          ([xshift=\sheetband]frame.north west) -- ([xshift=\sheetband]frame.south west);
      \end{tcbclipframe}},
  },
}

\newcommand{\titlelink}[3]{\href{#1}{#2\hspace{0.35em}#3}}

\begin{document}

\begin{tcolorbox}[
  spreadsheet frame,
  breakable,
]

\begingroup
\centering

{\LARGE\titleFont\color{mbaBrand}
\raisebox{-0.22em}{\mbabenchmark[1.15em]}\hspace{0.35em}\benchmarkname: \\[2pt]
Evaluating AI Agents on Constructing \\
Financial Spreadsheets Beyond Simple Verifiability \par}
\vspace{0.9em}

{\normalsize\authorFont\color{mbaBrandDeep}
\textbf{Julian Poeltl}\textsuperscript{*\,1}\quad
\textbf{Thomson Yen}\textsuperscript{*\,2}\quad
\textbf{Patrick Shea}\textsuperscript{*\,2}\\[3pt]
\textbf{Shubham Anil Varma}\textsuperscript{2}\quad
\textbf{Dhruv Rastogi}\textsuperscript{2}\quad
\textbf{Zihao Chen}\textsuperscript{3}\\[3pt]
\textbf{Mengshi Zhang}\textsuperscript{3}\quad
\textbf{Hongseok Namkoong}\textsuperscript{2}\par
}
\vspace{0.7em}

{\normalsize\authorFont\color{mbaTextSoft}
\textsuperscript{1}ESB Reutlingen\quad
\textsuperscript{2}Columbia Business School\quad
\textsuperscript{3}TensorBlock\par}
\vspace{0.6em}

{\normalsize\authorFont\hypersetup{urlcolor=mbaAccentStrong}%
\titlelink{https://mbabench.org/}{\faGlobe}{Website}\hspace{0.6em}$|$\hspace{0.6em}%
\titlelink{https://github.com/namkoong-lab/MBABench}{\faGithub}{Code}\par}
\vspace{0.5em}

\endgroup

{\color{mbaBorderStrong}\hrule height 0.4pt}
\vspace{0.8em}

\begingroup
\authorFont
\input{abstract}
\par
\endgroup

\vspace{0.9em}

\begingroup
\small\authorFont
\noindent
\begin{minipage}[c]{0.40\linewidth}
\color{mbaTextSoft}
\textbf{Correspondence:} \texttt{ty2531@columbia.edu}\\[0.5em]
\textsuperscript{*}Equal contribution
\end{minipage}\hfill
\begin{minipage}[c]{0.58\linewidth}
\raggedleft
\def\logowidth{6.2em}%
\raisebox{-0.5\height}{\includegraphics[width=\logowidth]{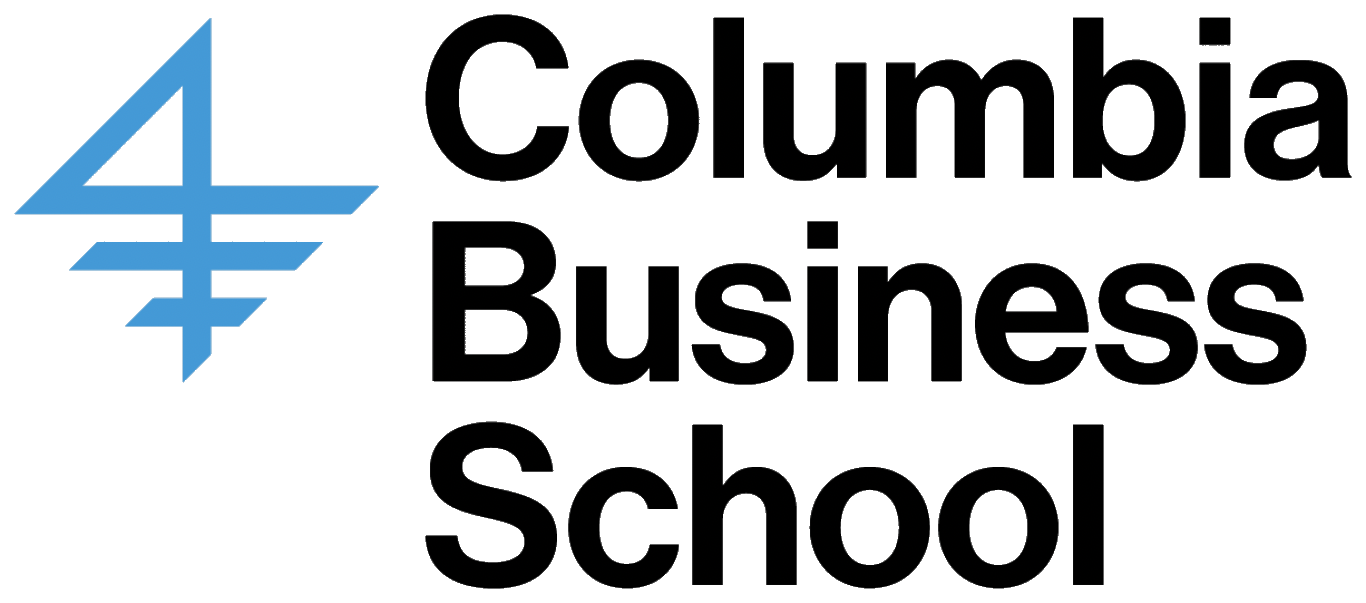}}%
\end{minipage}
\endgroup

\end{tcolorbox}

\vspace{1.0em}

\input{body}

\bibliographystyle{abbrvnat}
\setlength{\bibsep}{.7em}
\bibliography{bib}

\newpage
\appendix
\input{appendix}

\end{document}

%% file: math_commands.tex
\usepackage{amsmath,amsfonts,bm}

\def\eqref#1{equation~\ref{#1}}

\def\1{\bm{1}}

\DeclareMathAlphabet{\mathsfit}{\encodingdefault}{\sfdefault}{m}{sl}
\SetMathAlphabet{\mathsfit}{bold}{\encodingdefault}{\sfdefault}{bx}{n}

%% file: preamble.tex
\usepackage{float}
\usepackage{tabularx}          %
\usepackage{multirow}
\usepackage{makecell}          %
\usepackage{tikz}
\usepackage{tcolorbox}
\usepackage{xcolor}
\usepackage{amsmath}
\usepackage{subcaption}
\usepackage{graphicx}
\usepackage{booktabs}
\tcbuselibrary{skins, breakable, listings}

\usepackage{xspace} %
\usepackage{wrapfig}  %
\usepackage[htt]{hyphenat} %
\definecolor{lightciteblue}{RGB}{146, 197, 222}
\definecolor{darkerblue}{HTML}{2244AA}
\definecolor{lightcitered}{HTML}{CC3366}
\usepackage[colorlinks=true, citecolor=darkerblue, urlcolor=lightcitered]{hyperref}

\newcommand{\benchmarkname}{MBABench-2.0\xspace}
\newcommand{\prevbenchmarkname}{MBABench-1.0\xspace}
\newcommand{\tasknum}{{$101$}\xspace}
\newcommand{\tasktypenum}{{$16$}\xspace}
\newcommand{\rubricnum}{{$129$}\xspace}
\newcommand{\rubriccatnum}{{$12$}\xspace}
\newcommand{\agentnum}{{$20$}\xspace}
\newcommand{\modelnum}{{$8$}\xspace}
\newcommand{\harnessnum}{{$4$}\xspace}
\newcommand{\runnum}{{$2$K}\xspace}
\newcommand{\evalcost}{{\$$57$K}\xspace}
\newcommand{\highestpassrate}{{$16\%$}\xspace}
\newcommand{\judgefpr}{{$1.82\%$}\xspace}
\newcommand{\benchmarkcreationhours}{{$2{,}095$}\xspace}

\definecolor{classaccent}{HTML}{1F4E79}      %
\definecolor{classbg}{HTML}{F4F7FB}          %
\definecolor{classrule}{HTML}{D6DEE8}        %

\newif\ifclassfirst

\newtcolorbox{classbox}[2]{%
  enhanced, breakable,
  colback=classbg,
  colframe=classaccent,
  boxrule=0.6pt,
  arc=3pt,
  left=14pt, right=14pt, top=6pt, bottom=12pt,
  before skip=8pt, after skip=8pt,
  colbacktitle=classbg,
  coltitle=classaccent,
  titlerule=0.6pt,
  titlerule style={color=classaccent!40},
  toptitle=8pt,
  bottomtitle=6pt,
  title={%
    {\Large\bfseries\color{classaccent}#1}\par\vspace{1pt}%
    {\itshape\color{classaccent!75!black}\small #2}%
  },
  title after break={%
    {\bfseries\small\color{classaccent}#1}%
    {\itshape\small\color{classaccent!60}\ (continued)}%
  },
  before upper={\global\classfirsttrue},
}

\newcommand{\classterm}[2]{%
  \ifclassfirst
    \global\classfirstfalse
  \else
    \par\vspace{6pt}{\color{classrule}\hrule height 0.4pt}\vspace{6pt}%
  \fi
  \noindent\textbf{\color{classaccent}#1}\par\vspace{2pt}%
  \noindent #2%
}

\newtcolorbox{miniclassbox}{%
  enhanced,
  colback=classaccent!8,        %
  colframe=classaccent!45,      %
  boxrule=0.5pt,                %
  arc=2pt,                      %
  left=12pt, right=12pt, top=6pt, bottom=6pt,
  before skip=4pt, after skip=0pt,
}

\usepackage{xcolor}
\definecolor{changetext}{HTML}{1A5FA8}  %

\newif\ifshowchanges
\showchangestrue

\ifshowchanges
  
\else
  
\fi

%% file: abstract.tex
Spreadsheets are the lingua franca of modern finance, driving valuation, forecast, audit, and many other core workflows of the industry.
As LLM agents advance to automate knowledge work, spreadsheet construction is a natural and pertinent testbed.
However, most existing spreadsheet benchmarks evaluate agents on fill-in-the-blank style tasks, where an agent performs minor edits in otherwise complete spreadsheets.
This protocol fails to capture the most central and time-consuming part of the spreadsheet workflow: the modeling decisions, format design, and sheet layout regular knowledge workers need to toil through to build a workbook \textit{from scratch}.
We introduce \benchmarkname, a benchmark of \tasknum original spreadsheet construction tasks, the hardest of which is estimated to take $33$ hours for a modeler with $3$+ years of experience.
Unlike coding or math, where a unit test or a final answer can verify an agent's output, the quality of a financial spreadsheet that concerns readability, maintainability, or auditability, is not so simply verifiable. 
To holistically evaluate spreadsheet construction, we introduce \rubricnum criteria spanning \rubriccatnum categories that operationalize quality beyond the accuracy of the spreadsheet.
Evaluating over \agentnum agents across \modelnum models and \harnessnum harness types, we find that frontier agents rarely construct spreadsheets that are both accurate and meet our quality criteria:
the top agent achieves a pass rate of only \highestpassrate on the benchmark.
The benchmark demonstrates a substantial gap in spreadsheet building capability of current agents.

%% file: body.tex
\section{Introduction}
Finance is a major sector of the economy.
In the U.S., government agencies attribute \$$1.4$ trillion of GDP to the financial sector, at $4.9\%$ of the entire economy \citep{bea2026fininsgdp}.
Among the tools the sector relies on, spreadsheets are the lingua franca of modern finance: valuation, forecasts, audit, and numerous other key financial workflows all rest on spreadsheet software \citep{afp_spreadsheet, cfa_spreadsheet}.
As advances in large language model (LLM) agents accelerate, fueling the hope for agents to automate knowledge work, spreadsheet workflows emerge as a natural and pertinent testbed for their proficiency in financial work.

\begin{figure}[h]
\begin{center}
\includegraphics[scale=0.495]{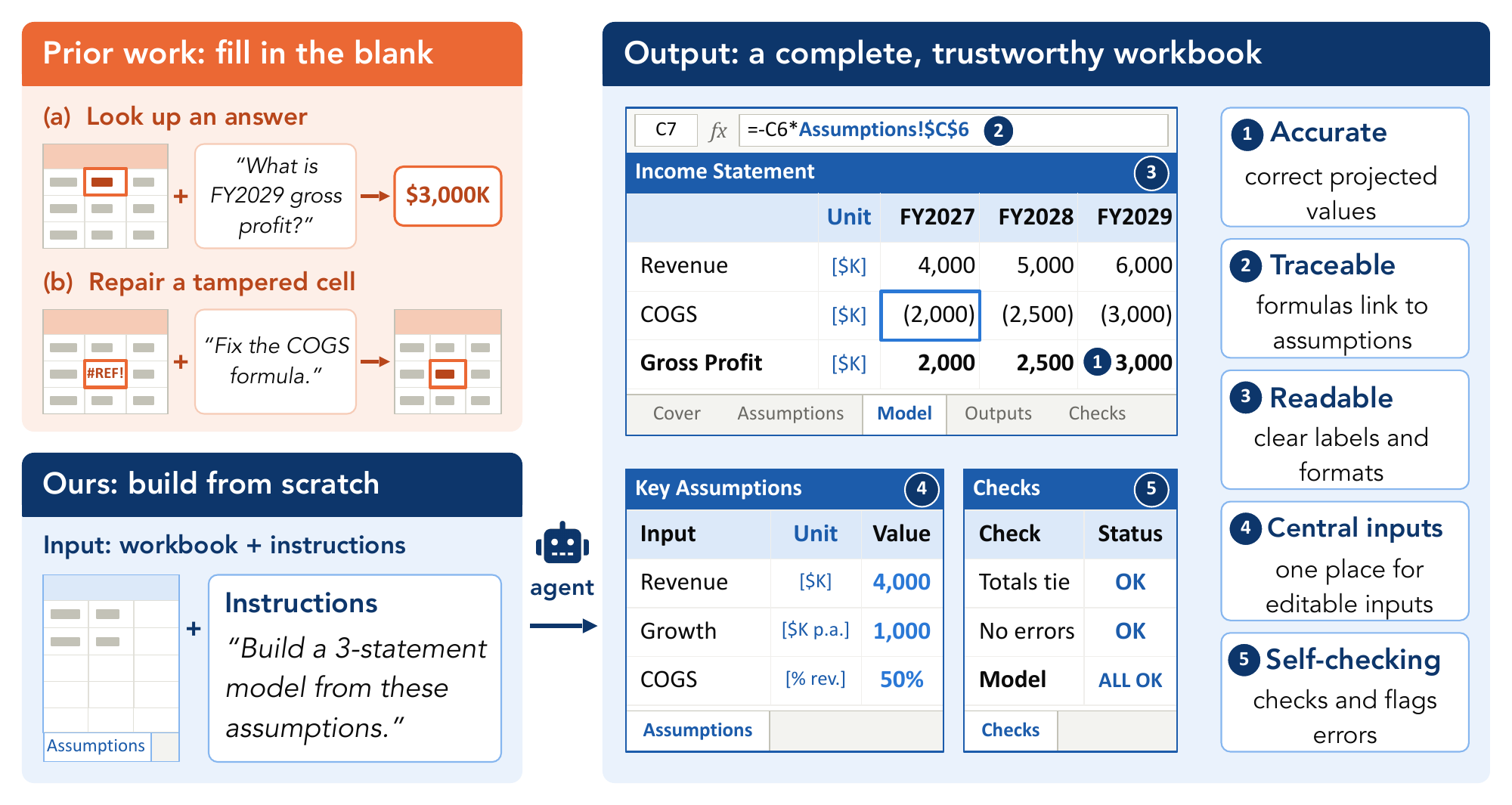}
\end{center}
\caption{
    \textbf{Focus of \benchmarkname} is to evaluate agents on constructing entire spreadsheets \textit{from scratch}.
    (\textit{Top Left}) Prior benchmarks ask agents to find answers in or correct otherwise finished workbooks.
    (\textit{Bottom Left \& Right})
    \benchmarkname instead gives agents only the inputs and instructions (like a real knowledge worker), and grades the delivered workbook on accuracy and on the quality criteria that make professional spreadsheets maintainable and reusable.
}
\label{fig:cartoon}
\end{figure}

A flurry of spreadsheet benchmarks have been proposed as a result.
The overwhelming majority, however, evaluate agents on fill-in-the-blank style tasks \citep{ma2024spreadsheetbench, krumdickbizbench, Choi_FinAgentBench, kundurthy2026bluefinbenchmarkingllmagents}.
There, agents are asked to extract answers to specific questions or to recover a few tampered formulas in an otherwise correct spreadsheet; even SpreadsheetBench~2 \citep{zhu2026spreadsheetbench2evaluatingagents} derives its tasks from complete workbooks with parts removed and largely grades whether the computed values match the reference.
This evaluation protocol misses the central, mundane, and time-consuming part of financial spreadsheet workflow --- the modeling decisions, format design, and sheet layout that experts toil through to build a high-quality spreadsheet \textit{from scratch}.
Figure~\ref{fig:cartoon} illustrates this difference.
Far fewer benchmarks \citep{patwardhan2025gdpval, lau2026bankertoolbenchevaluatingaiagents, yen2026mbabench} evaluate this \textit{spreadsheet construction} capability of LLM agents.

Evaluating this financial spreadsheet construction poses two main challenges.
The first concerns \textit{data availability}.
Finance is broad, encompassing investment banking, private equity, corporate finance, wealth management, and more, making it costly to assemble a sufficiently large and diverse set of tasks.
Exacerbating the issue, real financial data is often sensitive and proprietary, which precludes public release.
Reflecting these constraints, BankerToolBench
\citep{lau2026bankertoolbenchevaluatingaiagents} focuses exclusively on investment banking, whereas \prevbenchmarkname \citep{yen2026mbabench} covers finance broadly but releases only $38$ of its $202$ tasks, citing intellectual-property restrictions.
The second challenge concerns \textit{evaluation}.
In practice, spreadsheets serve not just as calculation devices but also as collaborative communication tools,
and miscommunication can be costly.
During Barclays' $2008$ acquisition of Lehman Brothers assets,
a junior associate accidentally led the bank to purchase $179$ additional assets due to lack of clarity in the spreadsheet \citep{barclays2008lehman_spreadsheetrisk}.
This communicative role of spreadsheets naturally demands additional quality criteria such as readability and auditability, beyond mere accuracy.
To capture these broader quality criteria, GDPval \citep{patwardhan2025gdpval} employs experts to manually annotate pairwise preferences, but this approach is costly to scale and hard to reproduce.
The community thus needs a public benchmark that offers both breadth and scale in its tasks, and evaluates quality criteria beyond accuracy in a transparent, scalable manner.

This paper presents \benchmarkname, a fully public benchmark comprising \tasknum spreadsheet construction tasks that cover key financial fields including investment banking, corporate financial planning and analysis (FP\&A), commercial banking, and wealth management.
While the spreadsheet quality beyond calculation accuracy may seem nebulous, the industry in fact largely agrees on principles that one should follow in a high-quality spreadsheet \citep{icaew2018code} (e.g., stating units, or consolidating key assumptions in one place). Where these conventions leave room for interpretation, firms typically codify their own house standards for employees to follow.
We therefore propose a taxonomy of \rubricnum criteria across \rubriccatnum categories, with the conventions left open fixed in house standards that every agent receives.
For instance, \texttt{No hardcoded numbers in formulas} requires all calculations to use one coherent set of assumptions, and \texttt{Numbers right-aligned} requires numerical values to be right-aligned.
Similar to \citet{lau2026bankertoolbenchevaluatingaiagents} and \citet{yen2026mbabench}, \benchmarkname employs a validated agentic judge to evaluate these quality criteria at scale.
Following the difficulty taxonomy of \citet{yen2026mbabench}, about half of our tasks are Medium-Hard ($30$) or Hard ($20$), the tiers where frontier agents still struggle.
In comparison, we have $7.5$\texttimes{} and $10$\texttimes{} as many difficult tasks as \prevbenchmarkname's public tasks ($4$ and $2$, respectively).
These tasks examine long-horizon model building ability, with estimated build times of up to $\mathbf{33}$ hours for a modeler with $3$+ years of experience.
Figure~\ref{fig:running_example} illustrates one such task:
only one agent answers all its questions correctly, and even the numerically correct workbook falls short of our quality criteria.
Finally, we conduct a large-scale evaluation of spreadsheet construction:
we evaluate over \agentnum agents (from selected combinations of \modelnum models $\times$ \harnessnum agentic harness types) for a total of \runnum long-horizon trajectories, exceeding \evalcost in agent inference cost alone.

We find that even the top agent passes only \highestpassrate of the tasks, and a substantial number of agents with weaker backbone models or harnesses pass none.
Agents fall short both in numerical accuracy and in the other quality criteria.
On numerical accuracy, agents perform surprisingly poorly with lower than $60\%$ accuracy on tasks rated Medium or harder.
Most of these numerical errors stem from agents misreading the instructions (e.g., paying the full monthly salary twice per month on a bi-weekly schedule), or from a lack of professional judgment (e.g., comparing a gross profit that includes other operating income against a budget gross profit that excludes it).
On the other quality criteria, agents' failures are more varied.
For example, their models are hard to modify because timelines are fixed to a specific horizon, and they often contain deeply nested formulas that are hard to audit.
These results suggest that current spreadsheet agents are still far from delivering professional spreadsheets.

Our major contributions are as follows:
\begin{itemize}[leftmargin=*, topsep=2pt, itemsep=2pt, parsep=0pt]
\item We create and release \tasknum original spreadsheet construction tasks, spanning \tasktypenum task types across key fields of finance.
\item We propose a taxonomy of spreadsheet quality consisting of \rubricnum criteria across \rubriccatnum categories. To enable transparent and scalable evaluation, we develop and validate a corresponding agentic judge system.
\item We conduct a large-scale evaluation of \agentnum agents.
Our results reveal several significant shortcomings; for example, agents still fail on harder tasks, and the harness appears to substantially shape the quality of the delivered spreadsheet.
\item Through qualitative analysis, we find common pitfalls in agents' attempts, revealing the ways in which agents fail in both numerical accuracy and other spreadsheet quality criteria.
\end{itemize}

\paragraph{Related Work}

\paragraph{Financial and Spreadsheet Benchmarks.}
Finance has long attracted interest from LLM benchmarks.
Early benchmarks are predominantly QA-based, testing models on simple derivation or finance knowledge \citep{xie2023pixiu, FinBen, XFinBench, mateega2025financeqa, lu2025bizfinbench}.
As LLM agents have improved, attention has shifted to spreadsheet manipulation that underpins several core finance workflows.
Before the long-horizon capabilities of agents matured, SpreadsheetBench \citep{ma2024spreadsheetbench}, widely used during this time, examined agents on targeted manipulations of existing spreadsheets, drawn from questions on Excel forums and graded on the resulting cell values.
Many subsequent evaluations follow the same pattern, focusing on summarization, QA, and a few formula edits \citep{krumdickbizbench, Choi_FinAgentBench, kundurthy2026bluefinbenchmarkingllmagents}.
SpreadsheetBench~2 \citep{zhu2026spreadsheetbench2evaluatingagents} extends this to end-to-end workflows on large workbooks, but derives its tasks from complete workbooks with parts removed and, except for visualization tasks, grades whether the computed cell values match the reference.
Constructing spreadsheets from scratch is part of GDPval \citep{patwardhan2025gdpval}, which includes such tasks among its occupational deliverables, and is evaluated by \citet{lau2026bankertoolbenchevaluatingaiagents} for investment banking and by \citet{yen2026mbabench} for finance in general.
We build on this line of work with a fully public benchmark consisting of \tasknum spreadsheet construction tasks across general finance.

\paragraph{Agents for Professional Work and Evaluating on Quality Criteria.}
Agents are increasingly expected to perform economically valuable work.
The first promising domain is coding \citep{chen2021codex_humaneval, jimenez2024swebench},
followed by computer use \citep{zhou2024webarena, xie2024osworld}, legal work \citep{harveylab2026}, and other occupations from manufacturing to animation \citep{patwardhan2025gdpval, sun2026agentsexam}.
As agents' tasks become more realistic, accuracy-based evaluation no longer captures the full spectrum of desiderata in real-world work product.
For example, despite advances in standard coding benchmarks, agents often produce verbose and difficult-to-maintain code \citep{metr_swebenchnotmerged, yang2026programbench}.
GDPval \citep{patwardhan2025gdpval} accounts for these high-level \textit{quality criteria} in evaluation by having experts annotate pairwise preferences, at the cost of scalable evaluation.
Recent coding \citep{cognition_frontiercode} and spreadsheet benchmarks \citep{lau2026bankertoolbenchevaluatingaiagents, yen2026mbabench} tackle this with a mix of automatic graders and LLM-based judges.
This work takes the latter approach: we first propose a taxonomy of quality criteria for financial spreadsheets, and then develop and validate a corresponding agentic judge system for reliable and scalable grading.

\section{\benchmarkname Tasks}
\label{sec:task_creation}

\subsection{Difficult and Long-Horizon Financial Spreadsheet Tasks}

\begin{figure}[!t]
    \centering
    \includegraphics[width=\linewidth]{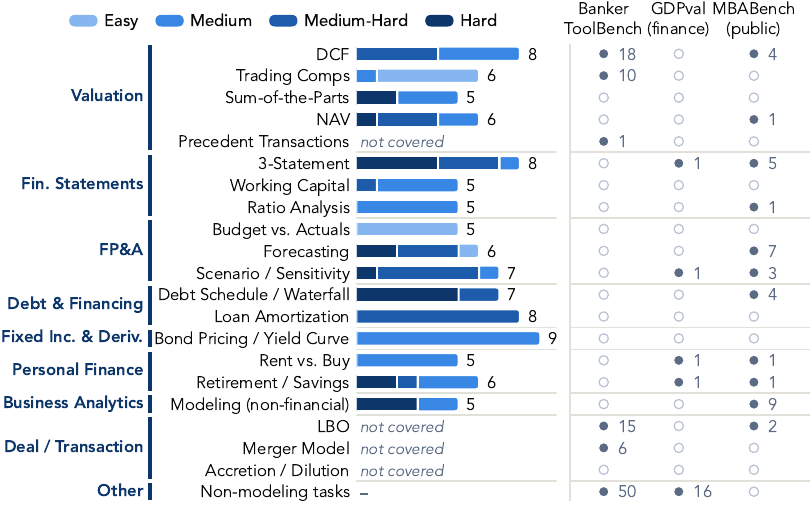}
    \caption{
        \textbf{Task Coverage of \benchmarkname.}
        Compared to BankerToolBench \citep{lau2026bankertoolbenchevaluatingaiagents} and GDPval \citep{patwardhan2025gdpval}, we cover significantly more modeling task types.
        Only MBABench-1.0 \citep{yen2026mbabench} is comparable in coverage, but they release only $38$ tasks, whereas we release all \tasknum tasks, with $50$ Medium-Hard+ challenging ones (compared to $6$ in MBABench public tasks).
    }
    \label{fig:coverage}
\end{figure}

The task set targets the core work of financial modelers, building a complete model from scratch, across a wide range of spreadsheet models used in finance, from company valuation and $3$-statement models to FP\&A, debt schedules, instrument pricing, personal finance, and business analytics.
With \tasktypenum task types across seven fields, it is broader than any comparable benchmark (Figure~\ref{fig:coverage}).
About half of the tasks ($50$ of $101$) are Medium-Hard or Hard, with estimated build times of up to $33$ hours (Appendix~\ref{app:TimeEstimate}).
We release the full task set, including the golden solutions.

\begin{wrapfigure}{r}{0.49\textwidth}
\centering
\includegraphics[width=\linewidth]{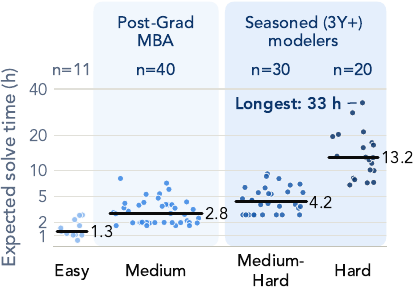}
\caption{
\textbf{Difficulty and Expected Build Time of the \tasknum Tasks in \benchmarkname}.
About half of the tasks are Medium-Hard or Hard, demanding in both modeling skill (x-axis) and estimated build time (y-axis).
}
\label{fig:difficulty}
\end{wrapfigure}

All tasks were created by finance experts rather than derived from existing materials, which reduces the risk of data contamination.
Since the tasks contain no client data or proprietary material, we can release all of them without privacy or intellectual-property restrictions.
Each task consists of an input workbook (the \emph{starting file} in our repository) and a \emph{golden solution} with the completed financial model.
The input workbook contains the task instructions, assumptions, supporting data where needed, and questions.
The instructions are phrased as practitioners would phrase such a request and were refined iteratively to remove the ambiguities we identified.
The questions are designed to examine whether the model is correct in all of its main results and intermediate calculations.
An attempt therefore satisfies the \texttt{Final calculation accuracy} criterion (Section~\ref{sec:rubric}) only if it answers all questions correctly.
All tasks passed a multi-stage quality assurance process with human and LLM-based review; Appendix~\ref{app:TaskDesign} describes the task design and quality assurance process in detail.

\paragraph{Difficulty and Task Types.}
To show what the benchmark covers, we manually annotated each task with a difficulty level and a task type (Figures~\ref{fig:difficulty} and~\ref{fig:coverage}).
The benchmark spans levels $2$--$5$ of a six-level scale:
\emph{Easy} tasks can typically be solved without deep financial expertise, whereas \emph{Hard} tasks are challenging even for seasoned modelers.
Task types range from discounted cash flow (DCF) valuations and $3$-statement models to debt schedules and personal-finance calculators.
Appendix~\ref{app:TaskAnnotation} details both annotations, including a description of each task type.

\subsection{An Illustrative Example - DayMonthYear}
\label{sec:example}
\begin{figure}[t]
    \centering
    \includegraphics[width=\linewidth]{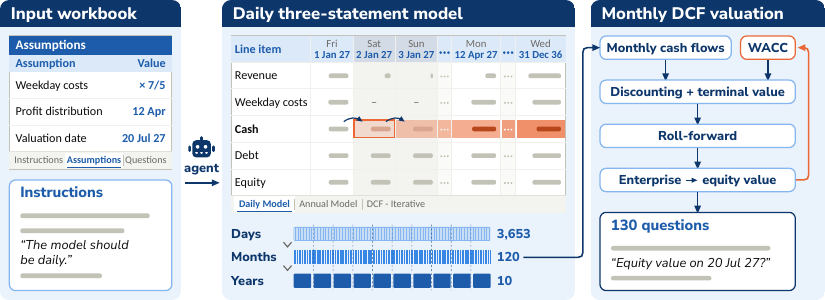}
    \caption{
    \textbf{Schematic view of the task \texttt{DayMonthYear} (Task $79$).}
    Orange cells trace an early error through all later days.
    The orange arrow marks the circular dependency between the WACC and the equity value.
    }
    \label{fig:running_example}
\end{figure}

To illustrate what a difficult task demands, we describe \texttt{DayMonthYear} (task~$79$; Medium-Hard; estimated build time of about $9$ hours) in detail here (see Figure~\ref{fig:running_example}).
The agent must build an integrated three-statement model (a standard tool for forecasting a company's financial statements) that runs daily over ten years, value the company with a monthly DCF (a standard method for estimating a company's value), and derive numerical solutions to $130$ questions.
The agent needs to perform each of the following calculations carefully to arrive at the correct answers:
\begin{enumerate}[leftmargin=*]
    \item \textbf{An Integrated Daily Model.}
    The three statements must agree on all $3{,}653$ days.
    \emph{Example mistake:} a \$$100$ expense that should be but is not deducted from cash flow, which results in an error that propagates to all later days.

    \item \textbf{Consistent Aggregates.} Every monthly and yearly figure must be derived from the daily values in that period, so that numbers from days, months, and years always agree.
    Shortcuts that bypass the daily values break this, because months have different numbers of days.
    \emph{Example mistake:} a cost of \$$100$ a day should total \$$3{,}100$ in January and \$$2{,}800$ in February; dividing the yearly \$$36{,}500$ cost across months evenly gives an incorrect cost of \$$3{,}041.67$ per month.

    \item \textbf{Calendar Conventions.} Some costs are defined as a daily average but are only incurred Monday to Friday.
    To keep the weekly total the same, each weekday must carry $7/5$ of the daily average and nothing on weekends.
    Other rules are tied to exact dates, such as annual increases on $1$~January and a profit distribution every $12$~April.
    \emph{Example mistake:} for a cost averaging \$$100$ a day, charging \$$100$ instead of \$$140$ on each weekday, resulting in \$$500$ instead of \$$700$ cost per week.

    \item \textbf{Valuation Steps.} Valuing the company takes four steps: converting each future month's cash into today's money (discounting at the weighted average cost of capital, or WACC), estimating the company's value for all years after the forecast (terminal value), shifting the result to the exact valuation date (roll-forward), and deducting debt and debt-like obligations while adding cash and investments (enterprise-to-equity bridge).
    Each step is simple, but skipping one still produces a reasonable-looking number.
    \emph{Example mistake:} Excel's \texttt{NPV} function implicitly values the cash flows as of $31$~December~$2026$, but the task asks for the value on $20$~July~$2027$, so the result fails to account for the growth in value over the $201$ days between the two dates.

    \item \textbf{Circular Dependencies.} A common approach to estimating company value is to discount its future cash flows at the opportunity cost of the capital provided by its shareholders and lenders (WACC).
    However, the value of the company depends on the WACC, while the WACC is itself estimated from the value of the company's equity.
    This circularity demands careful handling of circular references in the spreadsheet (e.g., starting from an initial guess and recomputing until the WACC and the company value agree).
    \emph{Example mistake:} guessing an equity value of \$$60$M to compute the WACC, obtaining a valuation of \$$80$M, and reporting the valuation without recomputing the WACC from \$$80$M, so the WACC and the company value disagree.

\end{enumerate}

None of these requirements is hard on its own; the difficulty comes from satisfying all of them across thousands of interdependent days, starting from a workbook that contains only the inputs.
Most prior benchmarks ask agents to look up an answer or repair a few formulas in a finished workbook.
For example, SpreadsheetBench~2 \citep{zhu2026spreadsheetbench2evaluatingagents} creates each task from a complete workbook by removing parts or injecting errors so that the structure of the workbook is largely provided.
In contrast, to solve tasks such as \texttt{DayMonthYear} in \benchmarkname, designing that structure and making sure the model is correct from the first daily balance to the final equity value is a central part of the task.

\section{Evaluation Method}
\paragraph{What \benchmarkname Evaluates.}

\label{sec:rubric}

\begin{figure}[t]
    \centering
    \includegraphics[width=\linewidth]{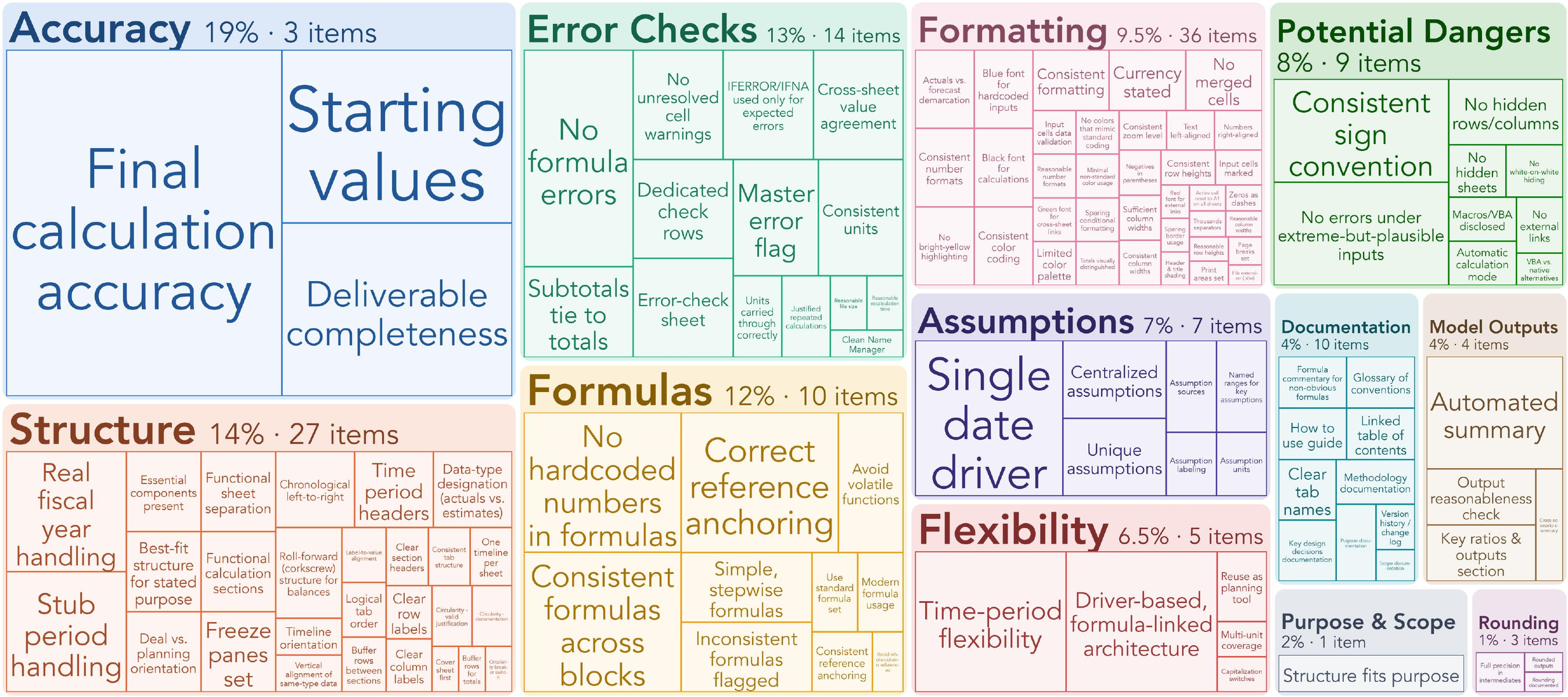}
    \caption{
        \textbf{The Full Rubric of \benchmarkname}.
        We introduce a taxonomy of \rubricnum criteria that captures what makes a financial spreadsheet trustworthy in practice.
    }
    \label{fig:rubric_map}
    \vspace{-1em}
\end{figure}

Financial modeling tasks vary widely: a $3$-statement model, a DCF valuation, and a working capital model serve drastically different purposes.
Yet beneath this variety, good financial models share a common and codifiable standard: they should separate inputs from calculations, use transparent formulas, follow consistent formatting conventions, and include error checks so that the model is \textit{clear} and \textit{maintainable} to others.
This commonality allows us to design a list of \textit{universal} criteria that we apply to all tasks (e.g., \texttt{No hardcoded numbers in formulas}).
In addition, we supplement the list with \textit{conditional} criteria whose applicability depends on the specific task (e.g., \texttt{Single date driver}).
This fixed criterion set keeps the grading scale consistent across tasks and types of financial model and allows the judge to be tested criterion by criterion (see Section~\ref{sec:rubric_judge_val}).
Our rubric thus consists of \rubricnum criteria in total ($93$ universal and $36$ conditional; see Figure~\ref{fig:rubric_map}).

\paragraph{Codifiable Standards and Agent Prompt.}
Public financial modeling codes agree on principles but leave certain conventions open, such as the sign convention (e.g., whether costs are shown as negative numbers) or number formats (e.g., whether negatives appear in parentheses) \citep{icaew2018code}.
Firms fix these in internal guidelines, and spreadsheet agents increasingly receive such guidelines as context \citep{microsoft2026copilotagents, anthropic2025financial}.
We mirror this setting by specifying the standards agents are supposed to follow:
every agent receives house standards, a compact markdown file written in the style of a firm's internal guidelines (Appendix~\ref{app:HouseStandards}).
Stating the target convention ensures that agents are not penalized for conventions they could not infer, and lets the judge check conformance to a known specification rather than decide whether an alternative is acceptable, which keeps verdicts comparable across agents.
The rubric itself is not disclosed.

\paragraph{Criteria Deep Dive.}
After annotating applicability, each task in \benchmarkname is graded on between $105$ and $126$ criteria.
Each criterion is further assigned to one of \rubriccatnum categories and one of four importance levels, from which its weight is derived.
We defer details to
Appendix~\ref{app:FullRubric}.
This design is \emph{comprehensive}, as every task is checked against the common core and against the conditional criteria its model requires, and \emph{efficient}, as a new task needs only a yes/no relevance decision per conditional criterion and a new model type only the few criteria specific to it.
It is also \emph{verifiable}: because each criterion is defined once for all tasks, the judge can be tested on each criterion individually (Section~\ref{sec:rubric_judge_val}).
Finally, although the applicable criteria differ between tasks, each category carries the same weight in every task, which keeps the grading scale consistent across tasks and types of financial models.

\begin{figure}[htbp]
     \centering
     \includegraphics[width=\textwidth]{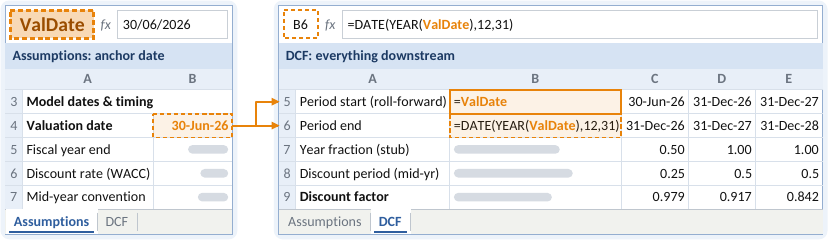}
     \caption{\textbf{Exemplar Criterion.}
     One labeled anchor date (\textit{left}, \textcolor{mbaAnchor}{\texttt{Assumptions!B4}}, named \textcolor{mbaAnchor}{\textbf{ValDate}}) drives the period ends, stub period, mid-year discount periods, and roll-forwards of the DCF sheet (\textit{right}).
     }
    \label{fig:single_date_driver}
\end{figure}

As an example, consider the conditional criterion \texttt{Single date driver} (Figure~\ref{fig:single_date_driver}).
This criterion applies to any task with a time component and requires a single, clearly labeled anchor date that drives all time periods and dependent calculations.
A spreadsheet that satisfies this criterion allows an analyst to easily shift the entire model timeline by changing a single cell, which is particularly useful when a planned transaction is postponed.
Without it, the analyst would have to painstakingly update all cells that depend on the anchor date, which risks introducing mistakes.

\paragraph{How \benchmarkname Evaluates.}
Each criterion is marked satisfied or violated.
\texttt{Final calculation accuracy} is graded deterministically by comparing the agent's answers to the task's questions with the correct answers (Appendix~\ref{app:Criteria}), as are $19$ other criteria that fixed rules can check reliably.
The remaining $109$ criteria are graded by an agentic judge using GPT-5.6 Sol as the backbone model, because grading them requires context-dependent judgment (e.g., whether row labels are clear and descriptive) and must accommodate every acceptable layout (e.g., units placed one to three cells away from the numbers they describe).
Since all tasks share one list of criteria, we were able to build the judge around this rubric, and validate it in Section~\ref{sec:judge_val}.
We defer further description of the judge to Appendix~\ref{app:Judge}.

\begin{wraptable}{r}{0.5\textwidth}
\vspace{-12pt}
\centering
\small
\caption{\textbf{Criteria by Importance.} Det.: graded deterministically; $k_i$: violations tolerated by the pass rule.}
\label{tab:criteria_breakdown}
\setlength{\tabcolsep}{4pt}
\begin{tabular}{@{}lrrrr@{}}
\toprule
Importance & Total & Det. & Judge & $k_i$ \\
\midrule
Ultra important & $12$ & $1$ & $11$ & $1^{\dagger}$ \\
Highly important & $41$ & $6$ & $35$ & $2$ \\
Important & $52$ & $6$ & $46$ & $3$ \\
Not absolutely crucial & $24$ & $7$ & $17$ & $2$ \\
\midrule
All & $129$ & $20$ & $109$ & -- \\
\bottomrule
\multicolumn{5}{@{}l}{\scriptsize $^{\dagger}$\texttt{Final calculation accuracy} must be satisfied.}
\end{tabular}
\vspace{-16pt}
\end{wraptable}

\paragraph{Metrics.}
Ideally, a spreadsheet satisfies all applicable criteria; we therefore report the \textbf{Pass Rate} of each agent across tasks, with a small tolerance defined below.
To account for the judge's false positive rate (FPR; i.e., incorrectly flagging a satisfied criterion as violated) in grading these criteria, we require the spreadsheet to satisfy the deterministically graded \texttt{Final calculation accuracy} and to violate no more than $k_i$ criteria for each of the four importance levels $i$ (see Table~\ref{tab:criteria_breakdown}, with further details in Appendix~\ref{app:FullRubric}).

\begin{figure}[t]
    \centering
    \includegraphics[width=\linewidth]{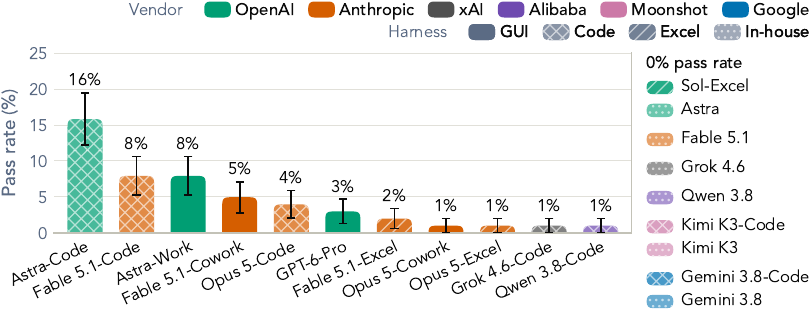}
    \caption{\textbf{Pass Rate Leaderboard.}
    Agents regularly fail on \benchmarkname.
    Notably, the agentic harness appears to strongly impact how agents perform (e.g., see Astra and Fable~5.1 across harnesses).
    }
    \label{fig:leaderboard}
\end{figure}

We choose $k_i$ by weighing criteria importance in practice, and so that a perfect spreadsheet would be deemed a pass with high probability.
With the empirical FPR of \judgefpr (see Section~\ref{sec:full_judge_val}) and the pass rule in Table~\ref{tab:criteria_breakdown}, we have a high rate of successfully recognizing a perfect workbook: 
\[
      \mathbb{P}(\mathsf{Pass} \mid \mathsf{Perfect})
      =
      \prod_{i=1}^4
      \mathbb{P} \left(
        \mathsf{NumViolated}_i
        \leq k_i
      \right)
      =
      \prod_{i=1}^{4} F_{\mathsf{Bin}}\!\left(k_i;\, \mathsf{NumJudgeCriteria}_i, \mathsf{FPR}\right)
      =
      0.946,
\]
where $F_{\mathsf{Bin}}(k; N, p) = \sum_{j=0}^{k} \binom{N}{j} p^{j} (1-p)^{N-j}$ is the binomial CDF.
Note that the pass rule above is intentionally lenient; combined with the judge's higher false negative rate (Section~\ref{sec:judge_val}), this means the pass rate will accept a workbook with genuine violations, and hence should be read as an upper bound on the percentage of acceptable workbooks.

Since a substantial number of agents have a $0\%$ pass rate, we also report the \textbf{Score}, the sum of the weights of the criteria an attempt satisfies (Appendix~\ref{app:Weighting}).
Note that since many rubric criteria are related to structure or format of the spreadsheet, it can be deceptively high even for an unusable spreadsheet; submitting the input workbook unchanged would generally score around $50\%$.
Appendix~\ref{app:HumanReference} details the pass rule and why we do not report a human baseline.

\paragraph{Agentic Harnesses.}
We evaluate each model through up to four harnesses.
\textbf{GUI}: the LLM vendor's own web pages (e.g., Claude Cowork).
\textbf{Code}: Claude Code for Anthropic models and the open-sourced Codex CLI for all others.
\textbf{Excel}: the proprietary Excel add-ins by OpenAI and Anthropic.
\textbf{In-house}: a custom agentic harness that lets LLMs operate on spreadsheets via \texttt{openpyxl}.
We defer details of these harnesses to Appendix~\ref{app:AgenticHarnesses}.
Note that since we do not have direct access to the designs of proprietary harnesses, we do not aim to disentangle the effect of model and harness quantitatively.
The results serve to ($1$) provide practically useful observations on performance of existing agents and ($2$) assess the effect of harnesses for a fixed underlying model.

\section{Judge Validation}
\label{sec:judge_val}

In this section, we validate the reliability of the agentic judge we employ for scalable evaluation.
We approach the validation by stress-testing the judge on grading two types of attempts:
(a) hand-crafted examples that satisfy or violate a specific criterion, so we can have a controlled and targeted examination of the judge's performance on individual criteria, and (b) actual attempts by evaluated agents, so we can examine the judge's performance on the attempt distribution that it sees at deploy time.

\subsection{Per-Criterion Controlled Validation}
\label{sec:rubric_judge_val}
We validated the agentic judge on each of the $109$ criteria graded by the judge.
For each criterion, we built a pair of examples by adapting a task's golden solution: a \emph{clean example} that clearly satisfies the criterion, and an \emph{erroneous example} into which the task's author manually introduced an unambiguous but not trivially detectable violation (Appendix~\ref{app:ValidationExamples}).
We ran the judge on each of the $218$ examples for three repetitions.
Of these, $207$ ($95.0\%$) are graded correctly in all three runs and $217$ ($99.5\%$) in at least two.

Over the $654$ individual verdicts, the judge achieves $98.2\%$ accuracy, a false positive rate of $0.31\%$ ($1$ of $327$ verdicts on clean examples), a false negative rate of $3.4\%$ ($11$ of $327$ verdicts on violations), and an $F_1$ of $98.1\%$.
No category falls below $93\%$ per-verdict accuracy on these examples.

\subsection{Validation on Agent Attempts}
\label{sec:full_judge_val}

\paragraph{Validation Against Experts.}
To evaluate the agentic judge's accuracy on real attempts of full tasks, we selected $12$ attempts for annotation by human experts.
Since expert annotation is extremely time-consuming, we selected the attempts from four agents spanning a range of capability (Astra-Work, Fable-Cowork, Fable-in-house, Gemini-3.8-Code).
The experts graded each attempt on all applicable rubric criteria by inspecting the agent's workbook alongside the golden solution; their annotations were afterward compared with the judge's verdicts (Figure~\ref{fig:judge_agreement}). Across them, the agentic judge made $1{,}152$ determinations on specific rubric criteria. Of these, the judge achieved $96.4\%$ accuracy, an FPR of \judgefpr,
an FNR of $11.1\%$
and an $F_1$ of $90.4\%$.
Note that the low FPR helps the pass rate identify acceptable workbooks at a high rate.

\paragraph{Judge Self-Consistency.}
We took $90$ attempts across $30$ difficulty-weighted tasks from three agents, Codex (GPT-6 Astra), In-house (Fable~5.1), and Codex (Gemini~3.8 Flash), and regraded each attempt three times with the judge.
Let $v_a$ denote the variance in score for each attempt across repeated gradings.
We observe the standard deviation $\sqrt{\sum_{a}^{} v_a/90} = 3.0$ out of $100$.
Of all (attempt, criterion) pairs, $90.0\%$ receive the same verdict in all three runs, close to the $\mathbb P(\mathsf{agree}) = \mathsf{acc}^3 + (1-\mathsf{acc})^3 = 89.7\%$ expected if each run were independently correct at the empirical accuracy $\mathsf{acc} = 96.4\%$.
On the pass verdict, pairs of runs disagree in only $1.5\%$ of cases.
These results complement the earlier validation by speaking to the consistency of the judge.

\paragraph{Judge Backbone Model.}
We chose GPT-5.6 Sol as the judge's backbone to balance the accuracy of the judge and the cost per grading.
Here, we examine how closely the Sol-based judge agrees with a stronger but more expensive model, Fable~5.1.
We regraded all $303$ attempts by the same set of three agents above under the Fable-based judge.
On both score and pass rate, all $3$ agents maintain their relative rankings.
The scores from both judges have a Pearson correlation of $0.901$ across attempts.

Beyond judge agreement, we share additional findings on the effect of the agentic judge.
($1$) \textit{The superior judge is stricter.} Fable flags a violation in $60.0\%$ of the assessments when the two judges disagree, scoring attempts $1.91 \pm 0.39$ points lower on average.
This finding aligns with the qualitative observations by the experts during the initial judge selection stage of the project, where stronger judge models flagged more violations in our runs.
($2$) \textit{No evidence of self-preference.} In-house (Fable~5.1) trails Codex (GPT-6 Astra) by $11.28$ points under the Sol judge and by $11.58$ under the Fable judge, which differs only by a statistically insignificant $0.30 \pm 0.65$ across $303$ attempts.

\section{Results and Discussion}
\label{sec:results}

In this section, we first discuss the main quantitative results of \benchmarkname and then qualitatively assess how agents fail, both in numerical accuracy and in the other $128$ quality criteria.

\subsection{Quantitative Analysis}

\paragraph{Pass Rate.}
We show the resulting leaderboard in Figure~\ref{fig:leaderboard}, and find that most attempts do not meet our pass bar.
Each agent has one recorded attempt per task, following common practice in end-to-end agentic evaluation (Appendix~\ref{app:AgenticHarnesses}); we therefore report error bars to reflect the resulting uncertainty.
Note that even with the same underlying model, the pass rate changes drastically (e.g., GPT-6 Astra in Codex vs.\ ChatGPT, or Fable~5.1 in Claude Code vs.\ Claude for Excel).
This suggests that the harness, or the product built around it, has a major effect on the quality of the resulting spreadsheet.

\begin{figure}[t]
    \centering
    \includegraphics[width=\linewidth]{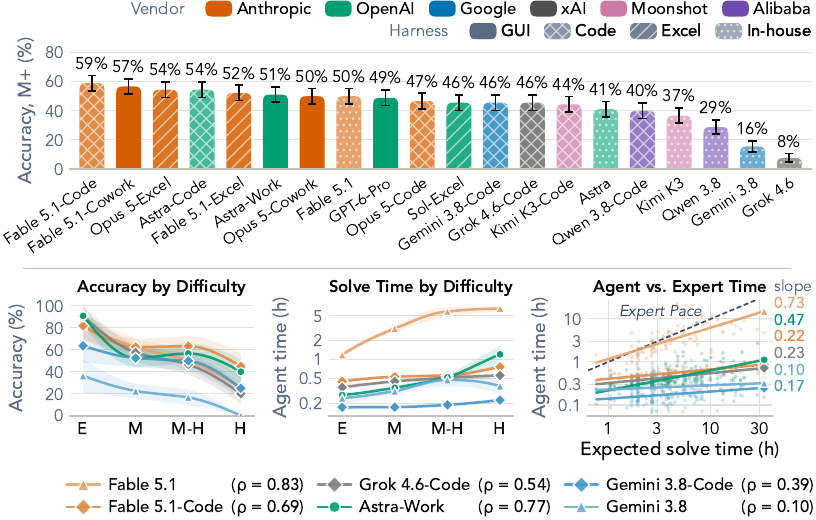}
    \caption{
        \texttt{Final Calculation Accuracy} \textbf{on More Difficult Subset.}
        \textit{Top:} purely in accuracy, the top agent solves only $59\%$ of the harder tasks.
        \textit{Bottom Left:} Accuracy decreases significantly for harder tasks.
        \textit{Bottom Mid/Right:} better agents tend to identify difficulty and spend more time accordingly. $\rho$ denotes Pearson correlation.
    }
    \label{fig:acc_diff}
\end{figure}

\paragraph{Numerical Accuracy on Difficult Tasks.}
Here we examine agents' performance purely on whether they arrive at the correct numerical solution.
Since Easy tasks are largely solved by the top agents, we primarily focus on tasks that are Medium and harder (see Figure~\ref{fig:acc_diff}).
Noting the conceptual simplicity of the computation as shown in Section~\ref{sec:example}, we surprisingly find that the top agent reaches only $59\%$ accuracy on Medium and harder tasks, and only $45\%$ on Hard tasks.
Another noteworthy finding is that numerical accuracy is largely stable across the commercial harnesses in contrast to pass rate, showing that harnesses impact the correctness of computation less.

\paragraph{Agent Solve Time.}
In general, wall-clock solve time increases with task difficulty for all agents.
Among the six agents shown, those whose solve time increases least with difficulty also perform worst (see Gemini agents in the bottom middle panel of Figure~\ref{fig:acc_diff}, and how Pearson correlation in the bottom right correlates with agent performance), which suggests that proficient agents could recognize how difficult a task is and spend more effort correspondingly.
Agents also take far less wall-clock time than the estimated human build times, although these savings must be weighed against the share of incorrect answers.

\paragraph{Reasoning Effort Ablation.}
To evaluate the impact of reasoning effort, we ran Claude Code (Fable~5.1) at three levels of reasoning effort: Max, High, and Low. We find a positive correlation between reasoning levels and scores, with scores averaging $90.1$ for Max, $88.3$ for High, and $86.9$ for Low.

\subsection{Qualitative Analysis}
\label{sec:qualitative}
In this section, we provide a complementary qualitative assessment of how agents fall short of professional standards.
We discuss here the main shortcomings shared across agents (with more details in Appendix~\ref{app:GeneralAnalysis}) and defer agent-specific failure modes to Appendix~\ref{app:ModelAnalysis}.

\subsubsection{Final Calculation Accuracy}

On more difficult tasks, the workbooks remain complete and structurally sound, but the underlying financial models are more often incorrect.
To understand why agents often build faulty financial models, we analyze every attempt that violates \texttt{Final calculation accuracy} and identify five main causes:
\begin{enumerate}[leftmargin=*]
    \item \textbf{Misreading of instructions} ($44\%$) includes cases when agents paid the full monthly salary twice per month on a bi-weekly payment schedule or used simple interest when the Canadian compound method was requested.
    \item \textbf{Lack of professional judgment} ($28\%$) arises when agents fail to follow standard practice that a human expert would apply by default.
    For example, agents repaid debt only from cash while a committed credit line remained unused, or compared a gross profit that includes other operating income with a budget figure that excludes it.
    \item \textbf{Delivery and technical failures} ($15\%$) cover workbooks that are set up but leave every answer cell blank and generated files whose formulas show \#NAME? errors once they are recalculated.
    \item \textbf{Implementation errors} ($11\%$) occur when clear instructions are provided but the financial model is nonetheless built incorrectly, as when an agent booked a payroll-tax payment as an inflow instead of an outflow or modeled $45$ quarterly loan payments instead of $46$, dropping the final coupon.
    \item \textbf{Missing sense checks} ($2\%$) let implausible results pass.
    In one task, an agent let the share of customers using a product keep growing past $100\%$, so the estimated market size came out about $50$\texttimes{} too large.
\end{enumerate}
Frontier agents (the $12$ Claude and GPT agents) violate \texttt{Final calculation accuracy} on fewer tasks than the eight non-frontier agents (Gemini, Grok, Kimi, and Qwen), and for different reasons:
$85\%$ of the frontier agents' erroneous attempts are due to misreading or a lack of professional judgment and only $13\%$ to errors in building or delivering the model (implementation errors or delivery and technical failures), compared with $52\%$ and $47\%$ for the non-frontier agents.
Stronger agents thus make far fewer errors in building or delivering the model, but not fewer misreadings or errors of professional judgment.
These two causes account for almost all erroneous attempts of the strongest agents.

For example, on \texttt{DayMonthYear} from Section~\ref{sec:example}, only Fable 5.1-Cowork answers all $130$ questions correctly.
Seven agents misread the question:
they report the balance of share capital and additional paid-in capital ($933.6$ million) when the question asks for its cash-flow movement ($26.4$ million).
Other errors trace back to the difficulties described there:
Sol-Excel builds the daily balance sheet incorrectly, so its cash is off by about $47$ million, and also implements the iterative WACC incorrectly; Kimi K3 in the in-house harness sidesteps the scale of the task, replacing the required daily model with a monthly approximation over $120$ months.

\subsubsection{Other Quality Criteria}
Unlike failures in \texttt{Final calculation accuracy}, issues in quality criteria exhibit less commonality across agents, so we defer a more detailed account of agent-specific failures to Appendix~\ref{app:ModelAnalysis}.
Agents often apply presentation conventions such as freeze panes and color coding inconsistently.
Beyond presentation, several failure modes on the quality criteria are shared across agents.

\paragraph{Poor Reusability.}
The resulting models are often hard to reuse.
Timelines are often fixed to the horizon specified in the task instructions, so rerunning a financial model with a later start date or a longer duration requires manual editing.
On \texttt{DayMonthYear}, $11$ of the $14$ agents outside the in-house harness violate \texttt{Time-period flexibility} in this way.
This includes Fable 5.1-Cowork, the only agent that answers all $130$ questions correctly, but its model hardcodes the dates and the ten-year horizon and breaks when key inputs change substantially, so it violates two ultra important criteria (\texttt{Time-period flexibility} and \texttt{No errors under extreme-\allowbreak but-\allowbreak plausible inputs}) and therefore does not pass the task.

\paragraph{Poor Auditability.}
On harder tasks, the logic evolves into long, nested formulas that are very hard to review, which makes it particularly difficult to pinpoint where a model goes wrong.
Compared with the same models in the GUI harness, the Excel add-ins produce workbooks with worse auditability: the workbooks often leave out what helps a reviewer follow the model, such as named ranges for key inputs, sources for assumptions, and notes explaining the formulas.

\paragraph{Superficial Checks.}

Agents often build sense checks from their own calculations, so the checks mainly confirm internal consistency.
For instance, on the task \texttt{TheLiquidityEngine}, Astra in Codex sets each forecast to the average of the past five years and then checks that $\min(\text{past}) \le \text{forecast} \le \max(\text{past})$.
The check uses only the given data, not an outside benchmark or judgment, and, therefore, the average always passes it.
The master error flag can likewise show ``OK'' while some of its checks are broken, which recurs in dozens of workbooks across several agents.
In Grok~4.6's attempt on the task \texttt{Momentum}, for instance, the flag \texttt{IF(COUNTIF(D7:D30,"FAIL")=0,"OK","FAIL")} relies on \texttt{COUNTIF}, which counts only cells showing ``FAIL'' and ignores error values such as \texttt{\#VALUE!}, so the flag shows ``OK'' even when a check is broken.

\paragraph{Top Agent.} Even the top-ranked agent, Codex (GPT-6 Astra), shows these weaknesses:
its sense checks compare outputs only with figures from the same workbook, it never names its key inputs, and it prefills even small models with thousands of inactive formula rows or columns (Appendix~\ref{app:model_gpt6astra}).

\section{Conclusion}
We introduce \benchmarkname, a fully public benchmark of \tasknum diverse and difficult financial spreadsheet construction tasks, 
developed over approximately \benchmarkcreationhours person-hours.
Beyond evaluating a spreadsheet on accuracy, we propose a comprehensive taxonomy of \rubricnum criteria that details what constitutes a quality spreadsheet as codified in the industry. 
To support grading these criteria at scale, we also develop a corresponding agentic judge and validate its reliability against human judgment. 
Across state-of-the-art models and all available product harnesses ($20$+ agents), we find even the best performing agent passes only a small fraction of tasks.
We characterize how agents fail financial spreadsheet tasks in both numerical accuracy and in the high-level quality criteria we taxonomized.
Overall, modern agents still fall short of delivering high-quality spreadsheets that professional settings demand. 
We hope \benchmarkname provides a transparent, scalable testbed for agent-built spreadsheets and highlights the gap between the professional standard that finance practitioners rely on and what agents currently deliver.

\section*{Acknowledgment}
We gratefully acknowledge the support of OpenAI, Anthropic, and the Laude Institute, whose generous backing made \benchmarkname possible.

%% file: appendix.tex
\section{Rubric}
\label{app:FullRubric}
\subsection{Rubric Design and Applicability Annotation}
\label{app:RubricAnnotation}

\paragraph{Motivation.}
Task-specific rubrics would have to restate the common core of financial models for every task; they are costly to write, inconsistent across tasks, and impractical to validate per criterion.
BankerToolBench \citep{lau2026bankertoolbenchevaluatingaiagents}, for instance, uses a separate rubric for each of its 100 tasks and validates its grader only in aggregate; HealthBench \citep{arora2025healthbench} validates its grader per criterion only on the 34 criteria shared across conversations, not on its tens of thousands of conversation-specific ones.
We instead designed a single rubric that combines universal criteria, which capture this common core, with conditional criteria, which capture what differs between models.

\paragraph{Criteria Creation.}
The rubric was developed by finance experts and covers the quality dimensions they consider essential when building and reviewing spreadsheet-based financial models.
It deliberately extends beyond numerical correctness: the Accuracy category accounts for only 19\% of the total weight, and the remainder goes to criteria assessing, among other things, adherence to industry-standard formatting conventions, adaptability and maintainability, and formula transparency.
Each criterion is scored as satisfied or violated: it is satisfied only if no mistake is found (Appendix~\ref{app:Weighting}).
In the version given to the judge, each criterion consists of a name, a description, and a characterization of good and bad practice; Appendix~\ref{app:Criteria} restates these criteria in a form intended to be read, and the verbatim rubric is available in our repository.
For criteria that proved prone to misinterpretation during judge validation (Section~\ref{sec:rubric_judge_val}), the judge additionally receives clarifying notes, for example that a cover sheet is judged by its contents rather than by its name.

\paragraph{Universal and Conditional Criteria.}
Of the \rubricnum criteria, 93 are universal and apply to every task; the remaining 36 are conditional and apply only to tasks whose models require the feature they check.
These features include circular references (e.g., \texttt{Circularity -- valid justification}), multiple business units (\texttt{Multi-unit coverage}), alternative accounting treatments (\texttt{Capitalization switches}), stub periods (\texttt{Stub period handling}), a non-calendar fiscal year (\texttt{Real fiscal year handling}), a time dimension (e.g., \texttt{Single date driver} or \texttt{Chronological left-to-right}), multiple scenarios (\texttt{Cross-scenario summary}), and a workbook large enough to need navigation (\texttt{Linked table of contents}).
Table~\ref{tab:conditional} lists all conditional criteria by category; Accuracy, Formulas, Purpose \& Scope, and Rounding contain only universal criteria.

\paragraph{Applicability Annotation.}
For each task, its author annotated which of the 36 conditional criteria apply to the model the task requires; universal criteria apply by definition.
The annotation was completed before grading and does not depend on any agent's output.
A criterion annotated as not applicable is neither presented to the judge nor scored for that task, and its weight is redistributed within its category (Appendix~\ref{app:Weighting}).
Between 12 and 33 conditional criteria apply to each task, so that each task is graded on between 105 and 126 criteria.
Because every category contains at least one universal criterion, all \rubriccatnum categories contribute to the score of every task.

\paragraph{Categories and Importance Levels.}
The experts also assigned each criterion to one of \rubriccatnum categories, which group criteria by the aspect of the workbook they assess, and to one of four importance levels, which reflect how strongly a violation impairs the workbook.
Appendix~\ref{app:Criteria} introduces each category, and Appendix~\ref{app:Weighting} describes how categories and importance levels are converted into weights.
Of the \rubricnum criteria, 12 are ultra important, 41 highly important, 52 important, and 24 not absolutely crucial (Table~\ref{tab:importance}).

\paragraph{Grading.}
\texttt{Final calculation accuracy} is graded deterministically by comparing the agent's answers on the questions tab with those of the golden solution; a single wrong, missing, or hardcoded answer violates the criterion.
The equivalence rules given in its description (Appendix~\ref{app:Criteria}) disregard differences of form: an answer counts as correct if it rounds to the correct value at the required precision, differs only in display (e.g., a percentage given as a fraction, a negative number in parentheses, or a different date format), or, for costs and other outflows, differs only in sign.
A different unit scale, such as thousands instead of millions, is never accepted.
Of the other criteria, 19 are checked deterministically by fixed rules and the remaining 109 by the agentic judge (Appendix~\ref{app:Judge}).

\paragraph{Relation to the House Standards.}
The rubric itself is not disclosed to the agents.
Instead, the house standards (Appendix~\ref{app:HouseStandards}) state the conventions behind several criteria that vary between organizations, such as the change log, the linked table of contents, the checks tab with a single flag on the cover, and the sign convention, without revealing the criteria or their weights.
The rubric encodes principles shared across public codes \citep{icaew2018code}; the house standards fix the conventions those codes leave open, so that criteria such as \texttt{Consistent sign convention} have a single correct target.

\paragraph{Advantages.}
This design combines the coverage of task-specific rubrics with the rigor of a single shared rubric.
\textbf{Comprehensive.} Universal criteria set a common quality bar, while conditional criteria capture model-specific requirements, so every applicable rubric criterion is checked for every task, whatever model it requires.
\textbf{Efficient and extensible.} A new task requires no new criteria, only a yes/no relevance decision per conditional criterion, and even a new model type needs only the few criteria specific to it.
\textbf{Verifiable.} Because each criterion is defined once for all tasks, the judge can be tested on each criterion individually (Section~\ref{sec:rubric_judge_val}); and because each category carries the same weight in every task, the grading scale stays consistent across tasks and types of financial models.

\begin{table}[htbp]
\centering
\begin{tabular}{lrrr}
\toprule
Level of importance & Universal & Conditional & Total \\
\midrule
ultra important        & 9  & 3  & 12 \\
highly important       & 29 & 12 & 41 \\
important              & 41 & 11 & 52 \\
not absolutely crucial & 14 & 10 & 24 \\
\midrule
Total                  & 93 & 36 & 129 \\
\bottomrule
\end{tabular}
\caption{Number of universal and conditional criteria per level of importance.}
\label{tab:importance}
\end{table}

\begin{table}[htbp]
\centering
\small
\begin{tabular}{@{}p{0.27\linewidth}p{0.68\linewidth}@{}}
\toprule
Category & Conditional criteria \\
\midrule
Assumptions & Named ranges for key assumptions; Single date driver \\
Documentation & Linked table of contents \\
Error Checks & Error-check sheet; Master error flag \\
Flexibility & Multi-unit coverage; Capitalization switches; Reuse as planning tool; Time-period flexibility \\
Formatting & Actuals vs.\ forecast demarcation; Input cells data validation; Green font for cross-sheet links; Red font for external links; Currency stated; Print areas set; Page breaks set \\
Model Outputs \& Executive Summary & Cross-scenario summary \\
Potential Dangers & Macros/VBA disclosed; VBA vs.\ native alternatives \\
Structure & Cover sheet first; Timeline orientation; Deal vs.\ planning orientation; Real fiscal year handling; Stub period handling; Functional sheet separation; Logical tab order; Consistent tab structure; Freeze panes set; One timeline per sheet; Chronological left-to-right; Time period headers; Data-type designation (actuals vs.\ estimates); Roll-forward (corkscrew) structure for balances; Circularity -- valid justification; Circularity -- documentation; Circularity breaker switch \\
\bottomrule
\end{tabular}
\caption{The 36 conditional criteria by category. Accuracy, Formulas, Purpose \& Scope, and Rounding contain only universal criteria.}
\label{tab:conditional}
\end{table}

\subsection{Weighting of the Rubric Criteria}
\label{app:Weighting}

The final score aggregates the \rubricnum rubric criteria of Section~\ref{sec:rubric} (93 universal and 36 conditionally applicable) into a single score, combining each criterion's score with its weight.
The weighting procedure itself is a fixed formula, but the constants that go into it were set by judgment.
All inputs described below, as well as the assignment of criteria to categories, were determined by our finance experts.

\paragraph{Base Weights.}
Each criterion was assigned one of four levels of importance, which determines its base weight $b_i$:

\begin{center}
\begin{tabular}{lc}
\toprule
Level of importance & Base weight $b_i$ \\
\midrule
ultra important       & 12 \\
highly important      & 5  \\
important             & 2  \\
not absolutely crucial& 1  \\
\bottomrule
\end{tabular}
\end{center}

\noindent
The scale is intentionally convex rather than linear: within a category, violating a single ultra important criterion outweighs violating up to five merely important ones, since the former generally renders the workbook significantly less usable while the latter only impair its quality.
Across categories, relative importance is mediated by the category weights $W_c$ introduced below.

\paragraph{Discrimination Discount.}
Nine criteria can largely be met by retaining or restating what the input workbook already provides, such as the labeled assumptions tab or the task instructions.
For these, satisfying the criterion indicates mainly that the agent preserved or carried over existing material, rather than demonstrating financial-modeling skill, so we expect them to discriminate less between strong and weak agents than criteria that must be satisfied from scratch.
We retain them, because violating them is still a genuine shortcoming, but multiply their base weight by a discount factor $d_i = 0.35$.
All nine are universal, so they apply to every task; the remaining 120 criteria keep their full weight ($d_i = 1$).

\paragraph{Normalization and Category Weights.}
Let $\mathcal{A}_c$ be the set of criteria in category $c$ that apply to a given task. Within each category the discounted base weights are normalized, and the resulting in-category weight is then scaled by the fixed category weight $W_c$.
The overall weight of criterion
$i \in \mathcal{A}_c$ is

\begin{equation}
w_i \;=\;
\underbrace{\frac{b_i\, d_i}{\sum_{j \in \mathcal{A}_c} b_j\, d_j}}
          _{\text{in-category weight}}
\;\times\; W_c ,
\label{eq:weight}
\end{equation}

\noindent
and the score of an attempt on a task is
\begin{equation}
G \;=\; \sum_{c} \sum_{i \in \mathcal{A}_c} w_i \, s_i ,
\qquad s_i \in \{0,1\},
\label{eq:score}
\end{equation}
where $s_i$ is the score of criterion $i$: $s_i = 1$ if the judge or, for deterministically graded criteria, the fixed rule finds no mistake, and 0 otherwise.

By construction $\sum_{i \in \mathcal{A}_c} w_i = W_c$ for every category, and the category weights sum to 100\% (Table~\ref{tab:catweights}).
Neither the discrimination discount nor the exclusion of inapplicable criteria can therefore move weight between categories; both act only on the relative weights inside a category.
In particular, when a conditional criterion does not apply to a task it is dropped from $\mathcal{A}_c$, the remaining criteria of that category absorb its share in proportion to their own weights, and the category still carries a total weight of exactly $W_c$ in the final score.
Every category contains at least one universal criterion, so $\mathcal{A}_c$ is never empty.
Because $W_c$ is fixed independently of how many criteria a category contains, a category with many fine-grained criteria divides its budget more finely rather than accumulating a larger share of the score: the category weights express the importance of the category itself, not of the individual checks within it.
When all criteria apply, the 93 universal criteria account for 77\% of the total weight, and individual criterion weights range from 0.08\% (\texttt{File extension (.xlsx)}) to 10.36\% (\texttt{Final calculation accuracy}).

\paragraph{Worked Example.}
Assumptions carries $W_c = 7.0\%$ and contains seven criteria (Table~\ref{tab:catweights}): three highly important, two important, and two not absolutely crucial.
Four are discounted---two of the highly important ones and both important ones---so they enter with $5 \times 0.35 = 1.75$ and $2 \times 0.35 = 0.70$ in place of 5 and 2.
If all seven apply to the task, the denominator of Equation~\ref{eq:weight} is
\[
\sum_{j \in \mathcal{A}_c} b_j d_j
= 1.75 + 1.75 + 0.70 + 0.70 + 1 + 1 + 5 = 11.90 ,
\]
so a discounted highly important criterion carries
$\tfrac{1.75}{11.90} \times 7.0\% = 1.03\%$ of the final score, a discounted important one $\tfrac{0.70}{11.90} \times 7.0\% = 0.41\%$, each not
absolutely crucial one $\tfrac{1}{11.90} \times 7.0\% = 0.59\%$, and
\texttt{Single date driver}, highly important but not discounted,
$\tfrac{5}{11.90} \times 7.0\% = 2.94\%$.

The discount is visible in the first and last of these figures: \texttt{Centralized assumptions} and \texttt{Single date driver} are both highly important, yet carry 1.03\% and 2.94\% respectively.
Had the discounted one kept its full base weight, the denominator would have been 15.15 and it would have carried 2.31\%---so the discount leaves it with about 45\% of the weight it would otherwise have, rather than 35\%, because the reduction also shrinks the denominator it is divided by.

Two of the seven criteria are conditional.
In a compact task with no global constants worth naming, \texttt{Named ranges for key assumptions} does not apply, the denominator falls to 10.90, and the remaining six absorb its share in proportion: the discounted highly important criteria rise to 1.12\% and \texttt{Single date driver} to 3.21\%.
In both cases the category's weights still sum to 7.0\%.

\begin{table}[htbp]
\centering
\begin{tabular}{lrr}
\toprule
Category & \# Criteria & $W_c$ \\
\midrule
Accuracy                           & 3  & 19.0\% \\
Assumptions                        & 7  & 7.0\%  \\
Documentation                      & 10 & 4.0\%  \\
Error Checks                       & 14 & 13.0\% \\
Flexibility                        & 5  & 6.5\%  \\
Formatting                         & 36 & 9.5\%  \\
Formulas                           & 10 & 12.0\% \\
Model Outputs \& Executive Summary & 4  & 4.0\%  \\
Potential Dangers                  & 9  & 8.0\%  \\
Purpose \& Scope                   & 1  & 2.0\%  \\
Rounding                           & 3  & 1.0\%  \\
Structure                          & 27 & 14.0\% \\
\midrule
Total                              & 129 & 100.0\% \\
\bottomrule
\end{tabular}
\caption{Number of criteria and category weights $W_c$, in alphabetical order.}
\label{tab:catweights}
\end{table}

\subsection{Pass Rule and Human Baseline}
\label{app:HumanReference}

This section describes the standard that the pass rule reflects and explains why we do not report a human baseline, i.e., the pass rate of human modelers on the same tasks.

\paragraph{Standard of the Pass Rule.}
The pass rule requires correct results (\texttt{Final calculation accuracy}) and adherence to the modeling conventions that practitioners are trained to follow, but tolerates a limited number of violations at each importance level; this tolerance also absorbs occasional judge errors (Section~\ref{sec:rubric}).
Several observations indicate that this standard is attainable:
each criterion graded by the judge is satisfied by a golden solution, modified where necessary (Appendix~\ref{app:ValidationExamples}); the golden solutions show that every task can be solved correctly (Appendix~\ref{app:TaskDesign}); and the top agents already pass some of the tasks (Section~\ref{sec:results}).

\paragraph{Why No Human Baseline.}
Like quality standards in practice, the pass rule is defined by what the work requires, not by how often a particular group of humans meets it.
A human pass rate would also depend on choices for which no canonical setting exists:
the experience of the modelers, the time they are given, and whether their work is reviewed before grading.
Different choices could yield very different pass rates for the same tasks, so no single baseline would provide a stable reference point.
Even for one setting, a reliable baseline would require several qualified modelers per task, and thus considerable resources.
Instead, the benchmark relates its tasks to human capability through its annotations:
the difficulty levels reflect the expertise each task demands (Appendix~\ref{app:TaskDifficulty}), and the time estimates state how long a human modeler would need to build each model (Appendix~\ref{app:TimeEstimate}).

\subsection{Evaluation Criteria}
\label{app:Criteria}
For readability, the criteria below paraphrase the criteria given to the judge; the verbatim rubric is available in our repository.

Each criterion is tagged with three attributes.
Its level of importance is one of ultra important, highly important, important, or not absolutely crucial.
A conditional criterion applies only to tasks for which it is relevant.
A discounted criterion carries reduced weight because it can largely be met with what the input workbook already provides.
Appendix~\ref{app:Weighting} explains how these attributes enter the weighting.

\subsubsection{Category: Accuracy}

Accuracy covers whether the workbook returns the correct numbers for the task at hand, separating three parts that are frequently treated as one:
correct carry-over of the task inputs, correct calculation on those inputs, and presence of every required output.

\begin{classbox}{Accuracy}{Correctness of outputs, inputs, and deliverable coverage}

\classterm{Final calculation accuracy}{\textit{(ultra important · Conditional: no · Discounted: no)}
The final calculations must be accurate given the inputs and task instructions.
Hardcoded results are not accepted, even if the value happens to be correct, because it is impossible to verify whether the result was calculated in the workbook or copied from another program; only text answers, such as Yes or No, may be entered directly.
The rubric item is scored strictly:
a single incorrect, missing, or hardcoded answer is enough to violate it.
Differences of form, however, are not counted as errors.
A numerical answer is correct if it rounds to the ground-truth value at the precision required for that answer, or deviates from it by no more than one part in a million; where both values are rounded, a deviation of one unit in the last decimal place is accepted.
Whether answers are rounded as instructed is assessed under the Rounding category rather than here.
A percentage may be given as a fraction or vice versa, and values are compared irrespective of their display, so that negatives in parentheses, currency symbols, thousands separators, numbers stored as text, a dash for zero, or a different date format do not matter.
Answers stating that a value lies outside the model horizon, such as ``N/A'', are treated as equivalent to one another.
For costs and other outflows, an answer that differs from the ground truth only in sign is accepted, since sign conventions vary and the house standards prescribe negative values for these items; for all other items, including net, balance, and difference figures, the sign is part of the answer.
Differences in the unit scale of amounts, such as thousands instead of millions, are never accepted.}

\classterm{Starting values}{\textit{(highly important · Conditional: no · Discounted: no)}
The values the model starts from must be those given in the input workbook, carried over with the correct unit and scale.
Any deviation violates the criterion, whether a value differs from the input workbook or is carried over at the wrong magnitude, since an error at this point propagates into every calculation that depends on it.}

\classterm{Deliverable completeness}{\textit{(highly important · Conditional: no · Discounted: no)}
Every output, schedule, and deliverable specified in the task instructions must be present and fully built out.
The rubric item addresses coverage rather than correctness:
a workbook that completes only part of the task, or leaves a required schedule half-built, violates the criterion regardless of how accurate the finished portions are.}

\end{classbox}

\subsubsection{Category: Assumptions}

Assumptions covers how the inputs that drive the model are held and presented.
It asks whether the assumptions are gathered in one place and flow in a single direction into the calculations, whether each one is defined exactly once, and whether labels, units, and sources allow a reader to interpret and verify an input without tracing the formulas that use it.

\begin{classbox}{Assumptions}{Central, uniquely defined, and documented model inputs}

\classterm{Centralized assumptions}{\textit{(highly important · Conditional: no · Discounted: yes)}
All assumptions should be summarized at the beginning of the model or in a separate assumptions sheet, rather than hidden in calculation cells or scattered throughout the model with no single source of truth.
Assumptions should flow in one direction, from the assumptions sheet into the calculations and never the reverse.
A clearly labeled, self-contained input block on a build sheet is permitted, provided that it is the sole source for that driver and is noted in the central assumptions sheet.}

\classterm{Unique assumptions}{\textit{(highly important · Conditional: no · Discounted: yes)}
Each assumption is defined only once.
Duplicates or repetitions in cells or worksheets violate the criterion, since a value stored in multiple locations may be updated in one of those locations but overlooked in the others.}

\classterm{Assumption labeling}{\textit{(important · Conditional: no · Discounted: yes)}
Each assumption must have a clear, descriptive label identifying what it represents.
Vague labels such as ``Rate 1'' or ``Input'' do not satisfy the criterion, as they name a cell without explaining the quantity it holds.}

\classterm{Assumption units}{\textit{(important · Conditional: no · Discounted: yes)}
Each assumption must display an explicit unit of measure, such as \%, \$, multiple, or days, so the reader can interpret values without guessing.}

\classterm{Assumption sources}{\textit{(not absolutely crucial · Conditional: no · Discounted: no)}
Each assumption references a specific, verifiable source, for example a numbered section of the task instructions or a page of a company filing, dated where a date is meaningful, so that reviewers can verify the input and judge whether it has gone stale.
Where no external source exists, the assumption must be labeled explicitly as a judgment or estimate and carry an owner and a date; a bare reference to management does not satisfy the criterion.}

\classterm{Named ranges for key assumptions}{\textit{(not absolutely crucial · Conditional: yes · Discounted: no)}
Truly global and stable inputs, such as the tax rate or the valuation date, should use named ranges so that formulas can be read without tracing cell references.
Names must be self-explanatory and follow a consistent convention.
The rubric item penalizes overuse as well as absence, since naming everything hides the underlying references and clutters the Name Manager.}

\classterm{Single date driver}{\textit{(highly important · Conditional: yes · Discounted: no)}
A single, clearly labeled anchor date, that is, a valuation date, as-of date, or model start date, should drive discount periods, stub periods, mid-year conventions, and roll-forwards across the model.}

\end{classbox}

\subsubsection{Category: Documentation}

Documentation covers what the workbook tells a reader who did not build it.
It spans the statements that frame the model as a whole, that is, its purpose, scope, methodology, design decisions, and revision history, and the navigational and explanatory material that makes the sheets usable, from tab names and a linked table of contents to a glossary of conventions and commentary on individual formulas.

\begin{classbox}{Documentation}{Recorded purpose, conventions, and guidance for the reader}

\classterm{Formula commentary for non-obvious formulas}{\textit{(important · Conditional: no · Discounted: no)}
Formulas that are not obvious or are complex should be accompanied by an explanation, such as a cell comment, a note, or an explanation in an adjacent comment column.
The explanation must give the business rationale or the methodology behind the formula; a note that restates what the formula computes, such as ``calculates margin'', does not satisfy the criterion.}

\classterm{Glossary of conventions}{\textit{(important · Conditional: no · Discounted: no)}
The workbook documents in one place the colors, abbreviations, and conventions it uses, so that a reader can decode the sheets rather than reconstruct the conventions from context.}

\classterm{How to use guide}{\textit{(important · Conditional: no · Discounted: no)}
A ``How to use this model'' tab or section should walk users through the inputs they may change, how to select scenarios, and where to read the outputs, explaining how to operate the model from the user's perspective.}

\classterm{Linked table of contents}{\textit{(important · Conditional: yes · Discounted: no)}
The model should include a table of contents with working hyperlinks to each worksheet.
For very small workbooks, roughly three sheets or fewer beyond the tabs provided in the input workbook, a table of contents may reasonably be omitted.}

\classterm{Clear tab names}{\textit{(important · Conditional: no · Discounted: no)}
Every tab has a clear, descriptive, professional name that states what the sheet contains.
Default names such as ``Sheet1'' or ``Sheet3'' are unacceptable, as are cryptic abbreviations not explained in a glossary.}

\classterm{Purpose documentation}{\textit{(highly important · Conditional: no · Discounted: yes)}
The model should include documentation that clearly articulates its intended purpose and the business question it answers.}

\classterm{Scope documentation}{\textit{(important · Conditional: no · Discounted: yes)}
Documentation must explicitly list what the model covers, what it excludes, and the boundary assumptions that justify each exclusion.}

\classterm{Key design decisions documentation}{\textit{(important · Conditional: no · Discounted: no)}
Documentation explains the key design decisions and the reasoning behind them, for instance the choice of valuation methodology, the granularity of the build, or the consolidation approach.}

\classterm{Version history / change log}{\textit{(not absolutely crucial · Conditional: no · Discounted: no)}
The model should include a version history or change log recording what changed, when, and by whom, either on a dedicated tab or in a clearly labeled section of the cover sheet.}

\classterm{Methodology documentation}{\textit{(important · Conditional: no · Discounted: no)}
Methodology documentation should explain the modeling approach, conventions, and assumptions for stakeholders and users.
It may sit outside the workbook, provided the model points to it.}

\end{classbox}

\subsubsection{Category: Error Checks}

Error Checks covers whether the model is internally consistent and whether it demonstrates that consistency to the reader.
It combines arithmetic consistency, that is, subtotals matching the totals they roll into and a given value holding wherever it appears, with the absence of visible errors and warnings, disciplined use of error suppression, consistent units across the workbook, and the dedicated check rows, check sheet, and master flag that surface failures without manual tracing.
A smaller set of issues relates to workbook hygiene, specifically file size, recalculation time, and the Name Manager.

\begin{classbox}{Error Checks}{Internal consistency, visible errors, and built-in validation}

\classterm{Subtotals tie to totals}{\textit{(highly important · Conditional: no · Discounted: no)}
Subtotals must be correct sums of their line items and must tie to the totals into which they roll.
The check runs against omission and double counting alike; a line item left outside a sum range, or captured twice within it, violates the criterion even where the resulting figure looks plausible.}

\classterm{No formula errors}{\textit{(ultra important · Conditional: no · Discounted: no)}
The model must be free of visible Excel errors like \#REF!, \#N/A, \#DIV/0!, \#VALUE!, \#NUM!, \#NULL!, \#NAME?, and, in modern Excel, \#SPILL! and \#CALC!.}

\classterm{No unresolved cell warnings}{\textit{(highly important · Conditional: no · Discounted: no)}
The model should contain no unresolved cell warnings.
Such warnings point to numbers stored as text, cells omitted from ranges, or inconsistent formulas, and left standing they undermine trust in the model even where the underlying figure is correct.
Warnings that are known false positives, such as a year deliberately stored as text, should be dismissed or explained rather than left visible.}

\classterm{IFERROR/IFNA used only for expected errors}{\textit{(highly important · Conditional: no · Discounted: no)}
Error handling such as IFERROR or IFNA should suppress only errors the model produces intentionally, for example a lookup returning \#N/A in early periods before the data exists.
Genuine errors must remain visible for investigation, so blanket suppression that masks them violates the criterion.}

\classterm{Reasonable file size}{\textit{(not absolutely crucial · Conditional: no · Discounted: no)}
File size should be proportionate to the model's content.
As a guideline, a typical single-entity operating or valuation model rarely exceeds 10 MB, and sizes above 25 MB usually indicate oversized used ranges, legacy formatting, embedded objects, or model bloat.
A larger file is acceptable where the content justifies it, such as an embedded data extract, provided the reason is documented.}

\classterm{Reasonable recalculation time}{\textit{(not absolutely crucial · Conditional: no · Discounted: no)}
A full recalculation in under five seconds is a reasonable benchmark for a typical model, and noticeable lag when inputs change impedes interactive use and scenario analysis.
Long recalculation times commonly follow from volatile functions such as INDIRECT, OFFSET, or NOW, whole-column references, array formulas over large ranges, or overly complex formula chains.}

\classterm{Cross-sheet value agreement}{\textit{(highly important · Conditional: no · Discounted: no)}
A value for a specific item must match everywhere that item appears across sheets and outputs.
Discrepancies of this kind, for instance net income differing between the income statement and the cash flow statement, are treated as serious errors rather than presentational inconsistencies.}

\classterm{Clean Name Manager}{\textit{(not absolutely crucial · Conditional: no · Discounted: no)}
The Name Manager should contain only current, actively referenced defined names, with orphaned, outdated, or broken entries removed before delivery.}

\classterm{Dedicated check rows}{\textit{(highly important · Conditional: no · Discounted: no)}
Dedicated check rows should validate key calculations, such as totals tying or the balance sheet balancing, so that internal consistency is visible at a glance.
The rubric item is violated where the reader must re-derive totals or trace formulas manually because no row reports whether its check holds.}

\classterm{Error-check sheet}{\textit{(highly important · Conditional: yes · Discounted: no)}
A dedicated error-check sheet or summary should centralize the integrity tests, among them balance checks, cash ties, and roll-forward checks, and flag failures automatically.
Checks that exist only on the calculation sheets, without being summarized centrally, do not satisfy the criterion, since a reviewer must then tab through the entire workbook to assess integrity.}

\classterm{Master error flag}{\textit{(highly important · Conditional: yes · Discounted: no)}
A master error flag on the cover sheet or summary sheet should immediately indicate whether a check has failed, and conditional formatting should be used to ensure that the error status is unmistakably clear.}

\classterm{Consistent units}{\textit{(highly important · Conditional: no · Discounted: no)}
Units must be consistent across the model, with no mixing of scales, for example all figures in millions or all in thousands.
Where a different unit is genuinely appropriate, such as a price per unit, a headcount, or a per-share figure, it must be labeled explicitly.}

\classterm{Units carried through correctly}{\textit{(important · Conditional: no · Discounted: no)}
Units defined on the assumptions sheet must carry through unchanged to the calculation and output sheets.
A scale change mid-model violates the criterion, except where the conversion is needed to bring all units on a calculation sheet to the same magnitude.}

\classterm{Justified repeated calculations}{\textit{(important · Conditional: no · Discounted: no)}
The same calculation should not be computed in more than one place without a clear reason.
Where a figure has to appear on several tabs, it is shown by cell reference rather than re-derived, so that each quantity retains a single point of calculation.}

\end{classbox}

\subsubsection{Category: Flexibility}

Flexibility describes the extent to which the model can accommodate changes without having to be rebuilt.
The key considerations here are whether the architecture is driver-based and formula-driven, whether time periods can be adjusted, extended, or re-granularized without manual effort, and whether the structure anticipates situations that would otherwise necessitate a redesign:
the consolidation of multiple business units, a switch between capitalization and expense recognition, and the use of the model beyond its original intended purpose.

\begin{classbox}{Flexibility}{Capacity to absorb changes without rebuilding}

\classterm{Multi-unit coverage}{\textit{(not absolutely crucial · Conditional: yes · Discounted: no)}
Where multiple business units, divisions, or locations are in scope, the model should use a segmented structure with consolidation logic and, where applicable, inter-company eliminations and transfer pricing.}

\classterm{Capitalization switches}{\textit{(not absolutely crucial · Conditional: yes · Discounted: no)}
Where the task warrants alternative accounting treatment, switches should allow expenditures to be capitalized or expensed without rebuilding the model.}

\classterm{Reuse as planning tool}{\textit{(important · Conditional: yes · Discounted: no)}
Where a model built to evaluate a business opportunity is likely to be reused afterwards, for instance as an ongoing planning tool, the design should support that second life rather than suit one-time evaluation alone.}

\classterm{Time-period flexibility}{\textit{(ultra important · Conditional: yes · Discounted: no)}
Assumptions and time periods must be able to be updated, extended, or subdivided without requiring manual rework.
The rubric item is violated where updating the periods means rebuilding parts of the model by hand.}

\classterm{Driver-based, formula-linked architecture}{\textit{(ultra important · Conditional: no · Discounted: no)}
The model should be driver-based and formulaic, so that inputs can be changed and new line items added quickly.
Reliance on hardcoded values or rigid structures that force manual rework to update or extend the model violates the criterion.}

\end{classbox}

\subsubsection{Category: Formatting}

Formatting refers to the visual conventions that the workbook uses to indicate the function of each cell and the magnitude of each number.
Many of these elements relate to industry-standard color coding, that is, blue for hardcoded values, black for calculations, green for cross-sheet links, and red for external references, as well as the need to exercise restraint so that additional colors, conditional formatting, borders, and shading do not conflict with this color coding.
The remaining points concern number formats and finance conventions for negative numbers, zeros, separators, and currencies; the alignment and dimensions of rows and columns; and sheet- and workbook-level settings, such as zoom level, merged cells, print settings, and file format.

\begin{classbox}{Formatting}{Color-coding conventions, number formats, and visual consistency}

\classterm{Actuals vs.\ forecast demarcation}{\textit{(highly important · Conditional: yes · Discounted: no)}
Actuals, which are historical and fixed, and forecasts, which are formula-driven from editable assumptions, should be separated by labeling, formatting, or a vertical divider, so that the point of transition between them is obvious.}

\classterm{Input cells marked}{\textit{(highly important · Conditional: no · Discounted: yes)}
Input cells must be visually marked, by color, shading, or border, so that a user sees immediately where values may safely be entered and which cells carry formulas.}

\classterm{Input cells data validation}{\textit{(important · Conditional: yes · Discounted: no)}
Key input cells should use data validation to constrain entries to sensible ranges, types, or lists, so that an out-of-range entry is rejected or flagged at the point of entry rather than breaking the model silently.}

\classterm{Reasonable number formats}{\textit{(important · Conditional: no · Discounted: no)}
Number formats should be sensible and professional:
no excess decimals, currency symbols where they are needed, and thousands separators applied.}

\classterm{Consistent number formats}{\textit{(highly important · Conditional: no · Discounted: no)}
Comparable values must carry the same number format throughout the model, including the number of decimals displayed.}

\classterm{No bright-yellow highlighting}{\textit{(highly important · Conditional: no · Discounted: no)}
Bright yellow highlighting used during construction to flag placeholder, TBD, or work-in-progress items must be removed from the final deliverable, since it signals unfinished work.
Yellow used intentionally for a defined purpose, such as conditional formatting on error-check rows, a documented convention for marking input cells, or flagging assumptions that require periodic review, remains acceptable provided the glossary or conventions guide explains it; undocumented yellow cannot be told apart from a leftover marker.}

\classterm{Blue font for hardcoded inputs}{\textit{(highly important · Conditional: no · Discounted: no)}
Hardcoded inputs must use blue font, following the industry-standard color-coding convention.}

\classterm{Black font for calculations}{\textit{(highly important · Conditional: no · Discounted: no)}
Formulas and calculations must use black font, following the industry-standard color-coding convention.}

\classterm{Green font for cross-sheet links}{\textit{(important · Conditional: yes · Discounted: no)}
Cells that link to other sheets should use green font, following the industry-standard color-coding convention.}

\classterm{Red font for external links}{\textit{(not absolutely crucial · Conditional: yes · Discounted: no)}
Cells containing external links, that is, references to other workbooks, should use red font, so that they cannot be mistaken for internal formulas or cross-sheet links.}

\classterm{Consistent color coding}{\textit{(highly important · Conditional: no · Discounted: no)}
The color-coding scheme (blue inputs, black formulas, green cross-sheet links, and red external links) must be applied consistently across the entire model, so that the purpose of a cell is clear from its color alone.
Inconsistent coding is worse than none, because it makes color an unreliable signal.}

\classterm{Consistent formatting}{\textit{(highly important · Conditional: no · Discounted: no)}
Fonts, sizes, alignment, borders, and section styling should be consistent for comparable content across all sheets.
The comparison is between sheets, not only within one.}

\classterm{Limited color palette}{\textit{(important · Conditional: no · Discounted: no)}
Colors that go beyond the usual standard colors of blue, black, green, and red---whether used as text color, background color, or tab color---should form a small, consistent color palette rather than a wide range of shades.}

\classterm{No colors that mimic standard coding}{\textit{(important · Conditional: no · Discounted: no)}
Additional colors must not closely resemble the blue, black, green, or red of the standard convention, since a near match misleads the reader about the type of the cell.}

\classterm{Minimal non-standard color usage}{\textit{(important · Conditional: no · Discounted: no)}
The number of cells and elements carrying color beyond the standard coding should be limited, so that additional formatting does not clutter the sheet or distract from the intentional coding scheme.}

\classterm{Sparing conditional formatting}{\textit{(important · Conditional: no · Discounted: no)}
Conditional formatting should be confined to error-check rows, data-validation indicators, error flags, and threshold alerts, and should not override the standard color-coding convention.}

\classterm{Sparing border usage}{\textit{(not absolutely crucial · Conditional: no · Discounted: no)}
Borders should be used sparingly, to partition sections and separate key areas of the model; applied throughout, they produce a cluttered appearance.}

\classterm{Header \& title shading}{\textit{(not absolutely crucial · Conditional: no · Discounted: no)}
Cell shading is reserved for headers, titles, and documented markers such as input cells, so that headings are visually distinct from data rows without shading spreading across whole sections.}

\classterm{Totals visually distinguished}{\textit{(important · Conditional: no · Discounted: no)}
Totals and subtotals should be bolded, underlined, or otherwise set apart from the individual items that make them up.}

\classterm{Consistent zoom level}{\textit{(important · Conditional: no · Discounted: no)}
Every sheet should use the same zoom level, commonly 75\%, 80\%, or 100\%, so that the model does not jump in scale as the reader moves between tabs.}

\classterm{Active cell reset to A1 on all sheets}{\textit{(not absolutely crucial · Conditional: no · Discounted: no)}
Before delivery, the cursor should be parked at cell A1, or at the designated home cell of the sheet, on every tab, so that each one opens at its intended starting position.}

\classterm{Text left-aligned}{\textit{(important · Conditional: no · Discounted: no)}
Body text and row labels should be left-aligned.
Column headers above numeric columns, such as period labels, are the standard exception and may be right-aligned or centered so that they sit over the values beneath them.}

\classterm{Numbers right-aligned}{\textit{(important · Conditional: no · Discounted: no)}
Numerical values should be right-aligned within their cells, consistently throughout the model.}

\classterm{Negatives in parentheses}{\textit{(important · Conditional: no · Discounted: no)}
Negative values should be displayed in parentheses rather than with a minus sign, following finance convention.}

\classterm{Zeros as dashes}{\textit{(not absolutely crucial · Conditional: no · Discounted: no)}
Zero values should be displayed as dashes rather than as the digit zero, following finance convention.}

\classterm{Thousands separators}{\textit{(not absolutely crucial · Conditional: no · Discounted: no)}
Monetary and other large quantity values should carry a thousands separator.
Values that conventionally carry none, namely years, percentages, and valuation multiples, are excluded, and applying a separator to them, as in a year shown as 2,024, is itself a violation.}

\classterm{Currency stated}{\textit{(highly important · Conditional: yes · Discounted: no)}
The currency used in the model must be stated explicitly as a unit, and stated per block or column, which matters in particular where FX is involved.
Values without a stated currency are ambiguous.}

\classterm{Sufficient column widths}{\textit{(important · Conditional: no · Discounted: no)}
Column widths must be sufficient to display all values in full, with no truncated text and no \#\#\# in place of a number.}

\classterm{Reasonable column widths}{\textit{(not absolutely crucial · Conditional: no · Discounted: no)}
Columns should be no wider than their content requires, so that several full-width columns fit on screen at once, and long text should be wrapped rather than accommodated by widening the column.}

\classterm{Consistent column widths}{\textit{(important · Conditional: no · Discounted: no)}
Calculation columns on a sheet should share a consistent width.
Header, text, and unit columns may be wider.}

\classterm{Consistent row heights}{\textit{(important · Conditional: no · Discounted: no)}
Row heights should be consistent within each tab, with header rows permitted to be taller.}

\classterm{Reasonable row heights}{\textit{(not absolutely crucial · Conditional: no · Discounted: no)}
Rows should by default be one line high.
Taller rows are acceptable but should be used sparingly.}

\classterm{No merged cells}{\textit{(highly important · Conditional: no · Discounted: no)}
Merged cells interfere with the spreadsheet's grid logic, resulting in incorrect formulas, failed sorting, and reference errors.
Center Across Selection should be used instead, as it achieves the same visual effect for horizontal merges while preserving grid integrity.}

\classterm{Print areas set}{\textit{(not absolutely crucial · Conditional: yes · Discounted: no)}
Print areas should be defined on each tab so that printing or export to PDF captures the relevant content and nothing else, rather than empty space or extraneous material.}

\classterm{Page breaks set}{\textit{(not absolutely crucial · Conditional: yes · Discounted: no)}
Page breaks should be placed so that each tab prints legibly, whether to PDF or hard copy, without splitting tables, sections, or key content awkwardly across pages.}

\classterm{File extension (.xlsx)}{\textit{(important · Conditional: no · Discounted: yes)}
The workbook should be saved in a standard Excel format, that is, .xlsx, or .xlsm where VBA or macros are present.
Legacy or non-standard formats, such as .xls, .csv, or an export from another spreadsheet application, violate the criterion.}

\end{classbox}

\subsubsection{Category: Formulas}

Formulas covers how the calculations themselves are written:
which functions are used, how much logic a single cell is allowed to carry, whether the same logic is expressed identically wherever it recurs, and whether references are anchored so that a copied formula pulls the intended cells.
The rubric places greater emphasis on readability and verifiability than on conciseness, since a formula that a reviewer cannot follow cannot be verified.

\begin{classbox}{Formulas}{Function choice, formula construction, and referencing}

\classterm{Use standard formula set}{\textit{(important · Conditional: no · Discounted: no)}
The model should keep to the formula set that is standard in financial modeling and avoid uncommon constructions such as FILTERXML, WEBSERVICE, or recursive LAMBDAs where standard alternatives exist.}

\classterm{Modern formula usage}{\textit{(important · Conditional: no · Discounted: no)}
Modern functions should be preferred where they improve robustness or readability and backward compatibility is not a constraint:
e.g., XLOOKUP over VLOOKUP, IFS or SWITCH over deeply nested IF, and LET for readable intermediate variables.
INDEX with (X)MATCH remains a fully acceptable professional standard and is not penalized.
Unless the task instructions state otherwise, the client is assumed to run a current version of Excel; where an older version must be supported, or where a legacy construction computes more efficiently in a very large model, the legacy equivalent is acceptable and the trade-off should be documented.
The typical violation is a fragile legacy construction with no such reason given, most often VLOOKUP with a hardcoded column index or IFs nested many levels deep.}

\classterm{Avoid volatile functions}{\textit{(highly important · Conditional: no · Discounted: no)}
Volatile functions, among them INDIRECT, OFFSET, TODAY, NOW, RAND, RANDBETWEEN, INFO, and CELL, recalculate on every change in the workbook and should be avoided where possible.
Their use is acceptable where clearly justified, for instance for displaying the current date.}

\classterm{No hardcoded numbers in formulas}{\textit{(ultra important · Conditional: no · Discounted: no)}
Formulas must reference assumption cells rather than embed constants, as when a tax rate is written directly into a multiplication.
The only accepted exceptions are fixed conversion constants, such as 12 months, 365 days, or 100 for a percentage conversion, and technical remnants that are not true inputs.}

\classterm{Simple, stepwise formulas}{\textit{(highly important · Conditional: no · Discounted: no)}
Formulas should be simple, readable, and broken into logical steps through helper columns, helper rows, or LET.
As a practical guideline (the three-clause rule), a formula nesting more than about three functions deep should be split, although some complexity is acceptable where splitting would bloat the structure of the model.
Monolithic formulas that obscure the logic and make debugging difficult violate the criterion, as does the absence of helper columns where they are clearly required.}

\classterm{Consistent formulas across blocks}{\textit{(ultra important · Conditional: no · Discounted: no)}
Formulas should be identical across a row, that is, along the timeline, and across similar blocks, so that copying works as intended and any exception is immediately obvious.}

\classterm{Inconsistent formulas flagged}{\textit{(highly important · Conditional: no · Discounted: no)}
Where formulas must be inconsistent, for example in the first period of a roll-forward, the cells should be flagged visually, whether by distinctive fill, an exception-marker column, a cell comment, or an error-check flag, so that a reviewer can distinguish an intentional break from an error.}

\classterm{Correct reference anchoring}{\textit{(ultra important · Conditional: no · Discounted: no)}
Absolute, relative, and mixed references (\$) must be anchored correctly.
Mis-anchoring is a dangerous defect because a copied formula then pulls the wrong data silently, without producing any visible Excel error.}

\classterm{Consistent reference anchoring}{\textit{(important · Conditional: no · Discounted: no)}
Anchoring conventions (\$) should be applied consistently across comparable formulas and blocks.}

\classterm{Avoid whole-column references}{\textit{(not absolutely crucial · Conditional: no · Discounted: no)}
Whole-column and whole-row references, especially in array formulas, SUMPRODUCT, or user-defined functions, can force Excel to evaluate all 1{,}048{,}576 rows for every formula.
Defined ranges, trimmed references, table columns, or named ranges should be used instead.}

\end{classbox}

\subsubsection{Category: Model Outputs \& Executive Summary}

Model Outputs \& Executive Summary covers the results layer of the model, that is, the metrics it reports and the page on which they are presented.
It asks whether the output metrics relevant to the model's purpose are computed in a dedicated section, whether the results are checked for plausibility against outside reference points as part of the model rather than alongside it, and whether the summary recalculates with the inputs rather than simply presenting a manually created snapshot.

\begin{classbox}{Model Outputs \& Executive Summary}{Reported results, key metrics, and their plausibility}

\classterm{Output reasonableness check}{\textit{(important · Conditional: no · Discounted: no)}
Outputs should hold up under a directional sense check against intuition, industry benchmarks, comparable companies or transactions, or third-party estimates, carried out in an explicit sense-check section of the model.
The metrics to be checked follow from the model type:
growth and margins for an operating model, IRR and money multiple for a transaction model, debt service coverage for project finance, payback and NPV for a capex case, implied multiples for a valuation.}

\classterm{Key ratios \& outputs section}{\textit{(important · Conditional: no · Discounted: no)}
The model should include a dedicated key-ratios section computing the output metrics relevant to its purpose, for example margins, returns, leverage and coverage, growth rates, valuation multiples, or transaction returns.}

\classterm{Cross-scenario summary}{\textit{(not absolutely crucial · Conditional: yes · Discounted: no)}
A summary should display the results of all scenarios or cases in one view, rather than leaving them buried in separate sheets.}

\classterm{Automated summary}{\textit{(highly important · Conditional: no · Discounted: no)}
The summary page must update automatically when inputs change.
Values that are hardcoded or pasted in rather than formula-linked do not propagate and leave the summary to go stale.}

\end{classbox}

\subsubsection{Category: Potential Dangers}

Potential Dangers covers the properties of a workbook that mislead a reviewer or break under use:
content that is hidden rather than shown, dependencies on files outside the workbook, code that adds risk without adding capability, and settings under which displayed values need not reflect the current inputs.
Two items go beyond the file itself to the model's behavior, namely the consistency of its sign convention and its stability when assumptions are pushed to the edge of the plausible range.

\begin{classbox}{Potential Dangers}{Concealment, fragility, and unsafe workbook settings}

\classterm{No hidden sheets}{\textit{(important · Conditional: no · Discounted: no)}
The file should contain no hidden or very hidden sheets; any legitimate one must be disclosed and justified.}

\classterm{No hidden rows/columns}{\textit{(highly important · Conditional: no · Discounted: no)}
The file should contain no hidden rows or columns other than those collapsed within a grouping.
Groupings should be used where content needs to be collapsed.}

\classterm{No white-on-white hiding}{\textit{(important · Conditional: no · Discounted: no)}
No content should be concealed by formatting, in particular white font on a white background or a number format that renders a populated cell blank.}

\classterm{No external links}{\textit{(highly important · Conditional: no · Discounted: yes)}
The file should be self-contained, without links to other workbooks, unless the task instructions require them.
Such links break when the linked file moves, produce \#REF! errors, and hide dependencies.}

\classterm{Macros/VBA disclosed}{\textit{(important · Conditional: yes · Discounted: no)}
Macros and VBA add complexity, security risk, and maintenance burden, and should be avoided. Any that are present must be disclosed, reviewed, and justified.}

\classterm{VBA vs.\ native alternatives}{\textit{(not absolutely crucial · Conditional: yes · Discounted: no)}
Where VBA is present, it should handle tasks that native Excel features cannot perform, rather than duplicating them---for example, by using loops where an array formula would suffice, or by creating custom dialog boxes where data validation would be enough.}

\classterm{Consistent sign convention}{\textit{(ultra important · Conditional: no · Discounted: no)}
One sign convention, such as costs negative and revenues positive, must hold across the entire model, with particular attention to EBITDA add-backs, cash flow statement conventions, and the treatment of expenses on the income statement.
The same item shown with different signs in comparable places, for instance a cost entered as positive on one sheet and negative on another, violates the criterion.}

\classterm{No errors under extreme-but-plausible inputs}{\textit{(ultra important · Conditional: no · Discounted: no)}
The model should still work when key assumptions are changed substantially, for instance a useful life tripled, a valuation date shifted by twelve months, or revenue halved.
Stress of this kind exposes hidden breaks, balance sheet imbalances, divide-by-zero errors, broken roll-forwards, and nonsensical results such as negative shares outstanding.}

\classterm{Automatic calculation mode}{\textit{(important · Conditional: no · Discounted: no)}
The workbook should be delivered with calculation set to automatic, since manual mode means the displayed values need not reflect the inputs currently shown.
Where intentional circularity requires it, iterative calculation must be enabled and that requirement stated.}

\end{classbox}

\subsubsection{Category: Purpose \& Scope}

Purpose \& Scope holds a single item, which judges the model as a whole against the job it was built to do.
Where the other categories examine components, this one asks whether the overall architecture is the right one, given that a model can satisfy every local convention and still be built at the wrong level of detail for its purpose.

\begin{classbox}{Purpose \& Scope}{Fit between model architecture and stated purpose}

\classterm{Structure fits purpose}{\textit{(ultra important · Conditional: no · Discounted: no)}
The overall structure, that is, the calculation flow, the number of worksheets, and the level of detail, should be appropriate to the model's stated purpose.
An illogical flow, too many or too few sheets, or a level of detail mismatched to the question being answered violates the criterion.}

\end{classbox}

\subsubsection{Category: Rounding}

Rounding covers where in the calculation chain precision may be given up and how the resulting convention is communicated.
It separates the internal treatment, where full precision must survive to the end of the chain, from the presentation of the final figures, where rounding to a sensible precision conveys the certainty the numbers actually carry.

\begin{classbox}{Rounding}{Precision retained, presented, and documented}

\classterm{Full precision in intermediates}{\textit{(highly important · Conditional: no · Discounted: no)}
Intermediate calculations must retain full precision, with rounding applied only to the final outputs presented to users.
Display formatting, such as showing no decimals, is acceptable because it leaves the underlying value untouched.
Rounding inside the model is legitimate only where the quantity is genuinely discrete, for instance headcount, units sold, or share counts, or where a contract or the task instructions specify it; unjustified rounding compounds precision errors through the model.}

\classterm{Rounded outputs}{\textit{(important · Conditional: no · Discounted: no)}
Final outputs should be rounded to a sensible, readable precision that communicates the appropriate level of certainty, rather than displayed at full, false precision.
The task may dictate the precision to be used.}

\classterm{Rounding documented}{\textit{(important · Conditional: no · Discounted: no)}
Rounding conventions should be stated in visible proximity to the rounded figures, in the manner of a heading that gives the unit and the rounding step.
A convention that is wrong, undocumented, or stated far from the figures it governs violates the criterion.}

\end{classbox}

\subsubsection{Category: Structure}

Structure covers the architecture of the workbook and the layout within each sheet.
At workbook level it asks which components are present, how functions are separated across tabs, how the tabs are ordered, and whether the architecture matches the entity's fiscal calendar and the model's deal or planning orientation.
At sheet level it asks how the timeline is laid out and labeled, how calculations are grouped and headed, how labels align with the values they describe, and how balances and circular references are constructed.

\begin{classbox}{Structure}{Sheet architecture, timeline layout, and section organization}

\classterm{Essential components present}{\textit{(highly important · Conditional: no · Discounted: no)}
The model should include all essential components:
cover sheet, table of contents, summary or dashboard, assumptions, scenarios where applicable, calculation sheets, financial statements where applicable, error-check sheet, and output data.
The components scale with the size and purpose of the model, so in a small model several may reasonably sit on fewer sheets, provided each function remains clearly delineated.}

\classterm{Cover sheet first}{\textit{(not absolutely crucial · Conditional: yes · Discounted: no)}
The cover sheet should be the first tab, setting the context and orienting the user on opening, and typically stating the model's title, purpose, owner, version, and date.}

\classterm{Timeline orientation}{\textit{(important · Conditional: yes · Discounted: no)}
Within calculation tabs the timeline should run horizontally as standard, with periods across columns and line items down rows, applied consistently.
Period columns should align across tabs, so that a given column refers to the same period everywhere in the workbook, absent a shift in period length; misalignment between sheets creates link risk.
A vertical timeline is acceptable where the content justifies it, typically a register or data table whose records need several attribute columns visible at once and whose periods are few enough to scan, but it should not be the default, since it forces vertical scrolling to compare periods and breaks the copy-across formula convention.}

\classterm{Vertical alignment of same-type data}{\textit{(important · Conditional: no · Discounted: no)}
Columns and rows carrying the same type of information should be aligned vertically across the sheet, and the arrangement of the different calculations should follow a logical flow.}

\classterm{Label-to-value alignment}{\textit{(important · Conditional: no · Discounted: no)}
Subtitles, categories, and calculation names should be aligned with the values they describe, so that the reader can associate each label with its data without ambiguity.}

\classterm{Best-fit structure for stated purpose}{\textit{(highly important · Conditional: no · Discounted: no)}
Beyond being acceptable, the structure should be the best fit for the model's specific requirements, its audience, and its foreseeable future uses.
The rubric item is violated only where a clearly better alternative exists and can be articulated, and where the chosen structure creates avoidable friction.}

\classterm{Deal vs.\ planning orientation}{\textit{(highly important · Conditional: yes · Discounted: no)}
The calculation architecture should reflect whether the model serves one-time deal evaluation or ongoing planning, since the two differ in data-refresh cadence, flexibility, and depth.}

\classterm{Real fiscal year handling}{\textit{(ultra important · Conditional: yes · Discounted: no)}
The model must follow the entity's real fiscal calendar, including non-calendar fiscal years such as July to June.
Where the task instructions specify no fiscal calendar, a calendar-year assumption is acceptable provided it is stated.}

\classterm{Stub period handling}{\textit{(ultra important · Conditional: yes · Discounted: no)}
Stub periods, that is, partial periods at the start or end of a forecast, must be calculated with a day-count convention, such as act/365, act/360, or 30/360, appropriate to the model's purpose, the task instructions, and the jurisdiction.}

\classterm{Functional sheet separation}{\textit{(highly important · Conditional: yes · Discounted: no)}
Cover, inputs, calculations, analyses, and outputs should each sit on their own sheet, with no mixing of functions on a single sheet unless there is a clear structural reason.}

\classterm{Logical tab order}{\textit{(important · Conditional: yes · Discounted: no)}
Tabs should be ordered to follow the flow of the model, from cover and table of contents, through assumptions and inputs, calculations, financial statements, analyses and scenarios, to outputs and summary, and finally the error checks.}

\classterm{Functional calculation sections}{\textit{(highly important · Conditional: no · Discounted: no)}
Calculations should be split into clearly labeled functional sections, for example revenue, COGS, operating expenses, capital expenditure, debt, and equity, rather than run together as one undifferentiated block.}

\classterm{Buffer rows for totals}{\textit{(not absolutely crucial · Conditional: no · Discounted: no)}
Buffer rows or columns should sit above or beside totals, so that new line items can be inserted without breaking the sum formulas.}

\classterm{Buffer rows between sections}{\textit{(important · Conditional: no · Discounted: no)}
Major sections should be separated by buffer rows rather than running together.}

\classterm{Clear section headers}{\textit{(important · Conditional: no · Discounted: no)}
Each major component of the model should carry a clear, descriptive section header.}

\classterm{Consistent tab structure}{\textit{(important · Conditional: yes · Discounted: no)}
Similar tabs should share a layout, using the same columns for headings, sub-headings, units, and subtotals.}

\classterm{Freeze panes set}{\textit{(highly important · Conditional: yes · Discounted: no)}
Freeze panes should keep row labels and the timeline visible while scrolling.
They should not be set excessively either:
no more than about two thirds of a regular screen view should be taken up by the frozen area, so that enough of the calculations remains visible.}

\classterm{One timeline per sheet}{\textit{(important · Conditional: yes · Discounted: no)}
Each worksheet should contain a single timeline, since several on one sheet create ambiguity about the period a calculation refers to.
Where a sheet genuinely needs a second periodicity, such as an annual summary beneath a quarterly build, the blocks must be clearly separated, labeled, and justified.}

\classterm{Chronological left-to-right}{\textit{(highly important · Conditional: yes · Discounted: no)}
The columns should be arranged chronologically from left to right, with the earliest time periods on the left and those furthest in the future on the right.}

\classterm{Time period headers}{\textit{(highly important · Conditional: yes · Discounted: no)}
Header rows should state the time period of each column, so that the reader knows immediately which period a column of data represents.}

\classterm{Data-type designation (actuals vs.\ estimates)}{\textit{(highly important · Conditional: yes · Discounted: no)}
Header rows should indicate whether the values in a column are actuals, estimates or forecasts, or another type of data, whether by appending A and E to the period label or by a separate designation row.
Without it the reader cannot separate historical from projected data.}

\classterm{Clear row labels}{\textit{(important · Conditional: no · Discounted: no)}
Row labels should be clear, descriptive, professional, correctly spelled, and expressed in standard financial terminology appropriate to the target audience.}

\classterm{Clear column labels}{\textit{(important · Conditional: no · Discounted: no)}
Column labels should be clear, descriptive, professional, correctly spelled, and expressed in standard financial terminology appropriate to the target audience.}

\classterm{Roll-forward (corkscrew) structure for balances}{\textit{(highly important · Conditional: yes · Discounted: no)}
Balance items that move over time, among them debt, cash, PP\&E, equity, provisions, and share count, should be built as explicit roll-forwards:
opening balance, additions, deductions, closing balance, with each movement on its own line and each period's closing balance feeding the next period's opening balance.
A compressed single formula, or an opening balance that does not equal the prior closing balance, violates the criterion.}

\classterm{Circularity -- valid justification}{\textit{(important · Conditional: yes · Discounted: no)}
Any circular reference in the model must rest on a clear and valid modeling reason, such as an average debt balance, a revolver sweep, an iterative WACC, or interest on average cash.
An unjustified circular reference points to a formula error or poor design.}

\classterm{Circularity -- documentation}{\textit{(important · Conditional: yes · Discounted: no)}
Any intentional circular reference must be documented, by cell comment or model note, explaining its purpose, the mechanics of the circularity, and how it resolves.
Undocumented circularity leaves a reviewer unable to tell the intentional from the accidental.}

\classterm{Circularity breaker switch}{\textit{(not absolutely crucial · Conditional: yes · Discounted: no)}
Each intentional circular reference should have a circuit-breaker switch for debugging, that is, a flag cell that toggles the circularity off and replaces the circular formula with zero or a static value.}

\end{classbox}

\section{Task Annotation}
\label{app:TaskAnnotation}
Tasks of this kind are costly to create:
each is authored from scratch by a finance expert, requires a long build time (Appendix~\ref{app:TimeEstimate}), and passes several review rounds (Appendix~\ref{app:TaskDesign}).
Individual difficulty levels and task types therefore contain limited numbers of tasks.
We accordingly focus difficulty-related findings on the harder levels jointly (Section~\ref{sec:results}) and use task types to describe the benchmark's coverage rather than to compare agents across types.

\subsection{Task Difficulty}
\label{app:TaskDifficulty}

\subsubsection{Procedure and Weighting}
\label{app:DifficultyProcedure}

Every task was scored in each of four categories: Scope, General Modeling, Finance Knowledge, and Implementation in Excel. For all four categories, a numerical score between 1 and 6 was assigned based on the classes described in the subsections below. The scores were then weighted, and the weighted sum was rounded to the nearest whole number to determine the overall difficulty.
Halves were rounded up, so a weighted sum of 3.5 yielded level 4 (Medium-Hard).
One task reached a rounded score of 6 (Very Hard); since a level containing a single task would not be representative, we reclassified it as Hard, so that no task is labeled Very Hard.

\begin{classbox}{Difficulty Category Weights}{Scoring procedure with four categories}

\medskip
\begin{center}
\begin{minipage}[c]{0.46\linewidth}
\centering
\begin{tabular}{lc}
\toprule
Category & Weight \\
\midrule
Scope & $\dfrac{1}{3}$ \\[8pt]
General Modeling & $\dfrac{1}{3}$ \\[8pt]
Finance Knowledge & $\dfrac{1}{6}$ \\[8pt]
Implementation in Excel & $\dfrac{1}{6}$ \\[4pt]
\bottomrule
\end{tabular}
\end{minipage}%
\hfill%
\begin{minipage}[c]{0.35\linewidth}
\centering
The rounded weighted sum\\[4pt] maps to a difficulty level:

\medskip
\begin{miniclassbox}
\centering
\begin{tabular}{cl}
1 & Very Easy \\
2 & Easy \\
3 & Medium \\
4 & Medium-Hard \\
5 & Hard \\
6 & Very Hard \\
\end{tabular}
\end{miniclassbox}
\end{minipage}
\end{center}
\medskip

\end{classbox}

\subsubsection{Scope}

Scope measures the breadth and structural complexity of the task (how many components, steps, or interdependent pieces the task contains). While it is correlated with the task solving duration, it is not derived from it alone, since it takes into account not only the extent but also the complexity.

\begin{classbox}{Scope Difficulty Levels}{Breadth and structural complexity of the task}

\classterm{1 --- Very Easy}{\textit{Single step or very small task, usually one output.} This class is used if the task is very small or even only a single calculation with a single number as the required output.}

\classterm{2 --- Easy}{\textit{Small but multi-step task; limited tabs or exhibits.} This class is used if the task is still not extensive, but at least involves multiple steps to reach the solution.}

\classterm{3 --- Medium}{\textit{Standard professional case exercise.} This class is used for a standard task involving multiple steps and at least 30 minutes of solving time in a focused context.}

\classterm{4 --- Medium-Hard}{\textit{Broader professional case exercise.} This class is used for more extensive tasks involving at least 1 hour of solving time in a focused context.}

\classterm{5 --- Hard}{\textit{Complex, multi-part case with interdependent components.} This class is used for problems that take longer than an hour to solve and are also highly complex, involving several interdependent parts.}

\classterm{6 --- Very Hard}{\textit{Very complex case with many moving parts and interdependencies.} This class is used for very complex tasks requiring several hours of deep work to solve as well as involving several complex, highly interdependent parts.}

\end{classbox}

\subsubsection{General Modeling}

General Modeling measures the technical difficulty of the quantitative work itself (the sophistication of the calculations, logic, and model architecture required). The scale captures the depth of analytical rigor: how linked, layered, and error-prone the calculations are, and how much modeling knowledge and experience are required to solve the task.

\begin{classbox}{General Modeling Difficulty Levels}{Technical difficulty of the quantitative work}

\classterm{1 --- Very Easy}{\textit{Basic arithmetic / direct formula application.} This class is used if the modeling is straightforward and might even be done efficiently with pen and paper or using a calculator.}

\classterm{2 --- Easy}{\textit{Straightforward financial calculations, limited dependencies.} This class is used if the modeling is still straightforward, but several calculations are based on each other.}

\classterm{3 --- Medium}{\textit{Multiple linked calculations and assumptions.} This class is used for a standard financial model that includes multiple interrelated calculations and assumptions.}

\classterm{4 --- Medium-Hard}{\textit{Non-trivial structure.} This class is used if the modeling is beyond standard and, for example, requires modeling across several interrelated tabs.}

\classterm{5 --- Hard}{\textit{Advanced model logic; high risk of errors without careful checks.} This class is used for complex tasks in which the calculations are highly interrelated.}

\classterm{6 --- Very Hard}{\textit{Sophisticated architecture and/or novel approach design.} This class is used for highly complex tasks that require extensive experience in financial modeling to solve, given the complexity and interdependence of the task.}

\end{classbox}

\subsubsection{Finance Knowledge}

The category Finance Knowledge determines how much (and to what depth) finance knowledge is needed to solve the task. This determination also incorporates potential assistance given in the task instructions, while not completely dismissing the need for finance knowledge if the instructions provide such assistance.

\begin{classbox}{Finance Knowledge Difficulty Levels}{Depth of finance knowledge required to solve the task}

\classterm{1 --- Very Easy}{\textit{No finance knowledge necessary.} This class is used if no finance knowledge is needed to solve the task.}

\classterm{2 --- Easy}{\textit{Basic finance knowledge necessary.} This class is used if very basic finance knowledge is needed, such as basic principles of compounding.}

\classterm{3 --- Medium}{\textit{Standard finance knowledge necessary.} This class is used if finance knowledge acquired within the regular scope of a business degree is needed, i.e., the finance knowledge is necessary but also sufficient to complete a trivial (integrated) 3-statement model.}

\classterm{4 --- Medium-Hard}{\textit{Extensive finance knowledge necessary (topic is still relatively regular).} This class is used if the finance knowledge is still not niche, but not acquired via a regular business degree (e.g., transaction models like LBOs).}

\classterm{5 --- Hard}{\textit{More exotic topic with needed prior finance knowledge.} This class is used if the topic is more exotic and work experience in the specialized field is necessary to gain the needed knowledge.}

\classterm{6 --- Very Hard}{\textit{Very niche topic on an advanced level (extended topic knowledge necessary).} This class is used if the topic is not only very specialized, but also the needed knowledge is deep, which is typically acquired only through extensive specialized work experience.}

\end{classbox}

\subsubsection{Implementation in Excel}

Regardless of the overall model architecture, converting the input data into a viable solution in conjunction with financial expertise requires a certain level of proficiency in Microsoft Excel. While some tasks require only basic arithmetic in Excel, others cannot be completed without a solid understanding of the latest features.

\begin{classbox}{Excel Implementation Difficulty Levels}{Level of Excel proficiency required}

\classterm{1 --- Very Easy}{\textit{No Excel necessary / very basic (arithmetic).} This class is used if only basic arithmetic in Excel is needed.}

\classterm{2 --- Easy}{\textit{Basic Excel knowledge necessary (e.g., lookups).} This class is used if foundational formulas like lookups are needed to solve the task.}

\classterm{3 --- Medium}{\textit{Moderate Excel knowledge necessary (nested formulas starting to get involved).} This class is used if functions are not only used but also combined into nested formulas. In general, this level corresponds to the standard proficiency of a seasoned and experienced Excel user in a business setting.}

\classterm{4 --- Medium-Hard}{\textit{Intermediate Excel knowledge necessary or knowledge of financial Excel formulas.} This class is used if either knowledge of the financial formulas in Excel is necessary to solve the task or if knowledge exceeding the standard proficiency of a business user is needed (such as the usage of Data Tables or the FILTER function).}

\classterm{5 --- Hard}{\textit{Advanced and modern Excel knowledge a plus (e.g., array manipulation).} This class is used if advanced and novel Excel knowledge is needed to solve the task properly and in an efficient manner, though the task might still be solvable without such knowledge.}

\classterm{6 --- Very Hard}{\textit{Advanced and modern Excel knowledge needed (e.g., array manipulation).} This class is used if the task is not solvable without knowledge of advanced functions.}

\end{classbox}

\subsubsection{Exemplary Tasks}

\begin{classbox}{Exemplary Tasks by Difficulty Level}{Illustrative tasks for each difficulty level}

\classterm{Very Easy}{No task was classified as Very Easy.}

\classterm{Easy --- \texttt{Pharma} (task 97)}{Small classical comparables valuation with given comparable companies and provided data. Involves only a few steps with linked but straightforward calculations. Only standard finance knowledge is required, as well as moderate Excel knowledge.}

\classterm{Medium --- \texttt{EasyDCF} (task 20)}{Small DCF model with given assumptions and iterative WACC determination. The task is Medium in all four categories: a standard case exercise with multiple linked calculations and assumptions, requiring standard finance knowledge and nested Excel formulas.}

\classterm{Medium-Hard --- \texttt{DayMonthYear} (task 79)}{A very extensive DCF model with several submodels and two ways of calculating the company value. The model requires building a fully fledged preceding 3-statement model. Complex with a non-trivial structure due to the combinatorial nature. Solving the task requires both extensive finance knowledge and intermediate Excel skills.}

\classterm{Hard --- \texttt{DailyDetail} (task 77)}{A very complex and extensive debt waterfall model as used in corporate treasury including interest, FCF forecasting, and specific payout rules. It involves several interdependencies and multiple scenarios. The interconnected model and the model type (financing waterfall) make it very prone to errors. Solving the task requires finance knowledge beyond the standard level, due to the interest rate conventions involved, and advanced knowledge of the newest Excel functions is a plus.}

\classterm{Very Hard}{No task is classified as Very Hard; the only task that reached this level was reclassified as Hard (see Appendix~\ref{app:DifficultyProcedure}).}

\end{classbox}

\subsection{Time Estimates}
\label{app:TimeEstimate}

We decided to use time estimates determined by humans after we concluded that initial attempts with AI-based time estimates were not reliable enough.
The (human) time required to complete each task was estimated immediately after its creation.

The estimate covers the total time required to complete the task, including answering the questions and performing initial plausibility checks and quality assurance. The reference was a modeler with moderate experience---approximately 3+ years in financial modeling---who had already created similar models but might still need to double-check some more complex aspects or consult how-to sources.

For every model, the time was broken down into the following pieces:

\begin{itemize}
    \item Reading and interpreting the instructions
    \item Understanding the given assumptions
    \item Understanding the structure the model should have
    \item Planning the layout / model architecture
    \item Building the model (see details below)
    \item Answering questions and cross-checking
    \item Debugging / sense checks (initial QA before handing off to superiors)
\end{itemize}

A step may also have an estimated duration of 0 minutes if it is not required for certain (particularly small) models.

The main part of the estimate---and generally the most time-consuming part---is model building. The time required for model building was estimated and broken down individually for each task. The model development process was divided into model-specific sub-steps; a time range was then estimated for each sub-step, with the average value of each range used as the basis for the time estimate, and all sub-step estimates were added together.

\subsection{Task Types}
\label{app:TaskTypes}

The following entries describe the different financial models that we created as tasks.
The task type was specified directly during the creation process.

\begin{classbox}{Valuation Models}{Methods for estimating the value of a company or asset}

\classterm{DCF (Discounted Cash Flow)}{Estimates the intrinsic value by forecasting future cash flows and discounting them to present value using a risk-adjusted discount rate. This method is highly dependent on assumptions and focuses on the company's intrinsic ability to generate cash flow over time. It is one of the most commonly used valuation methods.}

\classterm{Comparables (Trading Comps)}{Values a company based on the market valuation of similar publicly traded companies. It relies on metrics such as EV/EBITDA or P/E ratio to perform a relative valuation to determine the value of the company being evaluated.}

\classterm{Sum-of-the-Parts Valuation}{Breaks down a diversified company into individual business segments and values each separately. The total enterprise value is the sum of all segment valuations. The valuation methods used might be different for each part.}

\classterm{NAV (Net Asset Value) Modeling}{Values a company based on the fair value of its underlying assets minus liabilities. Frequently used for asset-intensive companies such as real estate or mining firms, and for investment funds.}

\end{classbox}

\begin{classbox}{Financial Statement Models}{Models integrating the three financial statements and related analyses}

\classterm{3-Statement Model}{Integrates the income statement, balance sheet, and cash flow statement into a fully interconnected financial model. It serves as the foundation for most financial analyses and other modeling tasks.}

\classterm{Working Capital Modeling}{Forecasts short-term operating assets and liabilities (receivables, payables, inventory) as a standalone liquidity analysis. Differs from the complete 3-statement model in that it does not necessarily include the income statement or long-term items.}

\classterm{Ratio Analysis}{Applies predefined financial ratios to existing financial statements to interpret performance rather than forecast it. Unlike a 3-statement model, it does not provide forecasts but is used for diagnosis and benchmarking.}

\end{classbox}

\begin{classbox}{FP\&A Models}{Financial planning and analysis models}

\classterm{Budget vs.\ Actuals Variance Analysis}{Compares planned financial performance with actual results to identify variances. The purpose is to help management understand the factors driving performance and improve forecast accuracy.}

\classterm{Forecasting Model}{Projects a company's future financial performance---revenue, expenses, cash flow, and other key metrics---based on historical trends, business drivers, and stated assumptions. It serves as the baseline plan for budgeting and strategic decision-making.}

\classterm{Scenario / Sensitivity Analysis}{Scenario analysis defines a set of specific, named cases (base/bull/bear scenarios) to delineate the range of possible outcomes. Sensitivity analysis isolates individual variables to demonstrate the extent to which a single assumption influences the outcome. Both are built on a base model and are often not created as standalone models.}

\end{classbox}

\begin{classbox}{Debt \& Financing Models}{Financing structure and capital models}

\classterm{Debt Schedule / Waterfall Modeling}{Tracks debt levels, interest, and repayments over time, often across multiple tranches. Waterfall structures show how cash is distributed among creditors and shareholders.}

\classterm{Loan Amortization Schedule}{Breaks down loan repayments over time into principal and interest. This provides transparency regarding the repayment structure and the outstanding balance over time.}

\end{classbox}

\begin{classbox}{Fixed Income \& Derivatives Pricing}{Valuation of financial instruments and options}

\classterm{Bond Pricing / Yield Curve Analysis}{Calculates the value of fixed-income securities based on interest rates and cash flows. Yield curves are used to analyze market expectations and discount rates across various maturities.}

\end{classbox}

\begin{classbox}{Personal Finance Models}{Personal finance and investment planning models}

\classterm{Rent vs.\ Buy Analysis}{Compares the financial implications of renting versus buying a property. This takes into account costs, appreciation, and the opportunity cost of capital.}

\classterm{Retirement / Savings Calculator}{Projects future savings based on contributions, returns, and time horizon. It helps individuals plan for long-term financial goals such as retirement, or plan their finances more generally.}

\end{classbox}

\begin{classbox}{Business \& Data Analytics}{Non-financial and operational analytical models}

\classterm{Non-Financial or Non-FP\&A Model}{An analytical model outside the traditional financial realm, such as operational, marketing, or product analyses. Focuses on data-driven insights rather than purely financial results, but is nevertheless model-driven.}

\end{classbox}

\section{Task Design}\label{app:TaskDesign}
All tasks in our benchmark were created from scratch by finance experts and are not derived from existing materials.
This reduces the risk of data contamination, i.e., the risk that tasks or their solutions appear in the training data of the models underlying the evaluated agents, although the public release may expose future models to them.
Each task consists of two files:
an input workbook and a golden solution.
The input workbook contains three or four sections on separate tabs:
task instructions (analogous to a prompt), assumptions, supporting data (for some tasks only), and the questions on which the evaluation of \texttt{Final calculation accuracy} is based.

The instructions are as detailed as necessary while remaining concise.
We refined the instructions iteratively to remove the ambiguities we identified and to support a fair comparison of agents.
We deliberately refrained from including examples or explicit formulas, as this would not reflect how practitioners currently formulate such requests.
Where terminology or conventions are not uniformly defined in finance, we followed prevailing industry practice in both the instructions and the golden solution.
None of the questions are open-ended:
every question used to assess \texttt{Final calculation accuracy} has a single, definitive answer, which is numerical in most cases.

The number of accompanying assumptions varies by task type and can reach several hundred.
They are provided in tabular form with clear labels and, in some cases, additional descriptions.
In accordance with industry practice, we expect agents to reference these assumptions in their financial models rather than hardcode their values.
Where necessary, tasks also include supporting data, such as sales prices, trading data, or financial statements; in all but a few cases, this data was generated synthetically.
The questions tab, whose answer cells are left blank, is used for our main rubric criterion:
\texttt{Final calculation accuracy}.
The questions were sampled randomly across the model's main results and, where applicable, their periods, so that they cover its different parts.
Consequently, some questions are easy to answer, while others require all instructions to be implemented correctly.

The golden solution contains the tabs of the input workbook as well as the completed financial model.
Depending on the task requirements and on what is reasonable for the model, the financial model spans one or more worksheets.
In the golden solution, the questions tab contains references to the corresponding answers in the financial model and serves as the ground truth for evaluating the agents' \texttt{Final calculation accuracy}.
The golden solutions define the ground truth for \texttt{Final calculation accuracy}; for reasons of efficiency, they are reference computations rather than presentation-ready deliverables and are therefore not graded against the remaining criteria.
Their scores and pass rates therefore cannot serve as a reference for the agents' attempts, since the golden solutions may, for instance, omit documentation or formatting that the house standards require of the agents.

All tasks underwent a multi-stage quality assurance process.
We iteratively refined the instructions and the tasks themselves to remove ambiguities, and verified the correctness of the golden solution models and answers through both human and LLM-based review.
For the latter, an LLM was given the golden solution together with the instructions and asked to check whether the instructions had been implemented and the model built correctly.
A task was included in the benchmark only after at least two review runs in which no errors were found.

\section{Score Leaderboard}
\label{app:ScoreLeaderboard}
\begin{figure}[h]
    \centering
    \includegraphics[width=\linewidth]{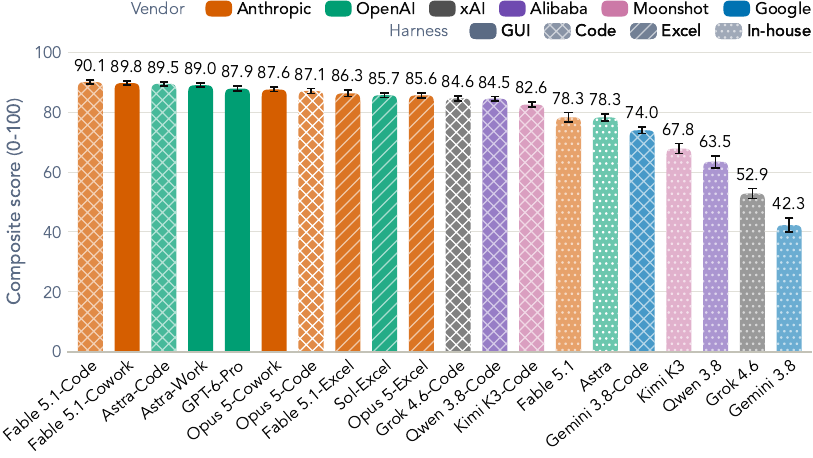}
    \caption{\textbf{Score Leaderboard.}
    Scores are much closer across agents than pass rates, since many criteria concern the structure or format of the spreadsheet and are satisfied even by attempts that fail the task.
    }
    \label{fig:leaderboard_score}
\end{figure}

Figure~\ref{fig:leaderboard_score} reports the score of each agent (Appendix~\ref{app:Weighting}), complementing the pass rate leaderboard in Figure~\ref{fig:leaderboard}.

\section{Expert Team and Effort}
\label{app:ExpertTeam}
Our experts comprise four people with finance training: three practitioners with industry experience and one accomplished competitor and task author in financial modeling and spreadsheet competitions.
All annotators have hands-on experience building the class of models the benchmark covers; the practitioners and the competitor in particular possess notably greater spreadsheet and industry expertise.
Developing the benchmark, including the rubric, the tasks, the judge, and the evaluation pipeline, took approximately \benchmarkcreationhours person-hours, excluding the agent runs themselves.

\section{Judge}
\label{app:Judge}

\subsection{Agentic Judge Details}
\label{app:agent_judge_details}

\begin{figure}
    \centering
    \includegraphics[width=\linewidth]{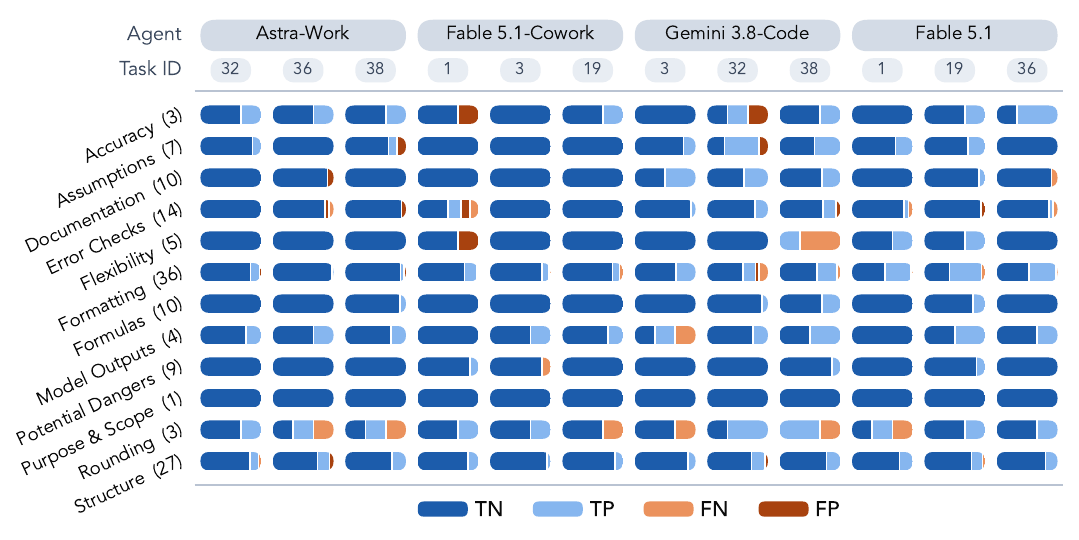}
    \caption{
    \textbf{Annotations on Judge Correctness.}
    TP: mistake caught correctly by judge.
    TN: attempt meeting the criterion recognized by judge.
    FN: mistake exists but not flagged.
    FP: mistake flagged but not present.
    The judge agrees with experts' annotations at high accuracy.
    In particular, FPR is far lower than FNR, which allows the judge to recognize a perfect attempt as passing at high rate.
    The higher FNR aligns with the pass rate design to be more lenient, but agents nonetheless fail to pass most tasks despite the leniency.
    }
    \label{fig:judge_agreement}
\end{figure}

The judge is an agentic grader built on GPT-5.6 Sol. For each grading it receives the attempt workbook, the task's golden solution, and the input workbook, serialized sheet by sheet with openpyxl, and can read any range of any of them on demand. It is given the judge-graded criteria applicable to the task (Appendix~\ref{app:RubricAnnotation}) and works through them in a single conversation, recording a pass or fail verdict with located mistakes for each, until every criterion has a verdict. \texttt{Final calculation accuracy} and 19 further criteria, such as \texttt{Sufficient column widths} and \texttt{No hidden sheets}, are not judged this way but checked deterministically (Appendix~\ref{app:RubricAnnotation}). Grading of the remaining criteria that would typically involve observation of the rendered workbook, such as \texttt{Totals visually distinguished}, relies on the creation of a properties block, which is included in what the judge receives. The properties block is created by leveraging openpyxl functionality to log stored settings and metadata about the state of the workbook that is more comprehensive than just values and formulas. The judge relies on this properties block to make inferences about the formatting of the workbook. For criteria that typically would require sensitivity-testing inputs, such as \texttt{No errors under extreme-but-plausible inputs}, the judge must reason about the architecture of the workbook and formula structure to determine if any potential criteria violations exist.

\subsection{Per-Criterion Validation Examples}
\label{app:ValidationExamples}
For each of the 109 criteria graded by the judge, we constructed one pair of validation examples from a single task to which the criterion applies.
Where necessary, we first modified the task's golden solution so that it clearly satisfies the criterion; the resulting workbook is the \emph{clean example}.
The author of the task then derived the \emph{erroneous example} by manually perturbing the clean example so that it violates the criterion.
Each perturbation was designed to make the violation unambiguous, so that the correct verdict is not in doubt, but not trivially detectable, so that the judge must inspect the workbook to find it.
We then ran the judge three times on each of the 218 examples, grading each only on the criterion it was built to test; Section~\ref{sec:rubric_judge_val} reports the results.
Based on initial validation runs, we iteratively refined the judging setup to obtain consistent grades, for instance by adjusting which file contents the judge receives.

\section{House Standards}
\label{app:HouseStandards}
\begin{classbox}{\# Financial Modelling --- House Standards}{September 2026}
\setlength{\parskip}{6pt}%
These are the conventions particular to us. Ordinary professional practice is assumed and not restated here. Where the engagement brief or instructions conflict with anything below, they govern; departures are fine with a reason noted on the cover.

-{}-{}-

\# 1 $\cdot$ Structure

Beyond the sheets any model needs, every model carries a linked contents page, a checks tab, and a change log recording what changed, when and by whom.

**The cover** states scope~--- including what the model deliberately leaves out~--- the legend of colours and abbreviations (glossary), the approach and the main design decisions behind it, and a short note on what may be changed, how a case is selected, and where the answer is read.

**Build for the second use.** Where continued use is foreseeable, build for it rather than for the first answer alone: extending or re-cutting the forecast means adding columns and changing inputs, and unless the instructions rule it out, adding or removing a business unit is a matter of inputs and switches rather than a structural rebuild.

**Periods.** A given period sits in the same column on every tab. Headers carry the period and its nature~--- FY2024A, FY2025E, Q1-26B. Periods follow the entity's fiscal year; where none is given, calendar year, stated on the cover. One labelled anchor date drives all date logic. Discount stubs run actual/365; instruments accrue on their own stated basis, recorded beside the input.

-{}-{}-

\# 2 $\cdot$ Building

**Units.** \$mm unless the instructions say otherwise, stated at the head of every block and column with the currency alongside. Anything in another unit~--- per-unit prices, headcount, per-share figures, multiples~--- is labelled where it appears.

**Rounding.** State the convention beside the figures it governs: ``\$mm, rounded to \$0.01mm'' at the head of the block.

**Signs.** Costs and outflows negative, revenues and inflows positive, everywhere~--- including add-backs in EBITDA bridges and movements in working capital.

**Assumptions.** An input block on a build sheet is acceptable only as the sole source for that driver, signposted from the assumptions tab. Each assumption carries a source precise enough to re-find~--- ``10-K FY24 p.45'', ``brief~\S2'', ``instructions line 23''~--- or is marked a judgment with an owner and date. Named ranges are reserved for a handful of global inputs.

**Formulas.** Prefer XLOOKUP\slash XMATCH to VLOOKUP, IFS or SWITCH to nested IFs, LET for repeated intermediates; INDEX\slash XMATCH is fine; legacy constructs where compatibility requires, with the reason noted. Use the `3-clause rule'. Mark any intentional break in an otherwise consistent row or block.

**Cases** run off one labelled selector on the assumptions tab that pulls the live set; structural switches~--- capitalise/expense, include/exclude a unit~--- work the same way, and no sheet carries a private toggle. Results for every case appear side by side in one view, not one case at a time.

**Checks** are consolidated on the checks tab, with a single flag on the cover that turns red on any failure. The sense check sits there too: key outputs set against benchmarks or comparables, in the workbook rather than in the covering note.

-{}-{}-

\# 3 $\cdot$ Presentation

**Colour.** Blue font: hardcoded inputs $\cdot$ black: formulas $\cdot$ green: links to another sheet $\cdot$ red: links to another workbook. Inputs also carry a pale blue fill; key inputs carry data validation. Beyond these, at most two or three additional colours across fonts, fills and tab colours, none of them resembling the four. Yellow means ``unfinished'' or ``review me'' and nothing else~--- nothing unfinished ships, and any yellow left in the delivered file is explained in the legend.

**Numbers.** Zeros as dashes. \$mm to two decimals, percentages to one, multiples to one (7.5x), counts to none~--- held constant across comparable values.

**Grid.** Arial 10 throughout. No merged cells~--- Center Across Selection instead. A buffer row inside every total range.

-{}-{}-

\# 4 $\cdot$ Delivery

Group rather than hide rows and columns, and nothing is hidden anywhere else either. Workbooks are self-contained: no links to other workbooks unless the brief or instructions require them, and no macros without disclosure on the cover and a reason native Excel cannot do the job.

Circularity only where the economics require it, each instance noted and fitted with a breaker switch, with iterative calculation enabled and stated.

Non-obvious formulas carry a note giving the reasoning~--- why this method~--- not a restatement of the formula.

One zoom level throughout, cursor at A1 on every sheet, print areas and page breaks set.
\end{classbox}

\section{Models}
\label{app:models}

\begin{table}[h]
\centering
\begin{tabular}{ll}
\toprule
\textbf{Vendor} & \textbf{Model} \\
\midrule
Anthropic               & Claude Fable~5.1 \\
Anthropic               & Claude Opus~5 \\
OpenAI                  & GPT-6 Astra \\
OpenAI                  & GPT-5.6 Sol \\
Google                  & Gemini 3.8 Flash \\
SpaceXAI (formerly xAI) & Grok 4.6 \\
Moonshot AI             & Kimi K3 \\
Alibaba                 & Qwen 3.8 Max \\
\bottomrule
\end{tabular}
\caption{Models evaluated. In agent names throughout the paper, GPT-6 Astra is shortened to Astra, GPT-5.6 Sol to Sol, and Gemini 3.8 Flash to Gemini 3.8; a suffix (-Cowork, -Work, -Pro, -Code, -Excel, -in-house) names the harness, and -in-house is omitted where the harness is clear from context.
GPT-5.6 Sol also serves as the judge backbone (Appendix~\ref{app:agent_judge_details}).}
\label{tab:models}
\end{table}

\section{Agentic Harnesses}
\label{app:AgenticHarnesses}
We use ``harness'' to mean everything around the model that a practitioner would receive as one product: the agent scaffold, its tools, how it reads and writes the workbook, its default settings, and its workflow (e.g., the Excel add-ins edit the open workbook in place). Our harness comparisons therefore measure the effect of the product as deployed, not of any single component.
Note that the main claims only state the harness has an effect, and we do not attribute the resulting differences to any specific component of the harness.

\paragraph{Common protocol.}
Every attempt starts from the task's input workbook, the house standards (Appendix~\ref{app:HouseStandards}), and a short instruction to build the model and record its answers on the questions tab.
The rubric is never shown.

Failures of our own infrastructure (browser, container, or provider outages) are retried and not recorded.
Each agent thus has one recorded attempt per task.
This follows common practice in end-to-end agentic evaluation:
SWE-bench \citep{jimenez2024swebench}, WebArena \citep{zhou2024webarena}, OSWorld \citep{xie2024osworld}, OSWorld~2.0 \citep{xie2026osworld2}, and SpreadsheetBench~2 \citep{zhu2026spreadsheetbench2evaluatingagents} likewise evaluate each model once per task, as long-horizon attempts are costly to repeat.
We therefore prioritized the number of agents evaluated over repeated runs.

All agent attempts are saved and graded as Excel files (.xlsx, or .xlsm where macros are present).
The GUI and Excel harnesses run proprietary products that expose no version identifiers; we therefore record the date of every attempt and release these dates with the benchmark, so results for these harnesses are dated snapshots of the products.
We evaluate models in several harnesses only to identify whether effects of model and harness exist, not to claim that the effect of any component is consistent and transferable (likely not the case, since model and harnesses are inherently coupled).
Note that when Gemini~3.8 Flash performs poorly in Codex, we do not conclude from this single result that the model itself is weak.
Instead, we rely on the observation that Gemini~3.8 Flash passes no task in either of the two harnesses it was evaluated in.

\paragraph{GUI.}
A Chrome session logged into the consumer product is controlled programmatically:
upload the task files, set the mode and reasoning effort, send the prompt, wait for the reply, download the produced workbook.
The products' usage caps are respected by pausing runs; a reply that is a quota message rather than a workbook is treated as an infrastructure failure and retried.

\paragraph{Code.}
Anthropic models are run in Claude Code. OpenAI and all other models with the exception of Anthropic models are run in Codex.
Claude Code and Codex run non-interactively inside a pinned container image with a firewall that resolves only the vendor's API endpoint.
The container holds the task files and the prompt as a text file; the agent may install and run any code it likes, and the image ships LibreOffice for formula recalculation.
Reasoning effort is pinned per agent at the highest tier the product offers for that model.
We recognize that Codex is likely tuned for OpenAI's models.
However, we note that our main claim is not that each harness has a specific and consistent effect, but merely that it has an effect, so this tuning does not constitute a confounding factor.
For the same reason, running Anthropic models in Claude Code and all other models in Codex is not a confound, since we compare harnesses only on the same model.

\paragraph{Excel.}
In the web-based version of Microsoft Excel, Excel Online, the pipeline opens a copy of the task's input workbook, opens the add-in panel offered by Anthropic or OpenAI, selects and verifies the model or reasoning-effort setting, attaches the remaining task files with the first prompt, and downloads the workbook the add-in edited in place.
It is the only harness in which the agent works directly inside the spreadsheet application.

\paragraph{In-house.}
This harness was built by the research team.
A single loop calls the model's API with 21 spreadsheet tools (file, worksheet, cell and formula edits, formatting, structure inspection, search); every formula edit triggers a LibreOffice recalculation whose results are returned to the model.
Each attempt is capped at 40 model calls to limit inference cost.
The API endpoints integrated into the harness include Anthropic, OpenAI, and TensorBlock.
Third-party libraries such as openpyxl are used to extract and serialize the spreadsheet data in a CSV format, which is fed to the LLM API. The prompt is included in the input sent to the API.

\section{Qualitative Analysis --- General Agent \& Harness Analysis}
\label{app:GeneralAnalysis}
This section describes patterns that apply to multiple agents, harnesses, or difficulty levels, rather than to individual agents; agents are mentioned by name only as examples.
It covers the 14 main agent--harness combinations at the highest reasoning-effort setting on all \tasknum tasks, graded with the full rubric.
\paragraph{Summary.}
Agents usually build structurally sound models, and their structural completeness holds even on the hardest tasks.
What falls with difficulty is correctness:
agents hand in complete, tidy workbooks with more wrong answers.
The integrity checks created by the agents tend to verify internal consistency rather than plausibility; industry-standard presentation conventions are applied inconsistently; and more complex tasks result in large, deeply nested workbooks.
Finally, the harness also plays a role:
while the Claude models in Claude for Excel match their Claude Cowork counterparts on answers but lose scaffolding and polish, the frontier models in the Code harness match their GUI counterparts of the same models on answers and score, though not necessarily on pass rate.

\subsection{What Agents Generally Do Well}

\paragraph{Sound structure.}
Agents deliver spreadsheets with live, driver-based formulas, correctly anchored references, centralized assumptions, and a cover stating the model's purpose.
This structural completeness holds as tasks get harder:
the agents keep handing in complete workbooks with live answers, proper roll-forwards, and dedicated check rows, and blank answer sheets remain rare at every difficulty level.

\paragraph{Near-perfect on standard mechanics.}
Accuracy is highest when a task is based on well-known mechanics: bond pricing, NAV roll-ups, comparables, ratio and budget-versus-actuals analysis, and rule-based non-financial tasks.
Here, most agents answer all or nearly all questions correctly, even on Hard tasks; on the Hard task \texttt{LogisticsCategorization} (task 34), for example, the stored answers of every agent match the golden solution.

\paragraph{Documenting purpose and architecture.}
Agents routinely open the workbook with a cover page or summary that outlines the purpose of the model and the business question and explains the methodology.
They use clear and descriptive tab names.
A reader who did not build the model can therefore find their way around without a walkthrough.
A live, formula-based output summary is consistently produced by the GUI agents, the Excel add-ins, and the two frontier coding agents, whereas the other Codex agents often leave it as hardcoded text.

\paragraph{Consistent signs and units, nothing hidden.}
Workbooks rarely contain hidden or sensitive elements:
no external links, no hidden sheets or rows, and no white-on-white text.
They keep one sign convention and one unit scale throughout and often document the convention; monetary blocks carry \$k, \$m, or USD tags, and scale conversions are explicit.
This reduces common causes of hidden errors and order-of-magnitude errors when a model is shared with another user.

\paragraph{Check sheets and master error flag as standard.}
Nearly every agent ships a dedicated check sheet of formula-driven pass/fail rows.
The rows are typically summarized under a master error flag on the cover page or in the summary.
The weak point is the roll-up:
a COUNTIF summary over the status column skips rows with errors, causing the master error flag to show ``OK'' even though the checks fail.
Such checks allow an auditor to immediately verify internal consistency, but a false ``all clear'' finding is worse for a client than no feedback at all.

\paragraph{Financial model style basics.}
Agents consistently follow the basic visual guidelines for a financial model:
hardcoded values in blue font, often on a pale background fill; thousands separators for large numbers; a uniform format for similar values; shaded title and header areas; and totals in bold.

\subsection{What Agents Generally Do Poorly}

\paragraph{Fixed-horizon builds that must be manually extended.}
Timelines are often tailored precisely to the length specified in the instructions, with a fixed number of rows or columns or a base year incorporated into the timeline formulas, so that any reasonable change in duration, start date, or rate results in the model being shortened or in errors.
The problem rarely occurs with Easy tasks but becomes common from the Medium level onward.
It follows the model more than the harness:
Kimi~K3 and Gemini~3.8 Flash in Codex often fix the horizon, and Fable~5.1 in Claude Code and Claude Cowork sometimes does, whereas GPT-6 in ChatGPT Pro, GPT-6 Astra in Codex, GPT-5.6 Sol in ChatGPT for Excel, and Opus~5 in Claude Cowork usually drive it from one anchor date.
This matters in practice, because models are routinely rerun with new data and conditions.

\begin{figure}[htbp]
    \centering
    \includegraphics[width=0.8\textwidth]{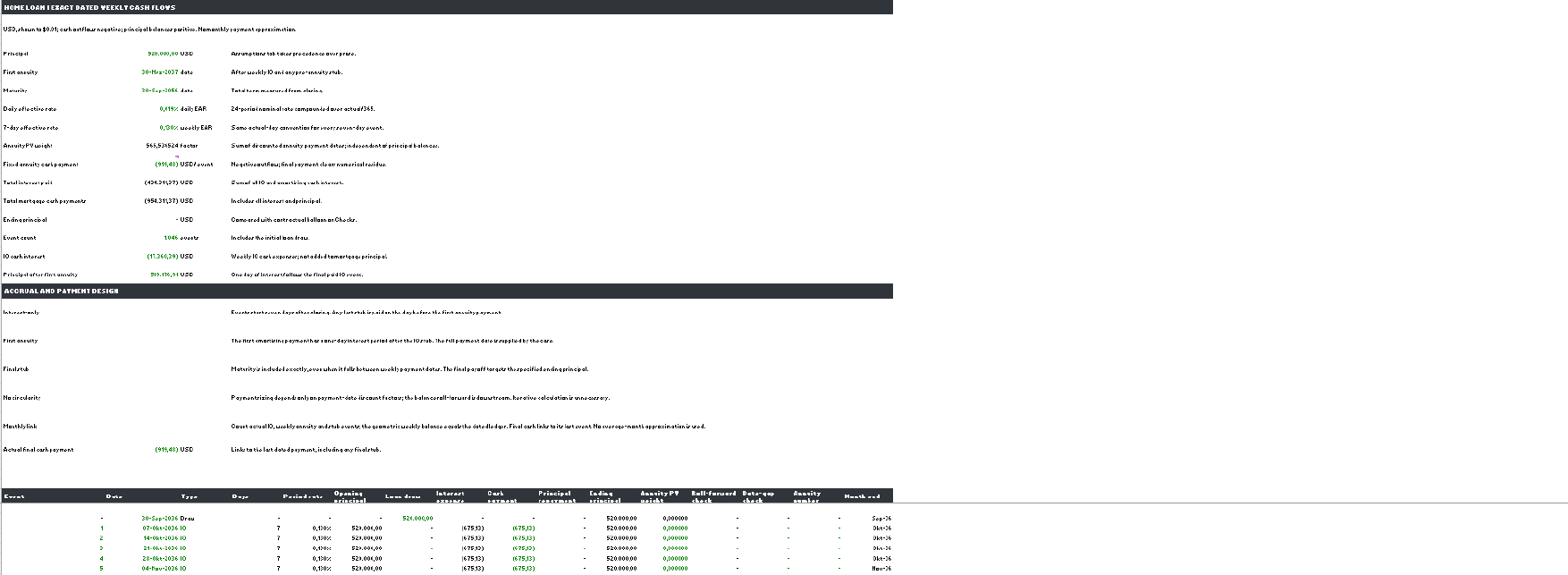}
    \caption{
        Excessive freeze panes: the frozen area ends only at the thin black line, so the sheet cannot be scrolled without zooming out, here to 40\%.
        \textit{ChatGPT Pro (GPT-6) on the task \texttt{DetailedLoan} (task 16, attempt 2064).}
    }
    \label{fig:FreezePanesExcess}
\end{figure}

\begin{figure}[htbp]
    \centering
    \includegraphics[width=0.8\textwidth]{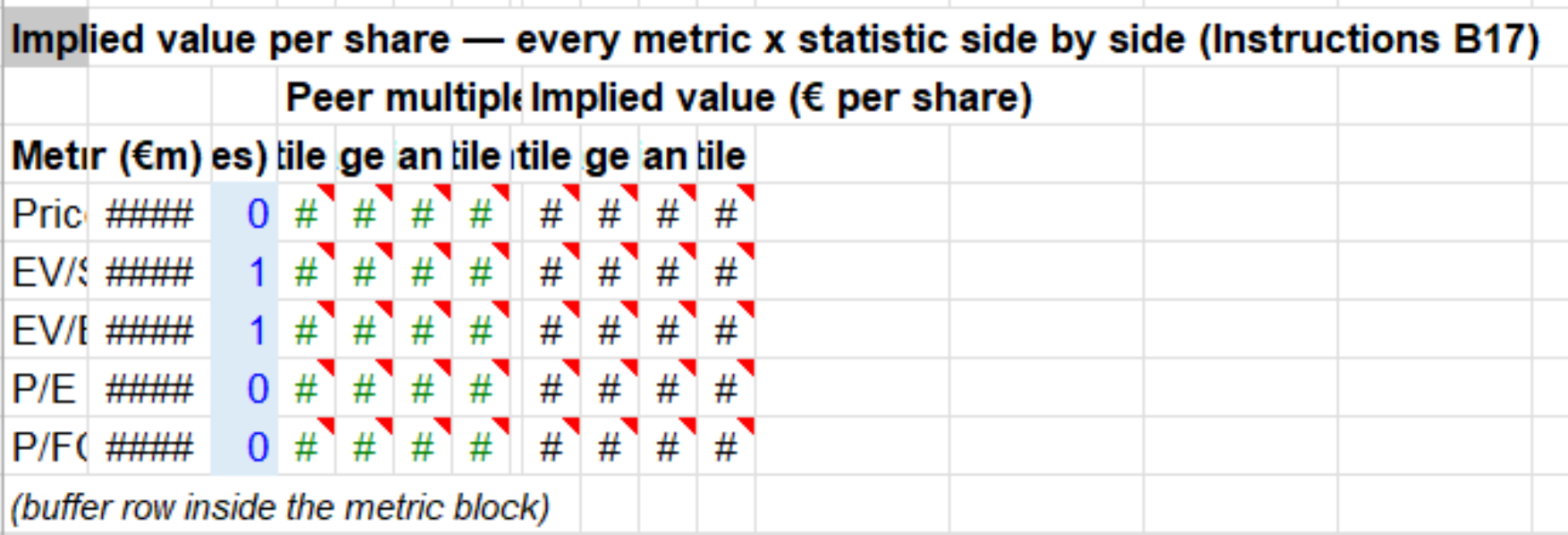}
    \caption{
        Columns that are too narrow display \#\#\# instead of their values.
        \textit{Claude for Excel (Opus~5) on the task \texttt{SAP} (task 51, attempt 2286).}
    }
    \label{fig:TruncationAsHT}
\end{figure}

\paragraph{Sheets not set up for reading.}
Workbooks are built for the agent, not for a reader who flips through them, prints them out, and forwards them.
Long schedules routinely ship without freeze panes or with so many rows frozen that they fill most of the screen (Figure~\ref{fig:FreezePanesExcess}), page breaks are rarely set outside the GPT-6 and GPT-6 Astra agents, and print areas are often missing.
On the long timelines of harder tasks, period columns are left too narrow, so whole blocks later display balances as \#\#\# (Figure~\ref{fig:TruncationAsHT}).

\paragraph{Sense checks reconcile but do not benchmark.}
Most agents add a block labeled ``Sense Checks'', but in almost all cases, the financial model is only validated against itself:
the results are compared with the underlying assumptions or with trivial constraints that any functioning model satisfies.
While these blocks confirm the arithmetic, they say nothing about plausibility; therefore, a workbook can pass every test and still yield an implausible valuation or cash position.

\paragraph{Inputs left unguarded.}
Important input fields are rarely protected in a way that actually prevents incorrect entries.
Although the GUI agents and most other agents in the Code harness add data validation rules, they disable error warnings, so the cell accepts any value; the Excel add-ins leave the warnings enabled but either skip validation or limit it to drop-down menus, thereby ignoring dates, interest rates, and maturities; and the weakest coding-CLI model does not add any validation at all.
Named ranges for global drivers are also rare outside a few agents.

\begin{figure}[htbp]
    \centering
    \includegraphics[width=0.8\textwidth]{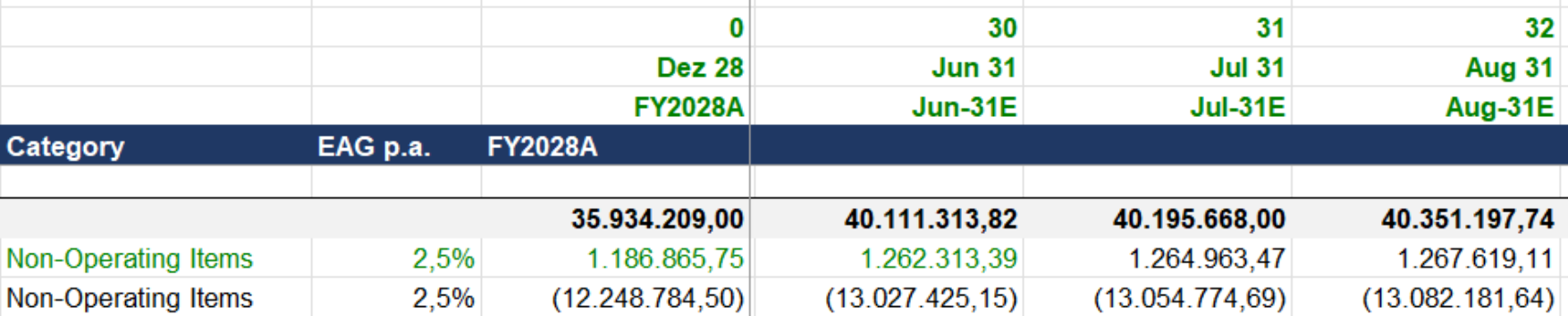}
    \caption{
        Font color drifting within a row: comparable same-sheet calculations switch from green to black partway across the row.
        \textit{Claude for Excel (Fable~5.1) on the task \texttt{RelativeDevelopment} (task 48, attempt 1556).}
    }
    \label{fig:ColorlegendDrifting}
\end{figure}

\paragraph{Color legend declared, then drifting.}
Nearly every agent uses the convention of blue, black, and green and documents it in a legend, but the distinction between black (calculations on the same sheet) and green (references to other sheets) drifts within a sheet (Figure~\ref{fig:ColorlegendDrifting}).
Once the color no longer corresponds to the cell type, a reviewer can no longer use it to track relationships, and a legend that contradicts the worksheet or is used inconsistently is more misleading than no legend at all.

\begin{figure}[htbp]
    \centering
    \includegraphics[width=0.85\textwidth]{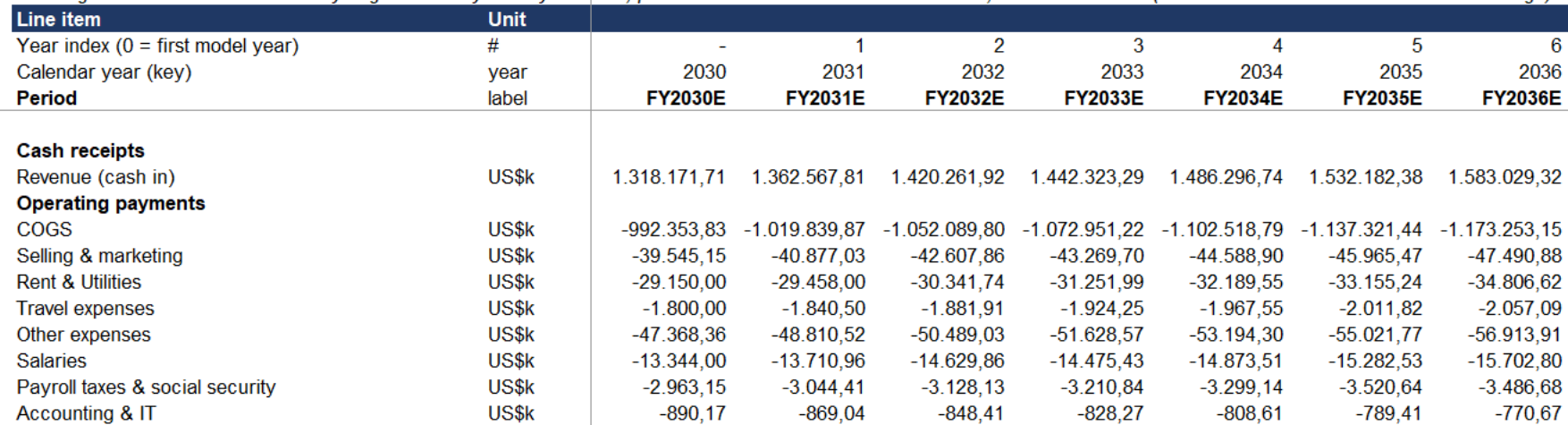}
    \caption{
        Negative numbers displayed with a minus sign instead of in parentheses.
        \textit{Claude Code (Fable~5.1) on the task \texttt{OnlyOperational} (task 94, attempt 1924).}
    }
    \label{fig:FinanceRulesIgnored}
\end{figure}

\paragraph{Finance display rules applied inconsistently.}
The basic number formatting is mostly correct, but each agent occasionally violates the financial presentation rules:
negative numbers are displayed with a minus sign instead of in parentheses (Figure~\ref{fig:FinanceRulesIgnored}), exact zeros are shown as 0.00 instead of as a dash, and blocks of numbers are centered or left-aligned instead of right-aligned.

\paragraph{Large, nested models on harder tasks.}
Starting with Medium-Hard tasks, the agents incorporate the logic into long single-cell formulas instead of helper rows and build brute-force engines; for example, Claude Cowork (Fable~5.1) answered the Hard task \texttt{FixedDetailed} (task 84, attempt 1624) correctly, but with nested IF, SUMIFS, and INDEX formulas across roughly 150,000 cells.
Such models are not only inefficient but also very hard to review.
At the same time, formula notes and output summaries become less frequent, and the answer cells of workbooks built in the Code harness increasingly display errors such as \#NAME? or \#VALUE!.

\section{Qualitative Analysis --- Agent-Specific Analysis}
\label{app:ModelAnalysis}
This section describes the strengths and weaknesses of selected models; patterns shared by nearly all agents are described in Appendix~\ref{app:GeneralAnalysis} and are not repeated.
Note that we focus our attention mostly on models actually performant on the benchmark, and this section is not meant to review all agents.
The points are selective:
each describes one of the most pronounced strengths or weaknesses of a model, measured against the model's own profile, against the other agents, or both.
They are therefore not a ranking across models:
a Good point for a lower-ranked model does not necessarily mean that it does better on that point than a higher-ranked model, unless the text says so explicitly.
Likewise, a point that is not listed for a model or harness does not mean that the model is free of that weakness or lacks that strength there, only that it is not among the model's most pronounced traits.
For models run in more than one harness, the parentheses after each point name the harnesses in which the model stands out on that point; the point may also apply to the other harnesses, only less prominently.
A concluding paragraph then separates the effects of the model from those of the harness: we attribute a pattern to the model if it stands out in several harnesses or if other models in the same harness behave differently, and to the harness only if the same model demonstrably behaves differently in another harness or if all models in that harness share the pattern.

\subsection{Claude}
\subsubsection{Fable~5.1 (Claude Code, Claude Cowork, Claude for Excel)}
\label{app:model_fable}
In Claude Code and Claude Cowork, Fable~5.1 is narrowly the most accurate model in the field and the one most likely to test its outputs against outside benchmarks, and it delivers well-sourced, well-documented workbooks.
In Claude for Excel, it answers nearly as accurately and builds some of the leanest engines in the field, but it falls behind on finishing.
Its weak spots are oversized text columns and rows, ``rounded to'' notes on figures that are only formatted, copied engines and deeply nested formulas, and timelines that are less flexible than those of the other leading agents.

\paragraph{Good:}
\begin{itemize}
\item \textbf{Sense checks against outside benchmarks} (Claude Code, Claude Cowork).
The model adds a sense-check block to its check sheet that tests headline outputs, such as margins, coverage ratios, or bond prices, against explicit external or judgment ranges, often with an owner and a date, whereas nearly every other model only reconciles its outputs with its own inputs.
However, the ranges are sometimes still derived from the financial model's own inputs.
In the task \texttt{CashNiagara} (Claude Code, task 7, attempt 1874), debt service, distributions, interest, cash, and repayments are compared with explicit judgment ranges, and in the task \texttt{DoctorvsTrade} (Claude Cowork, task 18, attempt 1327), the tax, debt, savings, net worth, and crossover outputs are checked against explicit benchmark ranges.

\item \textbf{Sourced assumptions and documented formulas} (Claude Code, Claude Cowork).
The model cites a source for almost every input, either a cell of the input workbook or the relevant part of the task instructions, gives its own judgment calls an owner and a date, and explains non-obvious formulas in cell comments or adjacent methodology notes.
For example, in the task \texttt{PlasticParts} (Claude Code, task 43, attempt 1495), every input cites its location in the input workbook or the task instructions, and every judgment input names an owner and a date; in the task \texttt{HouseorFlat} (task 86), both Claude Code (attempt 1861) and Claude Cowork (attempt 1625) explain their complex formulas in comments or adjacent methodology rows.

\item \textbf{Named key inputs} (Claude Code, Claude Cowork).
The model gives key global drivers, such as the anchor date, tax rate, and case selector, self-explanatory defined names and uses them in its formulas.
This is evident in the task \texttt{BasicGrowth} (Claude Cowork, task 2, attempt 1268), where the Name Manager lists \texttt{Anchor\_Date}, \texttt{Tax\_Rate}, \texttt{Model\_Years}, and \texttt{Case\_Selected}, and in the task \texttt{TheLiquidityLoop} (Claude Code, task 62, attempt 1814), where the key drivers are named and their input cells validated.

\item \textbf{Self-documenting workbooks with a master error flag} (Claude Code).
The workbook opens with a genuine cover stating purpose, owner, version, scope, and usage, followed by a hyperlinked table of contents, and it includes all standard components through to the checks and the change log; the check sheet rolls up into a formula-linked master error flag on the cover or summary that turns red on failure.
In the task \texttt{CustomerConcentration} (Claude Code, task 12, attempt 1876), for instance, the cover gives title, purpose, version, date, owner, scope, approach, usage, and a legend, and in the task \texttt{FiveAutomotive} (Claude Code, task 22, attempt 1775), the flags on both the cover and the summary link to the check sheet.

\item \textbf{Freeze panes and print areas} (Claude Code, Claude for Excel).
The model freezes the headers of long sheets at sensible positions without freezing too much, and Claude Code also defines print areas on every tab, including the tabs inherited from the input workbook.
The task \texttt{Bosch} (Claude Code, task 3, attempt 1252) illustrates this: every tab has a print area, and in the task \texttt{Nike} (Claude for Excel, task 93, attempt 1662), the top five rows of the long calculation sheets are frozen.

\item \textbf{Compact engines} (Claude for Excel).
The model solves a problem with one parameterized engine instead of copying logic:
native iterative calculation, or one shared FCFF build.
For example, in the task \texttt{RelativeDevelopment} (Claude for Excel, task 48, attempt 1556), native iteration replaces the 30 and 41 copied rows in which Claude Code and Claude Cowork unroll the WACC calculation.
\end{itemize}

\paragraph{Bad:}
\begin{itemize}
\item \textbf{Oversized text columns and rows} (Claude Code, Claude Cowork).
The model puts long texts, such as the cover, table of contents, checks, and change log, into extremely wide single columns or very tall wrapped rows; Claude Code, at the same time, often leaves some numeric and timeline columns too narrow, so they display \#\#\#.
This is evident in the task \texttt{PercentageDevelopment} (Claude Cowork, task 96, attempt 1633), where the cover column is 150 characters wide, the purpose column of the table of contents 120 characters, and the check-label column 92 characters, and in the task \texttt{DebtDeclineDynamics} (Claude Code, task 14, attempt 1935), where the text columns are 120 to 130 characters wide, while the loan amount on the schedule sheet displays as \#\#\#.

\item \textbf{``Rounded to'' notes for display formats} (all three harnesses).
The model labels sheets as ``rounded to \$0.01'' although the formulas carry full precision and only the number formats limit what is shown; the figures are displayed to, not rounded to, that precision.
In the task \texttt{CorpBond} (Claude Cowork, task 10, attempt 1298), ``rounded to \$0.01'' labels sit above thousands of schedule formulas that use no rounding function, and in the task \texttt{BreakingDowntheBalance} (Claude for Excel, task 4, attempt 1269), none of the 12,483 schedule formulas under such a label rounds.

\item \textbf{Copied engines and nested formulas} (Claude Code, Claude Cowork).
The model tends toward brute force:
it copies entire engines instead of creating a single parameterized engine or using a data table; furthermore, Claude Cowork writes monolithic formulas, even though helper rows would make the financial model easier to review.
In the task \texttt{CashNiagara} (Claude Code, task 7, attempt 1874), for instance, the entire $277 \times 86$ waterfall is copied three times, and in the Easy-level task \texttt{MeridianOutdoor} (Claude Cowork, task 88, attempt 1620), the commentary text formulas repeat nested IFERROR, TEXT, ABS, and IF calls.

\item \textbf{Fixed-width timelines} (Claude Code, Claude Cowork).
The model sometimes creates the timeline with a fixed number of columns or references fixed ranges, so extending the time period or changing the start date requires manually copying columns and editing ranges.
For example, in the task \texttt{DetailedLoan} (Claude Code, task 16, attempt 1893), the timeline is fixed at 1,008 columns, and the cover instructs users to copy the last column to the right for a longer horizon.

\item \textbf{Negatives with minus signs} (Claude Code).
The model often displays negative numbers with a leading minus sign instead of in parentheses, even for final answers and summary outputs.
This is evident in the task \texttt{Harrington} (Claude Code, task 31, attempt 1949), where answers are displayed as \textminus{}2,820.0000 and \textminus{}7.62\%.

\item \textbf{Unexplained formulas and unguarded circular loops} (Claude for Excel).
Non-obvious formulas, such as array MEDIAN(IF()), MMULT, and SUMPRODUCT/INDEX/MATCH blocks, carry no cell note or adjacent explanation, and some circular models switch on iterative calculation without a breaker switch, a stated reason for the circularity, or a note on how the loop resolves.
The task \texttt{ProductionPlanning} (Claude for Excel, task 44, attempt 1544) illustrates this: the SUMPRODUCT/INDEX/MATCH and MMULT blocks carry no explanatory note, and in the task \texttt{PlasticParts} (Claude for Excel, task 43, attempt 1541), the circular financing and cash block has no breaker, justification, or note.

\item \textbf{Misused font colors} (Claude for Excel).
The reserved blue input and green link fonts appear in cells that are neither input fields nor links, such as text in change logs, fixed labels, legends, and even entire blocks of same-sheet calculations.
This is evident in the task \texttt{FruitJuice} (Claude for Excel, task 28, attempt 1496), where same-sheet calculations are in the green reserved for cross-sheet links, and in the task \texttt{FundFun} (Claude for Excel, task 29, attempt 1502), where the entire change log is in input blue.
\end{itemize}

\paragraph{Model vs.\ harness:}
Most of these points stand out in both Claude Code and Claude Cowork and therefore belong to the model, including the ``rounded to'' notes; Claude Code merely documents rounding correctly more often and sets print areas more consistently.
The Excel add-in changes the picture: compared with the other two harnesses, the model builds leaner engines but explains fewer formulas and misuses font colors more often.
Since Opus~5 shows the same shift toward lean but sparsely documented engines in Claude for Excel, and all agents in the Excel harness blur the color code, these differences appear to stem largely from the harness.

\subsubsection{Opus~5 (Claude Cowork, Claude for Excel)}
\label{app:model_opus}
Opus~5 in Claude Cowork is a strong builder whose strengths are sense checks against outside benchmarks, named global drivers, and accurate rounding notes.
In Claude for Excel, its engines are unusually lean, but it leaves the finishing layer undone.
In both harnesses, its check flags are less reliable and its cell formatting is worse than those of the other leading models.

\paragraph{Good:}
\begin{itemize}
\item \textbf{Accurate rounding notes} (both harnesses).
The notes state that figures are ``shown to'' a given precision while full precision is carried, and they name the questions tab as the only place where ROUND is applied.
In the task \texttt{CashNiagara} (Claude Cowork, task 7, attempt 3260), the precision notes on the summary and annual cash-flow sheets read ``shown to'' and state that full precision is retained, and in the task \texttt{Telecom} (Claude for Excel, task 59, attempt 2343), a note separates display precision from the ROUND applied on the questions tab.

\item \textbf{Sense checks and named drivers} (Claude Cowork).
The check sheets often include a block that tests headline outputs against dated outside benchmarks or judgment ranges, and key global inputs, such as anchor dates, horizon length, case selector, and core rates, routinely get clear defined names.
For example, in the task \texttt{TheInterestGame} (Claude Cowork, task 60, attempt 3650), the installment, effective annual rate, interest burden, debt service, and yield are tested against outside benchmark ranges, and in the task \texttt{TBondII} (Claude Cowork, task 56, attempt 3635), ten global drivers are named.

\item \textbf{Insert-safe totals and an anchored horizon} (Claude Cowork).
Subtotal ranges include labeled buffer rows or columns, so that line items can be inserted without breaking the sums, and timelines are usually prebuilt across the required horizon and driven by one anchor date and a length driver.
In the task \texttt{DailyDetail} (Claude Cowork, task 77, attempt 3673), for instance, open buffer rows precede the totals, and in the task \texttt{CashConversionLab} (Claude Cowork, task 6, attempt 3235), the 60-quarter horizon runs from one anchor.

\item \textbf{Compact engines and freeze panes} (Claude for Excel).
Where other agents copy the same logic across dozens of case sheets or huge iteration grids, the model usually builds one compact engine and points everything else at it; it also freezes the header rows and label columns of long calculation and valuation sheets without freezing too much.
This is evident in the task \texttt{FundRebalancing} (Claude for Excel, task 85, attempt 2452), where one parameterized engine does the work, and in the task \texttt{ApfelInc} (Claude for Excel, task 1, attempt 2001), where the long tabs are frozen at sensible positions.
\end{itemize}

\paragraph{Bad:}
\begin{itemize}
\item \textbf{Check flags that cannot be trusted} (both harnesses).
Claude Cowork's check formulas often perform arithmetic on whole ranges or nest a multi-cell IF or ABS inside SUMPRODUCT or MAX without array entry; they return \#VALUE! and pass the error into the status cells meant to flag problems.
Claude for Excel's master error flag is often plain text or fixed green without conditional formatting, so it cannot turn red when a check fails.
For example, 18 check formulas return \#VALUE! in the task \texttt{Fixings} (Claude Cowork, task 23, attempt 3472), and the summary of the task \texttt{LiquiditySplatter} (Claude for Excel, task 33, attempt 2133) reads ``ALL CHECKS PASS'' while the debt waterfall contains \#NAME? errors.

\item \textbf{Loose cell formatting} (both harnesses).
Cross-sheet formulas appear in black despite the workbook's own green-link legend or switch between green and black; in Claude Cowork, numbers are often centered and titles merged.
For example, in the task \texttt{FlightPlan} (Claude Cowork, task 24, attempt 3375), each shortest-path sheet uses eight merged title ranges instead of Center Across Selection, and in the task \texttt{GermanLife} (Claude for Excel, task 30, attempt 2092), cross-sheet formulas are black on one sheet but green on the summary and tax sheets.

\item \textbf{Columns too narrow} (Claude Cowork).
Number and date columns are sometimes narrower than their content, so summary results, reconciliation values, or an entire date column display as \#\#\#, more often than for any other agent.
A case in point is the task \texttt{Bosch} (Claude Cowork, task 3, attempt 3290), where a column that is too narrow makes four summary outputs display as \#\#\#.

\item \textbf{Blocks below ``End Sheet'' markers} (Claude Cowork).
The model adds new model-control, scenario, and derived-driver blocks below the ``End Sheet'' marker of the inherited assumptions sheet instead of placing them in the declared structure of the sheet or on a separate sheet.
In the task \texttt{Office} (Claude Cowork, task 40, attempt 3636), for instance, the case and scenario controls and the anchor date follow the marker in row 121.

\item \textbf{Unfinished navigation, print setup, and formula notes} (Claude for Excel).
The table of contents on the summary sheet is plain text without hyperlinks, print areas are often undefined, and non-obvious logic, such as nested revolver waterfalls, array reconciliations, and nested lookups, often carries no explanation.
In the task \texttt{QuarterlyRatios} (Claude for Excel, task 47, attempt 2258), all eight sheets are listed, but none is linked; in the task \texttt{CoastalAggregates} (Claude for Excel, task 8, attempt 2020), no sheet has a print area; and in the task \texttt{DebtFlow} (Claude for Excel, task 15, attempt 2038), the revolver draw-and-repay and vesting-offset formulas are left unexplained.
\end{itemize}

\paragraph{Model vs.\ harness:}
The accurate rounding notes, the unreliable check flags, and the loose formatting stand out in both harnesses and belong to the model; that Fable~5.1 mislabels rounding in the same two harnesses underlines this.
Several differences are observed to follow the harness:
the sense checks, named drivers, and buffer rows that stand out in Claude Cowork are rare in the Excel add-ins, including Opus~5 in Claude for Excel;
its anchored timelines are also rare in Claude for Excel, although GPT-5.6 Sol builds such timelines in ChatGPT for Excel;
and its formula notes thin out there, as they do for Fable~5.1; its compact engines in Claude for Excel likewise mirror those of Fable~5.1 in that harness.

\subsection{GPT}
\subsubsection{GPT-6 Astra (ChatGPT Work, Codex)}
\label{app:model_gpt6astra}
In both harnesses, GPT-6 Astra builds polished, print-ready, well-documented workbooks with a working master error flag, but it never names its key inputs, places its own blocks below inherited ``End Sheet'' markers, and tests its outputs only against the workbook itself.

\paragraph{Good:}
\begin{itemize}
\item \textbf{Print-ready page setup} (both harnesses).
The model defines print areas on every tab and places manual page breaks at section or table boundaries on multi-page sheets.
The task \texttt{LogisticsCategorization} (ChatGPT Work, task 34, attempt 1335) illustrates this: every multi-page sheet has manual page breaks, and so do the long sheets in the task \texttt{8Options} (Codex, task 69, attempt 1748).

\item \textbf{Workbooks with source citations, notes, and a linked table of contents} (both harnesses).
Inputs cite their source sheet or cell, the model's own judgment calls carry an owner and a date, non-obvious formula blocks, such as WACC iteration, interpolation, asset vintages, and tax timing, carry cell notes or nearby methodology notes, and the table of contents holds working hyperlinks to every sheet.
For example, in the task \texttt{FiveAutomotive} (ChatGPT Work, task 22, attempt 1413), block notes and cell comments explain the WACC iteration, interpolation, debt, and asset-vintage blocks, and in the task \texttt{TwoNOne} (Codex, task 101, attempt 1736), notes cover debt timing, the 30/360 convention, tax, the waterfall, and the terminal value, and the contents tab links to every sheet.

\item \textbf{Master error flag wired to the checks} (both harnesses).
Every workbook gets a master error flag on the summary sheet that is linked to the roll-up of the check sheet and conditionally formatted to turn red on failure.
This is evident in the task \texttt{TwoNOne} (ChatGPT Work, task 101, attempt 1545), where the flag is tied to the count of failed checks, and in the task \texttt{CashConversionLab} (Codex, task 6, attempt 1842), where the formula-linked flag turns red on failure.

\item \textbf{Formulas that hold up under stress} (ChatGPT Work).
The formulas usually guard the edge cases, such as zero rates, schedule capacity, whole payment counts, selector validity, and zero-machine or negative-cash states, so plausible input changes rarely produce \#DIV/0!, \#N/A, or impossible balances.
In the task \texttt{LargeBalance} (ChatGPT Work, task 87, attempt 1593), no break was found when the duration and working-capital drivers were stressed.

\item \textbf{Enforced validation and one house style} (Codex).
The model puts validation with active error alerts and sensible bounds on key dates, rates, counts, and selectors, and it gives all built sheets one font and one title, header, and total style, with blue inputs, black formulas, green cross-sheet links, and negatives in parentheses.
In the task \texttt{ApfelInc} (Codex, task 1, attempt 1831), for instance, the key editable drivers carry validation with active alerts, and in the task \texttt{NextEraEnergy} (Codex, task 39, attempt 1719), the font colors follow the convention on every built sheet.

\item \textbf{Timelines that roll from one input} (Codex).
The model prebuilds the full required horizon from a central anchor date and period controls, so dates and periods roll forward without edits to the formulas.
A case in point is the task \texttt{OnlyOperational} (Codex, task 94, attempt 1826), where all dates of the 14-year timeline follow the central drivers, with no need for manual extensions to change the horizon.
\end{itemize}

\paragraph{Bad:}
\begin{itemize}
\item \textbf{No named ranges} (both harnesses).
The model never defines names, so global drivers, such as anchor dates, horizons, tax rates, WACC, and case selectors, are referenced only by cell address.
ChatGPT Work also adds data validation to key inputs but switches the error alerts off, so invalid entries pass without warning.
For example, in the task \texttt{ThreeCars} (task 65), both ChatGPT Work (attempt 1406) and Codex (attempt 1755) reference the valuation date, the tax rate, and WACC only by cell address.

\item \textbf{Content below ``End Sheet'' markers} (both harnesses).
The model appends its own controls, inputs, derived drivers, and notes below the ``End Sheet'' marker of the inherited assumptions sheet, so live inputs sit outside the sheet's declared structure; ChatGPT Work also stacks second timelines or date-helper series onto sheets that already have a main timeline.
The task \texttt{DoctorvsTrade} (task 18) illustrates this: ChatGPT Work (attempt 1403) continues live inputs and interpretation notes past the marker, and Codex (attempt 1845) adds 42 rows of controls and derived timing after it.

\item \textbf{Sense checks that never look outside} (both harnesses).
The model builds sense-check blocks, but these compare the outputs only with figures from the same workbook or with generic arithmetic bounds and often state outright that there is no external benchmark.
This is evident in the task \texttt{QuarterlyRatios} (ChatGPT Work, task 47, attempt 1468), where only internal break-even checks are run, next to a note that there are no external benchmarks, and in the task \texttt{EasyRev} (ChatGPT Work, task 81, attempt 1565), where the ``same-sample benchmark'' is another forecast built from the same inputs.

\item \textbf{``Rounded to'' labels for display formats} (ChatGPT Work).
Headers and notes often state that values are ``rounded to 0.01'' or ``to 0.1\%'', while the formulas keep full precision and only the displayed precision is as stated.
In the task \texttt{CFQuarterly} (ChatGPT Work, task 74, attempt 1425), for instance, a ``rounded to \$0.01mm'' label sits above thousands of formulas without ROUND.

\item \textbf{Oversized, formula-heavy files} (Codex).
The model prefills huge capacity grids with thousands of inactive formula rows or columns and adds duplicate copies of source data, so even small models end up relatively large and carry a heavy recalculation load.
In the task \texttt{LeaseorKeys} (Codex, task 32, attempt 1817), about 122,000 formulas are copied across 1,200 columns, and the checks contain very large ISERROR arrays.
\end{itemize}

\paragraph{Model vs.\ harness:}
The page setup, documentation, and master error flag, as well as the missing names, the blocks below ``End Sheet'' markers, and the inward-looking sense checks, stand out in both harnesses and belong to the model; ChatGPT Pro, running GPT-6, shows most of them as well.
Where the two harnesses are observed to differ, the difference concerns how strictly the model applies conventions:
in Codex, it keeps validation alerts on, describes rounding correctly, and prebuilds its timelines from one anchor, whereas in ChatGPT Work, it switches the alerts off, labels formatted figures as rounded, and sometimes fixes the length of its timelines.
In Codex, the prebuilt timelines come at the price of oversized files.
The robust formulas of ChatGPT Work and the house style of Codex, by contrast, stand out in one harness only.

\subsubsection{GPT-6 (ChatGPT Pro)}
\label{app:model_gpt6}
GPT-6 in ChatGPT Pro is a disciplined, convention-heavy builder that leads the field on flexible timelines, honest rounding statements, and a uniform house font, but it overbuilds its workbooks with brute-force grids and duplicated engines, skips freeze panes, and rarely tests its outputs against outside benchmarks.
Its page breaks, blocks below ``End Sheet'' markers, and inward-looking sense checks also mark GPT-6 Astra in both of its harnesses, which points to the model family rather than the harness.

\paragraph{Good:}
\begin{itemize}
\item \textbf{Flexible timelines.}
The model builds the full required horizon up front and drives it from one central anchor date and period inputs, so that changing the forecast length within that horizon requires no manual column copying or range edits.
For example, in the task \texttt{OnlyOperational} (task 94, attempt 2644), the 2030--2043 timeline follows the central date and period controls, and in the task \texttt{SingleForever} (task 53, attempt 2428), the formula-driven timeline responds to the horizon inputs without edits.

\item \textbf{Visible master error flag.}
The model almost always rolls every check up into one master pass/fail cell on the summary, linked to the check sheet and conditionally formatted to turn red on failure.
A case in point is the task \texttt{CashConversionLab} (task 6, attempt 2019), where the master flag is conditionally formatted.

\item \textbf{Honest rounding statements.}
The model keeps display precision separate from actual rounding:
calculation sheets are labeled ``shown to \$0.01'' and keep full-precision formulas, and only the final answer block applies ROUND.
The task \texttt{NestQuest} (task 38, attempt 2333) illustrates this: full-precision calculations, the ``shown to \$0.01'' display, and the explicitly rounded answers are kept distinct.

\item \textbf{Uniform, print-ready sheets.}
The model restyles every sheet it builds to a single font, Arial~10, with matching section bands, headers, and total styles, and it places manual page breaks at section or period boundaries on sheets that print to several pages.
In the task \texttt{ConstantFlow} (task 9, attempt 2036), for instance, all built sheets use Arial~10 throughout, and in the task \texttt{TheInterestGame} (task 60, attempt 2505), the multi-page sheets carry manual page breaks.

\item \textbf{Sourced assumptions and annotated formulas.}
The model adds a source column that cites the input workbook for each input, tags its own judgment calls with an owner and a date, and places methodology notes beside non-obvious blocks, such as WACC iteration, depreciation vintages, and cash-tax timing.
This is evident in the task \texttt{FruitJuice} (task 28, attempt 2237), where every input on the assumptions sheet carries a citation, and every judgment an owner and a date.
\end{itemize}

\paragraph{Bad:}
\begin{itemize}
\item \textbf{Overbuilt, bloated workbooks.}
The model tends toward brute force:
for small cases, it creates reserve calculation grids with more than a thousand rows, duplicates entire calculation modules instead of parameterizing a single one, expands logic into replicated matrices, and adds duplicate register and appendix blocks, resulting in inefficient workbooks that are difficult to review and modify.
For example, in the task \texttt{CorpBond} (task 10, attempt 2046), tens of thousands of reserve-grid formulas serve a case with 40 payments, and in the task \texttt{CashNiagara} (task 7, attempt 2039), the entire waterfall engine is duplicated (17,521 formulas) where one selector-driven engine would be sufficient.

\item \textbf{No sense checks against outside benchmarks.}
The model's sense-check sections contain only internal reconciliations, comparisons with the workbook's own historical figures, or trivial bounds.
In the task \texttt{EasyRev} (task 81, attempt 2610), the error metrics MAE and MAPE are checked only against another forecast built in the same workbook.

\item \textbf{Missing freeze panes.}
Very wide monthly schedules and long check and summary sheets usually ship without freeze panes, so row labels and period headers scroll out of view; when the model does freeze panes, it sometimes locks an excessive block of rows.
This is evident in the task \texttt{PastaInc} (task 41, attempt 2406), where the integrated statements span 149 columns with no frozen labels or headers, and in the task \texttt{CashConversionLab} (task 6, attempt 2019), where the frozen rows take up most of a regular screen view, leaving little room to see the actual financial model.

\item \textbf{Content below ``End Sheet'' markers.}
When extending an inherited assumptions sheet, the model keeps its ``End Sheet'' marker and appends new controls, drivers, and even live calculation blocks below it, so that this content sits outside the sheet's declared structure.
In the task \texttt{LeaseorKeys} (task 32, attempt 2319), live controls and calculations follow the inherited marker.

\item \textbf{Erratic row heights and misaligned text.}
The model occasionally generates tables whose body rows alternate between many custom heights, with tall detail rows, narrow spacer rows, and very tall outliers, and it sometimes aligns narrative text and labels to the right or centers them.
For example, in the task \texttt{MonthlyLiquidity} (task 37, attempt 2365), body rows range from 18 to 62 points, with one row at 310 points, and in the task \texttt{DoctorvsTrade} (task 18, attempt 2077), the labels and narrative text on the summary are centered instead of left-aligned.
\end{itemize}

\subsubsection{GPT-5.6 Sol (ChatGPT for Excel)}
\label{app:model_gpt56sol}
GPT-5.6 Sol in ChatGPT for Excel is a tidy, well-scaffolded builder that keeps its columns readable, sets print areas, and drives its timelines from start-date inputs, but it slips on several conventions:
rounding labels, font colors, buffer rows above totals, a glossary, and lean schedules.
Its drifting font colors and totals without buffer rows also mark both Claude for Excel agents (Fable~5.1 and Opus~5) and are thus likely effects of the Excel add-in harness; its readable column widths and complete print areas set it apart from them.

\paragraph{Good:}
\begin{itemize}
\item \textbf{Readable widths and print areas.}
The model wraps long narrative text instead of stretching columns, so that its own sheets stay within a readable width, and it usually defines a print area on every tab.
This is evident in the task \texttt{EasyDCF} (task 20, attempt 1324), where no built column is excessively wide, and all long text is wrapped.

\item \textbf{Timelines driven by start-date inputs.}
The model usually builds the full required horizon in advance and drives period labels and dates from a central start-date and period input, so that the timeline can be rolled forward without edits to the formulas.
In the task \texttt{DebtFlow} (task 15, attempt 1313), for instance, the full 30-year horizon is prebuilt with labels that follow the drivers.

\item \textbf{Complete scaffold with a live summary.}
The model reliably delivers the full workbook skeleton:
a summary as the first tab that doubles as a cover with owner, version, and date, followed by a table of contents, a check sheet, and a change log, and the figures on the summary are live formulas.
The task \texttt{CashConversionLab} (task 6, attempt 1280) illustrates this: the values and check status on the summary link to the assumptions, working-capital, and check sheets.
\end{itemize}

\paragraph{Bad:}
\begin{itemize}
\item \textbf{Wrong rounding notes.}
The model routinely labels sheets as ``rounded to \$0.01'' although the formulas keep full precision and only the display format is limited; on some tasks, it also wraps intermediate values in ROUND before they feed later calculations.
For example, in the task \texttt{BasicGrowth} (task 2, attempt 1267), sheets are labeled ``rounded to \$0.01'', yet none of their roughly 9,900 formulas uses a rounding function.

\item \textbf{Unreliable font colors.}
In a fair number of tasks, the model colors same-sheet calculations green, the color its own legend reserves for cross-sheet links, and mixes black and green for the same kind of formula, so that the font color no longer reveals to the reader what type of cell it is.
A case in point is the task \texttt{RevForecast} (task 49, attempt 1390), where same-sheet calculations are green in three columns and black in a fourth.

\item \textbf{No buffer rows above totals.}
The model puts totals directly below the last line item with fixed sum ranges, so that a row inserted above a total falls outside the sum.
In the task \texttt{Office} (task 40, attempt 1366), total revenues and gross asset value sit directly below fixed SUM ranges, although every other agent satisfied this criterion on the task.

\item \textbf{Oversized schedules that recalculate slowly.}
As the flip side of its prebuilt timelines, schedules run far past the active horizon, with thousands of mostly blank formula rows that SUMIFS, lookups, or whole-column references then rescan, which makes the recalculation load avoidably heavy.
In the task \texttt{PaydownPathways} (task 42, attempt 1368), for instance, a schedule of 5,200 periods, only 180 of them active, is rescanned by every yearly SUMIFS, LOOKUP, and AVERAGEIFS row.

\item \textbf{No glossary of conventions.}
The cover and summary explain scope and usage but sometimes give no glossary for the color conventions and abbreviations, such as FCF, EAR, and \$mm, although the house standards require one.
For example, in the task \texttt{SmallRiver} (task 98, attempt 1546), neither the blue/black/green convention nor abbreviations such as FCF, EAR, and \$mm are explained.
\end{itemize}

\subsection{SpaceXAI}
\subsubsection{Grok~4.6 (Codex)}
\label{app:model_grok}
Grok~4.6 in Codex is a structurally tidy builder that names its key inputs, freezes its long sheets, and documents rounding carefully, but it clutters its sheets with merged cells, centered numbers, and boxed grids, fills its summary tabs with hardcoded numbers, and occasionally ships workbooks broken by \#NAME? errors.

\paragraph{Good:}
\begin{itemize}
\item \textbf{Named key inputs.}
The model almost always gives key global inputs, such as dates, rates, case selectors, and unit conversions, short and descriptive defined names.
This is evident in the task \texttt{EasyDCF} (task 20, attempt 2353), where the Name Manager holds a full set of named DCF inputs, among them \texttt{ValDate}, \texttt{Rf}, \texttt{Beta}, \texttt{MRP}, and \texttt{TaxRate}.

\item \textbf{Freeze panes on long sheets.}
The model keeps titles, column headers, and period labels frozen at sensible positions on long calculation sheets.
The task \texttt{Proportionality} (task 46, attempt 2377) illustrates this: the headers and labels of the long documentation and calculation sheets are frozen, and none of the freezes is excessive.

\item \textbf{Honest rounding documentation.}
The model's notes separate figures that are only ``shown to'' a precision from answers that are actually rounded with ROUND, and they sit next to the outputs.
A case in point is the task \texttt{CashNiagara} (task 7, attempt 2259), where the annual sheet reads ``shown to \$0.01m, full precision held''.
\end{itemize}

\paragraph{Bad:}
\begin{itemize}
\item \textbf{Merged cells and centered blocks.}
The model routinely merges sheet titles, section headers, and whole cover or summary blocks instead of using Center Across Selection, although the house standards prohibit merged cells, and it centers whole blocks of numbers in statements, schedules, and ratio tables as well as body paragraphs on the cover and summary.
In the task \texttt{CorpBond} (task 10, attempt 2387), the cover contains 30 merged ranges and the summary 36, and in the task \texttt{RoadtoZero} (task 50, attempt 2329), long scope and methodology paragraphs are centered.

\item \textbf{Boxed grids and oversized rows.}
The model inserts thin borders on all four sides of nearly every cell within large calculation blocks, which duplicates the gridlines and creates a dense grid, and it leaves the narrative and header rows much taller than necessary, some even several times the regular row height.
For example, in the task \texttt{MarketBalanced} (task 35, attempt 2180), virtually every cell of the calculation engine is boxed, and in the task \texttt{ConstantFlow} (task 9, attempt 2249), a single row is 359.7 points tall.

\item \textbf{Summaries with hardcoded numbers.}
The model often fills the summary tab with typed snapshot figures without formulas, so the summary never updates when inputs change.
In the task \texttt{Bosch} (task 3, attempt 2088), for instance, the summary contains no formulas, and sales, margins, and balance-sheet figures are typed snapshots.

\item \textbf{Deeply nested formulas.}
The model writes long, monolithic formulas that nest IF, INDEX/MATCH, or SMALL chains four or more levels deep, notably in hand-built percentile interpolation and growth formulas, and leaves them as single blocks even where helper rows already exist.
This is evident in the task \texttt{TMobile} (task 57, attempt 2354), where a monolithic percentile formula chains IF, OR, ABS, INT, and SMALL, although helper rows exist.
\end{itemize}

\subsection{Moonshot AI}
\subsubsection{Kimi~K3 (Codex)}
\label{app:model_kimi}
Kimi~K3 in Codex is an agent that gets parts of the workbook plumbing right, with working table-of-contents links, clean sheet separation, honest rounding notes, and compact rows, but unlike the frontier agents, it ships hardcoded summary tabs and unsourced assumptions; it also builds hard-wired timelines and leaves inherited clutter and half-applied style rules behind.

\paragraph{Good:}
\begin{itemize}
\item \textbf{Honest rounding notes.}
The notes on the cover and sheets describe precision correctly:
intermediate values carry full precision, most figures are only displayed to two decimals, and ROUND is applied to the answer cells alone.
The task \texttt{TBondII} (task 56, attempt 2714) illustrates this: a note states that values are displayed to \$0.01 and only the answers are rounded.

\item \textbf{Working table of contents.}
The model puts HYPERLINK formulas to every worksheet on the cover or a separate contents tab.
For example, in the task \texttt{Bosch} (task 3, attempt 2085), functional HYPERLINK formulas open every sheet.

\item \textbf{Clean functional sheet separation.}
The model keeps the cover, inputs, calculations, outputs, checks, and answers on separate sheets or in clearly labeled blocks and does not append content below an inherited ``End Sheet'' marker.
A case in point is the task \texttt{CashNiagara} (task 7, attempt 2662), where each function has its own sheet.

\item \textbf{Compact rows.}
The model keeps the rows it builds at a sensible height; the only tall rows are inherited instruction rows.
In the task \texttt{CashNiagara} (task 7, attempt 2662), for instance, only the inherited instruction row reaches about 60 points; all other rows are lower.
\end{itemize}

\paragraph{Bad:}
\begin{itemize}
\item \textbf{Static summary tab.}
The tab named ``Summary'' is often typed narrative or a build plan without formulas, sometimes with input figures written into the prose, so nothing on it updates when inputs change.
In the task \texttt{DailyCash} (task 13, attempt 2633), the summary consists only of hardcoded constants.

\item \textbf{Unsourced assumptions.}
Input blocks, and assumptions the agent adds itself, such as benchmarks, screening bands, forecast periods, and growth drivers, often carry no source citation and no judgment label with an owner and a date.
The task \texttt{TheLiquidityEngine} (task 61, attempt 2972) illustrates this: the forecast-period, growth, averaging, and link assumptions have neither a source nor an owner or date.

\item \textbf{Hard-wired timelines.}
Timelines are often built from typed start dates, embedded years, constant year indices, or a fixed number of copied columns instead of flowing from one date or horizon driver, so rolling the model forward or changing its length requires manual edits.
For example, in the task \texttt{PastaInc} (task 41, attempt 2596), a hardcoded first date is copied across 145 fixed monthly columns, and in the task \texttt{ConstantFlow} (task 9, attempt 2683), the engine is filled to exactly 300 columns and does not size itself from the model-length driver on the assumptions sheet.

\item \textbf{Own style conventions not applied.}
The model documents green as the color for cross-sheet links but leaves whole blocks of cross-sheet formulas in black, and it uses number formats that show negatives with a minus sign rather than in parentheses.
This is evident in the task \texttt{OperationsWeekly} (task 95, attempt 2757), where whole aggregation blocks that link to the daily cash-flow sheet are black instead of the documented green.

\item \textbf{No freeze panes on large sheets.}
Long daily engines and very wide schedules usually have no frozen headers or label columns, and in one task, Kimi~K3 even removed freeze panes that the input workbook already had.
In the task \texttt{LogisticsCategorization} (task 34, attempt 2820), engines of 14,601 rows have no frozen header row.
\end{itemize}

\subsection{Google}
\subsubsection{Gemini~3.8 Flash (Codex)}
\label{app:model_gemini}
Gemini~3.8 Flash in Codex is a lean, bare-bones builder whose files stay compact and sensibly sized.
However, it ships static summaries, rigid hardcoded timelines, a master error flag that can never turn red, builder-only documentation, unsourced assumptions, and unfrozen, over-boxed sheets.

\paragraph{Good:}
\begin{itemize}
\item \textbf{Sensible column widths and row heights.}
The model's columns have a sensible width and its rows a normal height.
In the task \texttt{DailyDetail} (task 77, attempt 2104), for instance, no built column is excessively wide.

\item \textbf{Proportionate builds.}
The model sizes its schedules to the base case and avoids brute-force replication, such as copied WACC iteration blocks or thousands of prefilled spare rows, so its files stay proportionate to their content; the flip side is the rigid timelines listed below.
A case in point is the task \texttt{SpendingPlan} (task 54, attempt 2095), where the workbook takes only 488~KB for a scenario dataset of 7,500 rows.

\item \textbf{Totals clearly set off.}
Totals, subtotals, and closing balances are consistently bold, often with shading or double bottom borders.
For example, in the task \texttt{DetailedLoan} (task 16, attempt 2113), totals and ending balances are bold and often shaded.
\end{itemize}

\paragraph{Bad:}
\begin{itemize}
\item \textbf{Rigid, hardcoded timelines and drivers.}
Forecast years are typed labels and schedules are filled to a fixed length, so changing the start date, horizon, or payment frequency requires manual rework, and event months, thresholds, and weights are typed into formulas instead of being linked to a driver.
The task \texttt{DebtDeclineDynamics} (task 14, attempt 2119) illustrates this: the engine is fixed at 60 periods and does not resize when the frequency or duration changes, and in the task \texttt{Fixings} (task 23, attempt 2112), the years 2026--2035 are typed labels, and the period count on the assumptions sheet does not drive the schedule.

\item \textbf{Static summaries and check flags.}
The summary or cover is often a static build plan or narrative with almost no formulas, in which headline outputs, and even case inputs and dates, are retyped as constants; the master error flag on the check sheet has a static fill without conditional formatting, so a failing check can never trigger a visual alarm.
This is evident in the task \texttt{Proportionality} (task 46, attempt 2158), where the summary holds 61 constants and no formulas, and in the task \texttt{Bosch} (task 3, attempt 2082), where the master flag has a static blue fill and no conditional formatting rule.

\item \textbf{Formulas that break under plausible stress.}
Tax-rate, growth, and return formulas divide without a guard, so a plausible stress, such as pretax or net income falling to zero, throws \#DIV/0! into the valuation, and some logic is pinned to fixed levels and misbehaves when an input is halved.
In the task \texttt{EasyDCF} (task 20, attempt 2181), for instance, the unguarded ratio of tax to pretax income returns \#DIV/0! when pretax income is stressed to zero, and the error flows into the average tax rate and the valuation.

\item \textbf{Thin documentation and unsourced assumptions.}
The notes describe how the model was built, but not how to use it:
there is almost never a guide to which inputs are safe to edit or where to read the outputs, scope notes omit exclusions, complex nested formulas carry no comments, and assumption blocks carry no source, owner, or date, even for the model's own judgment calls.
In the task \texttt{BasicGrowth} (task 2, attempt 2141), the navigation section only lists the tabs, and nothing explains which inputs are editable or how to use the compounding selector, and in the task \texttt{BreakingDowntheBalance} (task 4, attempt 2147), the input block has no source field and no owner or date for any driver.

\item \textbf{Unfrozen, over-boxed, inconsistent sheets.}
The model almost never freezes panes or sets print areas, even on very long or wide calculation sheets, boxes every cell of long schedules in thin borders, switches fonts between its own Arial and the inherited Aptos Narrow, and is loose on color conventions.
For example, in the task \texttt{GermanLife} (task 30, attempt 2125), a calculation sheet of about 155 rows by 80 columns has no frozen panes.
\end{itemize}

\section{Finance Glossary}
\label{app:Glossary}
This appendix defines the finance and spreadsheet terms used in this paper for readers without a finance background.
Financial model types, such as DCF valuations or 3-statement models, and the fields of finance they belong to are described in Appendix~\ref{app:TaskTypes} and are not repeated here.

\begin{classbox}{Fields of Finance}{Fields of finance named in the paper beyond those in Appendix~\ref{app:TaskTypes}}

\classterm{Commercial banking}{Granting loans to and accepting deposits from businesses and individuals; these models include loan plans and credit analyses.}

\classterm{Corporate finance}{The management of a company's finances: its investment, financing, and dividend decisions.}

\classterm{Investment banking}{Advising companies on mergers, acquisitions, and capital raising, with valuation and transaction models playing a central role.}

\classterm{Private equity}{Investments in privately held companies, often as part of leveraged buyouts.}

\classterm{Wealth management}{Financial planning and investment advice for individuals and households, for example, with regard to retirement or home ownership.}

\end{classbox}

\begin{classbox}{Financial Statements and Metrics}{Accounting terms and key financial figures}

\classterm{Actuals, forecasts, and period labels}{Actuals are historical, reported figures; forecasts are projections calculated from assumptions over the forecast horizon, that is, the time period covered by a model.
Period labels indicate the type of period with a suffix, for example FY2024A (actual), FY2025E (estimate), or Q1-26B (budget).}

\classterm{Annual report (10-K)}{The annual financial report that publicly traded U.S. companies file with the Securities and Exchange Commission; a typical source of input data for financial models.}

\classterm{Capital expenditure (capex)}{Expenditures for long-lived assets, such as property, plant, and equipment (PP\&E), which are capitalized, that is, recognized as assets and depreciated over time, rather than expensed immediately.}

\classterm{Consolidation}{Combining the financial statements of several business units or companies into a single set of statements, eliminating the transactions between them (inter-company eliminations).}

\classterm{EBITDA}{Earnings before interest, taxes, depreciation, and amortization; a common measure of operating profitability.
Depreciation and amortization spread the costs of long-lived assets over their useful lives.}

\classterm{Financial statements}{The income statement (revenue, expenses, and profit over a period), the balance sheet (assets, liabilities, and equity at a point in time), and the cash flow statement (cash inflows and outflows over a period).}

\classterm{Fiscal year}{The 12-month period that a company uses for its financial statements, which does not necessarily coincide with the calendar year (e.g., July through June).}

\classterm{Free cash flow (FCF)}{The cash a company generates after taxes, capital expenditures, and investments in working capital; it is available to the company's investors.}

\classterm{Working capital}{Funds tied up in day-to-day operations, primarily accounts receivable and inventory, net of accounts payable.}

\end{classbox}

\begin{classbox}{Valuation and Financing}{Valuation methods, financing terms, and period conventions}

\classterm{Day-count convention}{The rule for counting days when calculating interest or discounting over a partial period, such as act/365 (actual days in a 365-day year) or 30/360 (30-day months in a 360-day year).}

\classterm{Deal models}{Models for transactions, not covered by \benchmarkname (Figure~\ref{fig:coverage}): a leveraged buyout (LBO) model for a predominantly debt-financed acquisition; a merger model with an accretion/dilution analysis, which shows how an acquisition affects the acquirer's earnings per share; and precedent transactions, in which a company is valued based on the prices paid in similar past transactions.}

\classterm{Deal vs.\ planning model}{A model designed to evaluate a one-time transaction, as opposed to a model that is used repeatedly for ongoing planning.}

\classterm{Enterprise value and multiples}{Enterprise value is the value of a company's business operations to all of its investors.
Multiples such as EV/EBITDA or the price-to-earnings ratio (P/E) relate a value to a financial metric and are used to compare similar companies.}

\classterm{Mid-year convention}{Discounting the cash flows of a period as if they occurred in the middle of the period rather than at its end.}

\classterm{NPV and IRR}{The net present value is the sum of an investment's discounted cash flows, including the initial investment; the internal rate of return is the discount rate at which the net present value is zero.}

\classterm{Revolver and cash sweep}{A revolving credit facility is a line of credit that a company draws on and repays as needed; with a cash sweep, excess cash is automatically used to repay debt.}

\classterm{Roll-forward (corkscrew)}{A schedule for a balance such as debt or cash: the opening balance plus additions minus reductions equals the closing balance, which becomes the opening balance of the next period.}

\classterm{Stub period}{A partial period at the beginning or end of a forecast, for example when the valuation date falls within a fiscal year; its length is determined with a day-count convention.}

\classterm{Tranche}{A separate portion of a company's debt with its own terms, such as interest rate, maturity, and priority in repayment.}

\classterm{WACC}{Weighted average cost of capital: the combined rate of return required by equity and debt investors, which is used as the discount rate in a DCF valuation.}

\end{classbox}

\begin{classbox}{Spreadsheet Modeling Practice}{Spreadsheet conventions and Excel features}

\classterm{Buffer row}{An empty row above a total or between sections, so that new line items can be inserted without breaking the sum formulas.}

\classterm{Cases and switches}{A case is a consistent set of assumptions, such as a base case or a worst case; a switch or selector is an input cell that selects the active case or turns a model feature on or off, for example whether expenditures are capitalized or expensed.
With a switch, a single parameterized calculation engine serves all cases instead of one copy of the engine per case.}

\classterm{Cell references and anchoring}{In a reference such as \texttt{\$B\$4}, the \$ signs fix the column and the row when a formula is copied (an absolute reference); if only one of the two is fixed, the reference is mixed.
If references are anchored incorrectly, copied formulas silently refer to the wrong cells.}

\classterm{Circular reference}{A formula that depends directly or indirectly on its own result, for example interest calculated on an average debt balance, which in turn depends on the interest paid.
Excel can resolve it through iterative calculation; a breaker switch turns the circularity off for debugging.}

\classterm{Color coding}{The industry-standard convention of showing hardcoded inputs in blue, calculations in black, links to other sheets in green, and links to other workbooks in red.}

\classterm{Cover sheet}{The first sheet of a workbook, which introduces the model: its title, purpose, scope, and conventions.}

\classterm{Data validation and conditional formatting}{Excel features that restrict what can be entered in a cell and that format cells automatically based on their values.}

\classterm{Error checks}{Formulas that test the integrity of a model, for example whether the balance sheet balances or whether the cash balance matches the cash flow statement; a master error flag shows in one cell whether any check has failed.}

\classterm{Error values}{Results such as \#REF! (invalid reference), \#N/A (value not available), or \#DIV/0! (division by zero) that Excel displays when a formula cannot be evaluated.}

\classterm{Freeze panes, print areas, and page breaks}{Settings that keep the headers visible while scrolling and that specify which cells are printed and where pages break.}

\classterm{Grouping}{Collapsing rows or columns behind an outline button; unlike hiding, grouping leaves a visible indication that content is there.}

\classterm{Hardcoding}{Entering a fixed value instead of a formula or cell reference.
Hardcoded inputs are allowed as long as they are marked as such; numbers entered directly into formulas or calculation cells are an error, since they cannot be traced back to or updated from the inputs.}

\classterm{Inputs, assumptions, and drivers}{Inputs and assumptions are values that are entered directly rather than calculated, such as a growth rate or a tax rate; a driver is an input on which many calculations depend, such as the valuation date, that is, the date as of which values are measured.}

\classterm{Merged cells}{Cells merged into a single cell, which interferes with formulas, sorting, and references; Center Across Selection achieves the same look without merging the cells.}

\classterm{Named range}{A descriptive name assigned to a cell or range (e.g., \texttt{TaxRate}) that formulas can use instead of a cell address.}

\classterm{Sense check}{Comparing key results with external reference points, such as industry benchmarks or comparable companies, to identify implausible results.}

\classterm{Sign convention}{A consistent rule for signs, for example costs and outflows negative, revenues and inflows positive.}

\classterm{Units and currency}{The scale and currency in which amounts are stated, for example \$mm for millions of U.S. dollars.}

\classterm{VBA and macros}{Visual Basic for Applications, Excel's programming language, and the automated routines written in it.}

\classterm{Volatile functions}{Functions such as OFFSET, INDIRECT, or NOW that recalculate on every change in the workbook, which slows down large models.}

\end{classbox}